\documentclass[structabstract]{aa}  
\usepackage{graphicx}
\usepackage{txfonts}
\usepackage{lscape}
\usepackage{natbib}
\usepackage{rotating}
\usepackage{txfonts}

\def\ie{i.e.}
\def\eg{e.g.}
\def\deg{\ifmmode^\circ\else$^\circ$\fi}
\def\Msun{\ifmmode{\mathcal{M}_\odot}\else{$\mathcal{M}_\odot$}\fi}
\def\Lsun{\ifmmode{\mathrm{L}_\odot}\else{L$_\odot$}\fi}
\def\Massstar{\ifmmode{\mathcal{M}_{\ast}}\else{$\mathcal{M}_\ast$}\fi}

\def\zf{\ifmmode{z_{\rm f}}\else{$z_{\rm f}$}\fi}
\def\Ks{\ifmmode{K_{\rm s}}\else{$K_{\rm s}$}\fi}
\def\Mcut{\ifmmode{M_{\rm cut}}\else{$M_{\rm cut}$}\fi}
\def\ncmod{{\tt NCMOD\/}}

\begin{document}

\title{On the buildup of massive early-type galaxies at $z\lesssim 1$}
\subtitle{I- Reconciling their hierarchical assembly with mass-downsizing}

\author{M.~Carmen Eliche-Moral\inst{1}, Mercedes Prieto\inst{2,3}, Jes\'{u}s Gallego\inst{1}, Guillermo Barro\inst{1}, Jaime Zamorano\inst{1}, Carlos L\'{o}pez-Sanjuan\inst{2,4}, Marc Balcells\inst{2,5}, Rafael Guzm\'{a}n\inst{6}, and Juan Carlos Mu\~{n}oz-Mateos\inst{1}
 }

\institute{Departamento de Astrof\'{\i}sica, Universidad Complutense de Madrid, E-28040 Madrid, Spain \\ \email{mceliche@fis.ucm.es}
\and
Instituto de Astrof\'{\i}sica de Canarias, C/ V\'{\i}a L\'actea, E-38200 La Laguna, Tenerife, Spain
  \and
Departamento de Astrof\'{\i}sica, Universidad de La Laguna, 
Avda.\ Astrof\'{\i}sico Fco.\ S\'anchez, E-38200 La Laguna (Tenerife), Spain
  \and
Laboratoire d'Astrophysique de Marseille, P\^ole de l'Etoile Site de Ch\^ateau-Gombert 38, F-13388,  Marseille, France
  \and
Isaac Newton Group of Telescopes, Apartado 321, 38700 Santa Cruz de La Palma, Canary Islands, Spain
  \and
Department of Astronomy, 477 Bryant Space Center, University of Florida, Gainesville, FL, USA
  }

   \date{Received April 12, 2010; accepted May 13, 2010}

\abstract
{Several studies have tried to ascertain whether or not the increase in abundance of the early-type galaxies (E-S0a's) with time is mainly due to major mergers, reaching opposite conclusions.} 
{We have tested it directly through semi-analytical modelling, by studying how the massive early-type galaxies with $\log(\mathcal{M}_*/\Msun)>11$ at $z\sim 0$ (mETGs) would have evolved backwards-in-time, under the hypothesis that each major merger gives place to an early-type galaxy.} 
{The study was carried out just considering the major mergers strictly reported by observations at each redshift, and assuming that gas-rich major mergers experience transitory phases as dust-reddened, star-forming galaxies (DSFs).} 
{The model is able to reproduce the observed evolution of the galaxy LFs at $z\lesssim 1$, simultaneously for different rest-frame bands ($B$, $I$, and $K$) and for different selection criteria on color and morphology. It also provides a framework in which apparently-contradictory results on the recent evolution of the luminosity function (LF) of massive, red galaxies can be reconciled, just considering that observational samples of red galaxies can be significantly contaminated by DSFs. The model proves that it is feasible to build up $\sim 50$-60\% of the present-day number density of mETGs at $z\lesssim 1$ through the coordinated action of wet, mixed, and dry major mergers, fulfilling global trends that are in general agreement with mass-downsizing. The bulk of this assembly takes place during $\sim 1$\,Gyr elapsed at $0.8<z<1$, providing a straightforward explanation to the observational fact that redshift $z\sim 0.8$ is a transition epoch in the formation of mETGs. The gas-rich progenitors of these recently-assembled mETGs reproduce naturally the observational excess by a factor of $\sim 4$-5 of late-type galaxies at $0.8<z<1$, as compared to pure luminosity evolution (PLE) models.}
{The model suggests that major mergers have been the main driver for the observational migration of mass from the massive-end of the blue galaxy cloud to that of the red sequence in the last $\sim 8$\,Gyr.}


\titlerunning{Reconciling the hierarchical buildup of mETGs at $z\lesssim 1$ with mass-downsizing}

\authorrunning{Eliche-Moral et al.}

   \maketitle

\section{Introduction}
\label{sec:introduction}

Early-type galaxies (ETGs) have become cornerstones in our understanding of the mass assembly in the Universe, as they dominate the massive end of the galaxy mass function at $z<0.8$, hosting more than a half of the total local stellar content \citep[][]{2004ApJ...608..752B,2010ApJ...709..644I}. Although hierarchical models are the most successful in reproducing the general properties of the Universe at the present \citep[see][]{2000ApJ...528..607N}, several observational results related to the formation of the mETGs call it into question \citep{1996AJ....112..839C,2004Natur.428..625H,2006ApJ...651..120B,2007ApJ...669..947J,2009MNRAS.395L..76D}. 

In particular, hierarchical and primordial-collapse theories of galaxy formation derive extremely-different assembly epochs for the mETGs. Actual $\Lambda$CDM models predict that these systems are the final remnants of the richest merging sequences in the Universe, and thus, the latest to be completely in place into the cosmic scenario (at $z<0.5$), whereas monolithic collapse theories push their formation towards much earlier epochs \citep[at $z>2.5$, see][]{1962ApJ...136..748E,1977egsp.conf..401T,1978MNRAS.183..341W,1993MNRAS.261..921K,2005Natur.435..629S}. In this sense, the observational phenomenon of galaxy mass-downsizing has become a challenge for hierarchical theories. Consisting on that the most massive galaxies ($\mathcal{M}_* > 5\times10^{11}\Msun$) seem to have have been in place since $z\sim 2-3$ \citep[][]{2007MNRAS.377.1717K,2008ApJ...680...41D,2009ApJ...691L..33K}, whereas the less-massive systems get their actual volume densities at later epochs \citep[][]{2004Natur.430..181G,2005ApJ...630...82P,2008ApJ...675..234P,2008A&A...491..713W}, mass-downsizing seems to favour a monolithic collapse origin for mETGs. This scenario is also supported by the substantial amounts of mETGs that are being detected at $z>2$ in the last years \citep[\eg, ][]{1988ApJ...331L..77E,1996ApJ...457..490F,1998Natur.394..241H,2002MNRAS.336.1342W,2004Natur.430..184C,2009A&A...501...15F}, and by the negligible number evolution that several authors report for them up to $z\sim 1.2$ \citep[][]{2003A&A...402..837P,2004cgpc.sympE..38N,2006A&A...453L..29C,2009MNRAS.392..718S,2009MNRAS.395..554F,2009MNRAS.396.1573F}. 

However, there are conflicting views on the amount of this number evolution, as the fraction of mETGs assembled since $z\sim 1$ can vary between $\sim 20$\% and $\sim 60$\%, depending on the author \citep[][F07 hereandafter]{2004ApJ...608..752B,2006A&A...453..809I,2007ApJ...665..265F,2007ApJS..172..406S,2008ApJ...682..919C,2010MNRAS.tmp..534M}. Moreover, the traces of past interactions exhibited by most of these systems and the merger fractions derived up to $z\sim 1$ suggest that major mergers must have contributed significantly to the evolution of mETGs in the last $\sim 8$\,Gyr \citep{2000MNRAS.311..565L,2002ApJ...565..208P,2003AJ....126.1183C,2005AJ....130.2647V,2008ApJ...684.1062F,2009MNRAS.394.1956C}. This is also supported by recent observations that prove the existence of a link between the AGN activity in low-redshift ETGs and a significant merger event in their recent past \citep{2007MNRAS.382.1415S,2010ApJ...714L.108S}.

\citet{2000ApJ...541...95V} showed that ETGs at high-redshift seem to form a homogeneous old population just because the progenitors of the youngest present-day ETGs are usually not included in the samples under study. This effect, known as \emph{progenitor bias}, can extremely affect the conclusions derived on the evolution of the ETGs in those studies that exclude late-type progenitors from their samples. In fact, a pioneering study by \citet{2009A&A...503..445K} has proven that, when the true progenitor set of the present-day ETGs is considered in the standard $\Lambda$CDM framework (regardless of their morphologies), less than 50\% of the stellar mass which ends up in ETGs today is actually in early-type progenitors at $z\sim 1$, in agreement with recent observations. This result means that observations have not established yet how or even when the mETGs formed, nor if their apparently unchanged properties until $z \sim 1$ are really due to their evolution or to observational uncertainties and progenitor bias.

Several studies have tried to ascertain whether or not the increase in abundance of ETGs with time arises primarily from major mergers, deriving opposite conclusions \citep{2006ApJ...648..268B,2007ApJ...665L...5B,2008ApJ...680...41D,2009ApJ...697.1369B,2010arXiv1002.3257C}. Most of these studies use merger trees from $\Lambda$CDM cosmological simulations, and some of them prove the feasibility of reproducing observations within a theoretical $\Lambda$CDM framework \citep[see, \eg,][]{2003ApJ...597L.117K,2005MNRAS.361..369L,2009A&A...503..445K}. But the real contribution of the observed major mergers to the evolution of the high-mass end of the galaxy LFs is still unknown. So, we propose to test the hierarchical origin of mETGs by studying how the population of mETGs would have evolved backwards in time, under the hypothesis that they derive from the major mergers that are strictly reported by observations at each redshift. We intend to test if major mergers are really the main contributors to the buildup of mETGs or if additional processes are required to explain it. This is the first paper of a series summarizing the main results of this model. In the present paper (paper I), we describe the model and compare its predicted LFs with observational data. Paper II analyses in detail the role of the observational major mergers (wet, mixed, and dry) in the buildup of mETGs since $z\sim 1$.

The present paper is organized as follows. Section \S\ref{sec:model} is devoted to the model description. In \S\ref{sec:lf}, we compare the evolution of the LFs predicted by the model for different galaxy types with real data, using different galaxy-selection criteria (on color, spectral type, morphology). The compatibility of the results with the mass-downsizing scenario and with other recent observational results is discussed in \S\ref{sec:discussion}. In this section, we also comment on the model limitations.  Finally, a brief summary of results and some conclusions are addressed in \S\ref{sec:conclusions}. We will use a $\Omega_M = 0.3$, $\Omega_\Lambda = 0.7$, $H_0 = 70$ km s$^{-1}$ Mpc$^{-1}$ concordant cosmology throughout the paper \citep{2000ApJ...528..607N,2007astro.ph..1490X}. All magnitudes are given in the Vega system.

\section{The model}
\label{sec:model}

The model adopts the backwards-in-time technique first introduced by \citet{1980ApJ...241...41T}. It traces back-in-time the evolution of the local galaxy populations considering two different sources of evolution: 1) the number evolution that observational merger rates imply at each redshift, and 2) the typical luminosity evolution (L-evolution) of each galaxy type due to its star formation history (SFH). The evolution of the volume element derived from the assumed cosmology is also considered. All model parameters and assumptions are strictly based on robust observational and computational results. The main novelty of the model is the realistic treatment of the effects of major mergers on the LFs. It considers the different phases and time-scales in a major merger, as stated by observations and N-body simulations. As the fate of a merger remnant depends strongly on the gas-content of the progenitor galaxies, we have also had into account the different properties of these progenitors depending on whether the merger is dry (between gas-poor galaxies), wet (both progenitors are gas-rich), or mixed (between a gas-poor galaxy and a gas-rich one), as well as the relative fraction of each merger type at each redshift according to observations. Two key assumptions relative to major merging are adopted in the model: 1) the fact that a major merger generates an E-S0a galaxy, and 2) that the gas-rich galaxies involved in a major merger experiment transitory phases in which they become DSFs. 

We use observationally-determined merger fractions to account for the number density of merger events at each redshift. The merger fraction at a certain redshift is defined as the number of mergers taking place per galaxy at that redshift. Considering the volume enclosed at the considered redshift interval, we can estimate the number density of major mergers at redshift $z$, $\phi_\mathrm{m}(z)$, as: 

\vspace{-0.2cm}\begin{equation}\label{eq:mergerratevol2}
\begin{array}{rcl}
\phi_\mathrm{m}(z) &=& f_\mathrm{m} (z) \cdot \phi_\mathrm{total}(z)  ,
\end{array}
\end{equation}
\noindent being $f_\mathrm{m} (z)$ the merger fraction and $\phi_\mathrm{total}(z)$ the total number density of galaxies at $z$ (\ie, the total galaxy LF at that redshift).

\subsection{The original \ncmod\/ code}
\label{sec:originalcode}

We have used the code \ncmod, created by \citet[][G98 hereandafter]{1998PASP..110..291G} for generating predictions on galaxy number counts, as a basis for the model. The author kindly made the code available at his web site, and permited its use, modification, and distribution without charge for the purposes of scientific research. Briefly, the code evolves the local LF back-in-time, for a number of galaxy types, using the spectral energy distributions (SEDs) from the Galaxy Isochrone Synthesis Spectral Evolution Library model \citep[GISSEL96, see][]{1993ApJ...405..538B,1996PASP..108..996L}. The code derives a matrix, $N[m(F),z,t]$, containing the number of galaxies at each redshift $z$ (per square degree and apparent magnitude interval) that contributes to the galaxy number counts at apparent magnitude $m$, for a given band $F$, and for the morphological type $t$. K- and e-corrections (evolutive corrections) corresponding to each galaxy type $t$ at each redshift $z$ are considered \citep{1988ApJ...326....1Y}, as well as dust-extinction effects \citep[][see \S\,2.6 by G98 for more details]{1988ApJ...333..673B,1991ApJ...383L..37W}. The original code implementation considers that galaxies do not experience any morphological evolution, so it just estimates the luminosity evolution backwards-in-time of a given present-day morphological type, according to the SFH assumed for this type.

G98 also inserted a procedure to account for number evolution in his model. Its merging procedure simply removes a given fraction of galaxies at each redshift from those galaxy types that the user considers to be affected by mergers. In this regard, the procedure is undoing the remnant galaxies into their original progenitors (we will refer to this as \emph{reversed merging}). This procedure has some problems in the counting method (as shown in \S\ref{sec:countingmethod}) and the merging procedures are too simplistic to account realistically for mergers. In particular, the assumptions that mergers can only occur between galaxies of the same galaxy type, that the remnant must also be of this type, and that the merger is instantaneous are clearly contrary to observations (see assumption \#2 in \S\ref{sec:assumptions}). Moreover, these procedures do not account for the cumulative effects of mergers on the LFs either. So, we have improved the original code to account more realistically for the effects of major mergers in galaxy evolution (see \S\ref{sec:merging}). For more details on the original code, the reader is referred to G98.

\subsection{General improvements to the model}
\label{sec:improvementstothemodel}

We have modified the original \ncmod\/ code to derive the LF of each galaxy type at each redshift, $\phi[M(F),z,t]$. In our improvements to the code, we have studied the morphological evolution  \emph{only driven by the major mergers strictly-reported by observational studies}. We are just interested in testing if the major mergers reported by observations can really provide an explanation to the observed evolution of LFs, or if additional processes are required to explain it. As a major merger basically converts the two progenitor merging galaxies into an E-S0a system (see assumption \#2 in \S\ref{sec:assumptions}), our code traces the effects of major mergers on galaxy evolution backwards-in-time just decomposing present-day early-type galaxies (ETGs) into their progenitors (which are of different morphological type). So, the only morphological transformation between types allowed in our code is through \emph{reversed major mergers}, \ie, from a ETG to its progenitors. 

This means that we have not accounted for other processes that can lead to a transformation between types (such as secular evolution through bars or minor mergers). If their contribution to the buildup of massive ETGs were really significant, our model should be unable of reproducing the observed evolution of the galaxy LFs just accounting for the effects of major mergers. However, observations pose that these processes have been more relevant in the formation of intermediate-mass disks of types later than Sa than in the buildup of ETGs \citep{2009A&A...501..119A,2010ApJ...710.1170L}. As the massive end of the galaxy LFs is controlled by ETGs up to $z\sim 1$, the assumption that Sa's-Irr's do not experience any morphological transformation backwards-in-time does not affect to the results derived for massive systems. We comment below the improvements implemented to the original code in detail.

\subsubsection{Counting method}
\label{sec:countingmethod}

Most of the authors using the backwards-in-time technique assume that the number density of galaxies of type $t$ at $z$, with absolute magnitude in the $F$-band $M(F)$, is equal to the number density of this galaxy type at $z=0$ exhibiting the same absolute magnitude, \ie, $\phi[M(F),z,t]=\phi[M(F),z=0,t]$ \citep[see][and eq.\,3 in G98]{1984Ap&SS.100..407M,1988ApJ...326....1Y,1991ApJ...383L..37W,2006MNRAS.366..858K}. However, this assumption does not account for the effects of the L-evolution of galaxies. In fact, the SED of a certain galaxy type at redshift $z$ has changed with respect to that at $z=0$, because of its SFH. Therefore, galaxies exhibiting $M(F)$ at $z$ do not exhibit the same $M(F)$ at $z=0$, unless no L-evolution is assumed, and thus: $\phi[M(F),z,t]\neq \phi[M(F),z=0,t]$ in general. 

The relation between $M(F)$ at redshift $z$ and the absolute magnitude that a galaxy of type $t$ exhibit at $z=0$, $M(F,z=0)$, is given by the following expression:

\vspace{-0.2cm}\begin{equation}\label{eq:colorrestframe}
\begin{array}{rcl}
M(F) - M(F,z=0) &=& -2.5 \log\frac{\int_{0}^{\infty} f_{\lambda'}[t_G(z):t] F(\lambda') d\lambda'}{\int_{0}^{\infty} f_{\lambda'}[t_G(0);t] F(\lambda') d\lambda'},
\end{array}
\end{equation}
\noindent where $F(\lambda)$ is the filter throughput, and $f_\lambda[t_G(z);t]$ and $f_\lambda[t_G(0);t]$ are the SEDs of a galaxy of type $t$ at redshift $z$ and at present, respectively. So, the galaxies observed at $z$ with $M(F)$ would be those that appear in the local LF with absolute magnitude $M(F,z=0)$ at $z=0$ (according to the previous equation), and not those with $M(F)$. This means that, in case of PLE:

\vspace{-0.2cm}\begin{equation}\label{eq:correctcounting1}
\phi[M(F), z, t] = \phi[M(F,z=0), z=0, t].
\end{equation}

If we also have number evolution, the previous equation is not valid, as galaxies from the original LFs at $z=0$ are dissapearing backwards-in-time through mergers. In this case, the cumulative effects of number evolution from $z=0$ to $z$ must be considered:

\vspace{-0.2cm}\begin{equation}\label{eq:correctcounting2}
\phi[M(F), z, t] = \phi_\mathrm{m}[M(F,z=0),z=0\rightarrow z, t],
\end{equation}

\noindent where $\phi_\mathrm{m}[M(F,z=0),z=0\rightarrow z, t]$ represents the number density of galaxies of type $t$ still existing at $z$, having into account the backwards-in-time effects of major mergers since $z=0$ up to $z$ in this population. Tests performed to estimate the effects of the different counting methods have shown that the old procedure underestimates the galaxy number counts at faint magnitudes as compared to our method, by a factor of $\sim 25$ at $m(B)=25$\,mag, and twice more at $m(K)=16-18$\,mag. Effects start to be significant at $z>0.8$ in all bands (from $U$ to $K$).

\subsubsection{Local luminosity functions by morphological type}
\label{sec:lfbymorphologicaltype}

Galaxies within the same morphological class also exhibit similar colors. In fact, \citet{2004MNRAS.353..713K} show that galaxy morphology is strongly related to star formation (SF). Therefore, we have decided to follow the evolution of the different morphological types existing at $z\sim 0$ backwards-in-time, assuming \emph{a priori} that all the galaxies within each type share assembly processes and SFHs. 

The code requires as input the local LFs of each galaxy type. Two studies on the local LFs given by morphological types were available: 1) the LFs derived by \citet[][N03 from now]{2003AJ....125.1682N} in the $r^\ast$-band, provided in four coarse morphological classes (E-S0, S0a-Sb, Sbc-Sd, Im); and 2) the LFs by \citet[][C03 hereandafter]{2003A&A...405..917C} in the $B_Z$ filter, for eleven thin morphological classes (E, S0, S0a, Sa, Sab, Sb, Sbc, Sc, Scd, Sd, Im). None of these two parametrizations fitted our requirements, as the last one gives better predictions on those observables associated to star-forming systems, whereas the former was better tracer of the mass evolution. Moreover, the S0a-Sb class by N03 is too coarse as to be represented with the same local SED and SFH \citep[][]{2006ApJ...639..644E}. 

\begin{figure}[t]
\begin{center}
\includegraphics*[width=0.5\textwidth,angle=0]{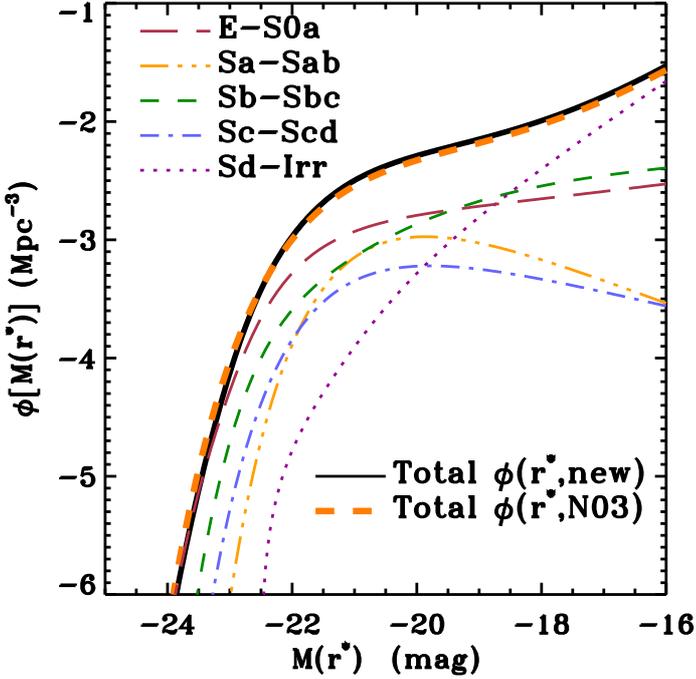}
\caption{Decomposition of the local LFs by N03 in $r^*$ into thinner morphological types, using the local LFs in $B$ by C03. The LFs plotted correspond to the following morphological classes: E-S0a, Sa-Sab, Sb-Sbc, Sc-Scd, and Sd-Irr (consult the legend in the figure). \emph{Solid line}: Total LF resulting from the performed decomposition. \emph{Thick, dashed line}: Original total LF by N03 in $r^*$-band. [\emph{A color version of this plot is available at the electronic edition}].
}\label{fig:lfs}
\end{center}
\end{figure}

So, we have decomposed the $r^*$-band LFs by N03 into thinner classes, using the information provided by the $B$-band LFs by C03, and trying to group more accordingly galaxies with similar spectrophotometric and morphological properties (E-S0a, Sa-Sab, Sb-Sbc, Sc-Scd, and Sd-Irr). The method is explained in detail in Appendix\,\ref{append:LFdistribution}. Figure\,\ref{fig:lfs} shows the resulting local LFs in $r^*$, as derived using this procedure. Notice that the LF obtained for the E-S0a class exhibits an increasing slope at the faint-end. This does not contradict studies reporting a decreasing faint-end slope ($\alpha>-1.0$) for the E-S0 class \citep{2003A&A...408..845D}, as our class also includes the S0a's. Schechter fits have been performed to these decomposed LFs for each morphological type, obtaining the parameters and $\chi^ 2$ residuals listed in Table\,\ref{tab:LFs4}. These LFs in the $r^*$-band have been considered as the input local LFs in the model. They trace the SF better than the original LFs by N03, with the advantage of a better correspondence with the stellar mass, as compared to the original LFs in the $B$-band. Several tests to check the robustness of this decomposition are commented in Appendix\,\ref{append:LFdistribution}.

\begin{table}
\caption{Schechter Parameters of the decomposed LFs$^\mathrm{a}$ in $r^*$ } 
\label{tab:LFs4} 
\centering 
\begin{tabular}{l c c c c} 
\hline\hline 
\multicolumn{1}{c}{Type} & \multicolumn{1}{c}{$\phi^*(0.01\,{\rm Mpc^{-3}})$} &  \multicolumn{1}{c}{$M^*(r^*)$} & \multicolumn{1}{c}{$\alpha$}   & \multicolumn{1}{c}{$\chi_\mathrm{fit} ^2$}\\
\multicolumn{1}{c}{(1)} & \multicolumn{1}{c}{(2)} & \multicolumn{1}{c}{(3)} & \multicolumn{1}{c}{(4)} & \multicolumn{1}{c}{(5)}\\
\hline 
E-S0a 	     & 0.20$\pm$0.22 & -21.51$\pm$0.20 & -1.0$^\mathrm{b}$ & 7.8e-11\\  
Sa-Sab    & 0.3$\pm$0.4  & -20.62$\pm$0.03  &  -0.49$\pm$0.14 & 1.3e-10\\ 
Sb-Sbc	&  0.169$\pm$0.007 &  -21.27$\pm$0.04 &  -1.13$\pm$0.03 & 8.7e-10\\
Sc-Scd 	&  0.1$\pm$0.4  & -21.06$\pm$0.04 &  -0.69$\pm$0.15 & 1.4e-10\\ 
Sd-Irr      &  0.037$\pm$0.020  & -20.94$\pm$0.07 &   -1.90$\pm$0.03 & 2.0e-9  \\
\hline 
\end{tabular}
\begin{list}{}{}
\item[$^{\mathrm{a}}$] Provided for $h\equiv H_0/100 = 1$. Magnitudes in the Vega system.
\item[$^{\mathrm{b}}$] Fixed $\alpha$ value (see Appendix\,\ref{append:LFdistribution}).
\end{list}
\end{table}

\subsubsection{Star formation histories}
\label{sec:sfh}

The SFH of each galaxy type in the model is fixed through the star formation rate (SFR) assigned to each type and the redshift at which the SF activity in each galaxy class starts ($z_\mathrm{f,*}$). We have used an exponential function to represent the SFRs of spiral galaxies \citep{1989ApJ...344..685K}, a single stellar population (SSP) for ETGs, and a constant SFR for Sd-Irr's. Different e-folding timescales have been considered for spirals, depending on their types and according to observations: $\tau_*=4$\,Gyr for Sa-Sab, 7\,Gyr for Sb-Sbc, and 9\,Gyr for Sc-Scd \citep{2005MNRAS.362...41G}. 

We have assumed metallicities that decrease from early to late types, in agreement with the observational results by \citet{2005MNRAS.362...41G}. The final parametrizations of the SFR considered for each galaxy type in the model are listed in Table\,\ref{tab:sfh}. In fact, these values are quite standard in population synthesis models \citep{2005MNRAS.362...41G,2009arXiv0901.1090T,2009ApJ...694.1099M}, and similar to those used in \citet{2006ApJ...639..644E}. The ages and metallicities considered in the model are coherent with the median values of these magnitudes derived for SDSS galaxies in different mass ranges \citep{2005MNRAS.362...41G}.
 
\begin{table}
\caption{SFRs, metallicities, and internal dust extinction per galaxy type} 
\label{tab:sfh} 
\centering 
\begin{tabular}{l c c r c} 
\hline\hline 
\multicolumn{1}{c}{Galaxy Type} & \multicolumn{1}{c}{SFR}  & \multicolumn{1}{c}{$\tau_*^\mathrm{a}$}  & \multicolumn{1}{c}{Z$^\mathrm{b}$}  & \multicolumn{1}{c}{$\tau_\mathrm{dust,star}(B,\mathrm{@}R_{25}/2)^\mathrm{c}$}\\
\multicolumn{1}{c}{(1)} & \multicolumn{1}{c}{(2)} & \multicolumn{1}{c}{(3)} & \multicolumn{1}{c}{(4)} & \multicolumn{1}{c}{(5)}\\
\hline 
E-S0a & SSP & ...&0.02 & 0.10\\
Sa-Sab & exponential& 4 & 0.02 & 0.55\\ 
Sb-Sbc & exponential & 7 & 0.008 & 0.70 \\ 
Sc-Scd & exponential & 9 & 0.004 & 0.37 \\ 
Sd-Irr & constant & ... & 0.004 & 0.09\\
\hline 
\end{tabular}
\begin{list}{}{}
\item[$^{\mathrm{a}}$] e-folding timescale in Gyr.
\item[$^{\mathrm{b}}$] Metallicity.
\item[$^{\mathrm{c}}$]Optical depth in $B$ due to internal dust extinction (M09). See the text for more details.
\end{list}
\end{table}

For all galaxy types, we assume that the SF starts at $z_\mathrm{f,*}= 3$, where the SFH of the Universe exhibits its peak \citep{2006ApJ...651..142H}. Notice that this redshift represents the \emph{formation redshift of the bulk of the SPs} for all the morphological types. This is different to the redshift at which the galaxy is assembled, as stars can be formed in different systems to those where they end up at $z=0$, because of merging. 

To simulate the spectral evolution of the composite SPs we have updated the library of evolutionary SP synthesis models of the original code to {\tt GALAXEV\/}, the Isochrone synthesis code by \citet[][B\&C03 hereandafter]{2003MNRAS.344.1000B}. Padova 1994 evolutionary tracks and the Chabrier initial mass function have been considered \citep[][]{2001ApJ...554.1274C,2002ApJ...567..304C,2003PASP..115..763C}. 

\subsubsection{Dust extinction}
\label{sec:dust}

G98 considered all the galaxy types to be equally extinguished, using a $B$-band optical depth $\tau_\mathrm{dust,star}(B)=0.2$ for local $L^ *$ galaxies of all types at 4500\AA. This assumption is a simplistic approximation to reality \citep[][]{2002A&A...383..801B,2002A&A...385..454B}. So, we have inserted of a different $\tau_\mathrm{dust,star}(B)$ for each morphological type. We have assumed the average values of dust extinction derived from the observational extinction profiles obtained by \citet[][M09 hereafter]{2009ApJ...701.1965M} for each morphological type. We have adopted the extinction value at its isophotal radius $R_{25}/2$ as the representative total internal extinction of that galaxy type (see Table\,\ref{tab:sfh}). The assumed $\tau_\mathrm{dust,star}(B)$ values are consistent with those derived by other authors within the typical observational errors \citep[$\pm 0.1$, see][]{1996A&A...306...61B,1998MNRAS.297..807T}. 

\subsection{Merging}
\label{sec:merging}

In the model, we are going to consider the effects of major mergers on the galaxy populations, but not those of minor mergers. Although their role on the galaxy evolution is poorly understood at present \citep{2004A&A...428..837F,2007AJ....133.2327G,2009ApJ...692..955M}, we know that major mergers drive much more dramatic structural changes, violent starbursts, and higher mass increments in a galaxy than minor mergers \citep{1991ApJ...370L..65B,2000MNRAS.316..315B,2003ApJ...597..893N,2005MNRAS.357..753G,2006A&A...457...91E,2006MNRAS.372L..78G,2007A&A...476.1179B}. Even considering that minor mergers have probably been more numerous than major mergers \citep[at least, by a factor of $\sim 2$, see \eg,][]{2006A&A...457...91E,2006MNRAS.370.1905H,2007A&A...476.1179B,2008IAUS..245...63C,2008ASPC..396..243K,2008ApJ...683..597S,2009ApJ...697.1971J,2009MNRAS.394.1713K,2009ApJ...699L.178N,2009ApJ...702..307S,2009ApJ...702.1005S,2010ApJ...710.1170L}, observations indicate that their contribution seems to have been significant only in low-mass systems \citep[with $\mathcal{M}_*/\Msun <10^{10}$, see][]{2009ApJ...697.1369B,2009arXiv0911.1126O,2010ApJ...710.1170L}. Therefore, we can neglect their effects in the model.

\subsubsection{Evolutionary stages in a major merger}
\label{sec:evolutionphases}

Depending on the gas content of the progenitors in a major merger, their colors can redden noticeably due to strong dust extinction during intermediate phases of the merger \citep{2010MNRAS.401.1552D}. According to \citet{2008MNRAS.391.1137L} and \citet{2008MNRAS.384..386C}, a wet major merger reddens noticeably due to strong dust extinction during two epochs of strong merger-induced starbursts: 1) in the pre-merger stage or tidal phase (enclosed between the first pericenter passage and the moment at which both progenitors have merged into one body), and 2) in the merging-nuclei stage or nuclear coalescence phase (during which both nuclei merge generating a unique remnant). In both phases, the merging galaxies also exhibit high morphological distortions \citep{2000ESASP.445...37L,2001ASPC..245..390L,2006ApJ...652..270B,2006MNRAS.373.1389C,2006A&A...454..125E,2006AJ....132...71H,2008MNRAS.384..386C,2008A&A...492...31D,2008ApJ...672..177L,2008MNRAS.391.1137L,2009MNRAS.399L..16C,2009MNRAS.395L..62G,2009ApJ...694L.123K}. During the last stage of the merger (post-merger or late remnant phase), the SF in the remnant is completely quenched by the galactic superwinds associated to the recently-formed stars or by AGN feedback, the SPs in the remnant fade passively, and the system experiences a strong relaxation \citep[see][]{1999ApJ...517..130H,2006A&A...454..125E,2004ApJ...607L..87C,2008MNRAS.384..386C,2008MNRAS.391.1137L,2007MNRAS.382.1415S,2010ApJ...714L.108S}. If gas ejection from the remnant is efficient enough and gas infall is negligible (as at it seems to be usual at $z<1$), the remnant evolves passively into an ETG in $\tau_3\sim 0.5$-1\,Gyr after the SF quenching \citep{2003ApJ...597..893N,2005ASSL..329..143S,2006ApJ...636L..81N,2006ApJ...641...90R,2007MNRAS.379..401E,2007AJ....134.2124R,2007AJ....133.2132S}. 

A mixed major merger presents the same structural stages as a wet merger, although the gas content and SF involved are lower. As it involves a gas-rich, spiral progenitor, appreciable morphological distortions make it detectable to asymmetry indices studies during the nuclei-merging phase \citep{2008MNRAS.391.1137L}. The starbursts driven by the interaction can also redden noticeably the remnant center \citep[][]{2001ApJ...551..651H,2002ApJ...581.1032S,2006AJ....132...71H,2009ApJ...690..802J}.

A dry major merger imprints negligible morphological distortions and entails very little SF, and thus, the remnant and the progenitors can be considered typical ETGs during the whole process  \citep{2006ApJ...636L..81N,2006AJ....132...71H,2006ApJ...640..241B,2007ApJ...665L...9R,2007AJ....134.1118D,2008ApJ...684.1062F}. 

The nominal timescales assumed for the pre-merger and merging-nuclei phases correspond to the average values derived from the whole range of values reported by \citet{2008MNRAS.391.1137L} and \citet{2008MNRAS.384..386C}: $\tau_1\sim 0.7$ and  $\tau_2\sim 1$\,Gyr, respectively. However, we have also accounted for the uncertainties inserted in the model due to these timescales, considering the extreme values reported by these authors for each timescale (see \S\ref{sec:lf} for more details).

\subsubsection{Assumptions on the merging events}
\label{sec:assumptions}

Concerning to major mergers, we have assumed in the model that:\\[-0.2cm]

%1%
\indent 1. \emph{A major merger is a collision between two galaxies, involving stellar mass ratios ranging between 1:1 to 1:3}.--- This is supported by the observational fact that major mergers involving more than two galaxies simultaneously are extremely rare in the Universe \citep[][]{2008MNRAS.384..886V}, and by recent studies stating that merger fractions determined using concentration-asymmetry-clumpiness (CAS) methods are only sensitive to major mergers involving stellar mass ratios between 1:1 and 1:3 \citep{2006ApJ...638..686C,2006MNRAS.373.1389C,2004AJ....128..163L,2008MNRAS.391.1137L}. In hierarchical models, unequal-mass mergers are more frequent than equal-mass mergers \citep{2001ApJ...561..517K,2006MNRAS.370..902K}, but as the distribution of mass ratios in real mergers for a given mass is unknown, we have assigned a mass ratio $0.3<mr<1$ to each merger event according to a random distribution. 

In order to establish a direct relation between stellar masses and luminosities in galaxies, we have performed all the computations in a NIR band ($K$). This assures that the stellar mass $\mathcal{M}_*$ of a galaxy is nearly proportional to the luminosity $L$ in that band, whatever the morphological type of the galaxy is. Neglecting the effects of merger-induced SF in first approximation, the absolute $K$-band magnitudes of both progenitors can be derived as:

\vspace{-0.2cm}\begin{equation}\label{eq:masses3}
\begin{array}{rcl}
M_{1}(K) - M(K) &=& -2.5 \log [1/(mr +1)]\\
M_{2}(K) - M(K) &=& - 2.5 \log [(mr /(mr +1)],
\end{array}
\end{equation}
\noindent being $M_{1}(K)$, $M_{2}(K)$, and $M(K)$ the absolute $K$-band magnitudes of the two progenitors and of the resulting ETG, respectively. \\[-0.3cm]

%2%
\indent 2. \emph{A major merger gives place to an ETG, independently of the morphological types of the progenitors}.--- There are numerous observational results that point to a major merger origin for most elliptical galaxies \citep{2005ASSL..329..143S,2007AJ....134.2124R,2008ApJ...684.1062F,2009ApJ...693..112P,2009MNRAS.397.1940F}. Although numerical simulations show that the final remnant structure depends strongly on the mass ratio, gas content, and orbit of the encounter, they state that a major merger between galaxies with typical gas contents basically gives place to an E-S0a \citep[][]{1996ApJ...471..115B,2004A&A...418L..27B,2005A&A...437...69B,2003ApJ...597..893N,2006ApJ...636L..81N}. 

This assumption does not contradict the disk rebuilding scenario proposed by \citet{2005A&A...430..115H} \citep[see also][]{2005ApJ...622L...9S,2009ApJ...691.1168H}. According to it, a large disk could be rebuilt after a major merger in $\sim 2$-3\,Gyr, in the case that the progenitor galaxies contain a large gas reservoir  \citep[amounting to $\sim 50$\% of their masses, see][]{2009arXiv0903.3962H,2009arXiv0903.3961P}. This scenario is supported by observational examples at $0.6<z<1$  \citep{2005A&A...430..115H,2009MNRAS.398..312G,2009A&A...496..381H,2009A&A...493..899P,2009A&A...501..437Y} and by numerical and cosmological simulations \citep[see numerous references in the introduction by][]{2009ApJ...702.1005S}. However, a close inspection to the sufficiently-relaxed observational examples shows that those cases related with major mergers are basically S0-S0a galaxies, whereas those exhibiting lower bulge-to-disk ratios (and thus, not being ETGs according to our definition) are associated to minor mergers instead \citep[][]{2006A&A...457...91E,2008A&A...483L..39C,2009ApJ...693..112P}. Therefore, the disk rebuilding does not conflict with our main hypothesis at $z\lesssim 1$ \citep[see an interesting discussion on the topic in][]{2009arXiv0911.1126O}. \\[-0.3cm]

\begin{figure}[t]
\begin{center}
\includegraphics*[width=0.5\textwidth,angle=0]{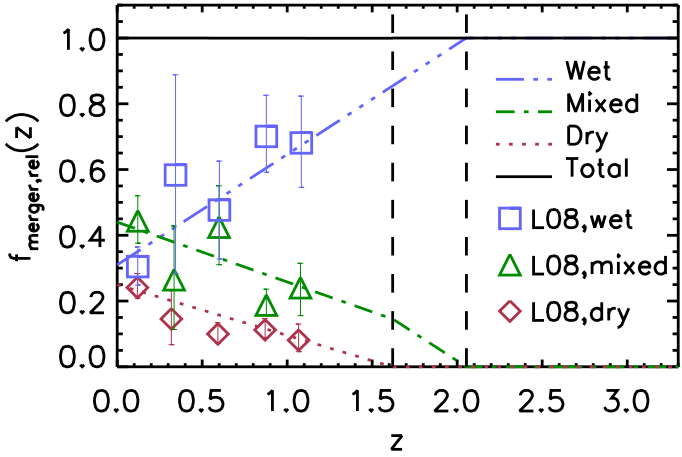}
\caption{Relative fractions of wet, mixed, and dry mergers as a function of redshift reported by L08 (\emph{squares}, \emph{triangles}, and \emph{diamonds}, respectively). \emph{Lines}: Linear extrapolations of the values at $z=0$ and $z=1$. Consult the legend in the figure. \emph{Vertical dashed lines}: Transition redshifts at which the fraction of one of the merger types becomes null, according to the extrapolations. [\emph{A color version of this plot is available at the electronic edition}].
}\label{fig:drywetmixedmergers}
\end{center}
\end{figure}

%3%
\indent 3. \emph{Major merger fractions vary as a function of redshift}.--- Observational errors usually associated to typical CAS methods lead to a possible overestimation of merger fractions by a factor up to $\sim 2$ \citep[][]{2008ApJ...672..177L,2009ApJ...697.1971J}. In order to account for these uncertainties, \citet{2008PASP..120..571L} have developed a maximum likelihood (ML) method that considers the biasses linked to the redshift degradation of the morphological information. This technique provides lower merger fractions than typical CAS methods \citep[][LSJ09 hereandafter]{2009ApJ...694..643L,2009A&A...501..505L}, quite in agreement with visual and pair counting estimates \citep{2009ApJ...697.1971J}. Therefore, we have adopted LSJ09 merger fractions up to $z\sim 1$ for the model, keeping in mind that they are limited to galaxies brighter that $M(B) = -20$ (Vega magnitudes). This corresponds roughly to galaxies with $M\lesssim M^\ast$ for $z<1$. According to these authors, the merger fraction evolves as a power-law at $z\leq 1$ as:
\begin{equation}\label{eq:LSJ09}
f_{\rm m}(z) = f_{\rm m}(0)(1+z)^{m},
\end{equation}
\noindent where $m = 1.8\pm 0.5$ and $f_{\rm m} (0) = 0.014 \pm 0.003$. The uncertainties of these parameters have been considered for error assessment (see \S\ref{sec:lf}). The extrapolation of LSJ09 trends at $z>1$ provides similar results to those derived by \citet{2008MNRAS.386..909C} at $1<z<1.5$. However, the high observational errors of merger fractions at $z>1$ make the model results at these redshifts uncertain. \\[-0.3cm]
 
%4%
\indent 4. \emph{Merger fractions in the $B$-band trace reliably the number of merger events at $z\lesssim 1$}.--- Some studies report that merger fractions derived in NIR are lower than those estimated from optical bands, probably because the star-forming features of galaxies seem more asymmetrical as we move to shorter rest-frame wavelengths \citep{2004ApJ...601L.123B,2009A&A...497..743H,2009ApJ...695.1315R}. However, \citet{2009ApJ...694..643L} find that the morphological merger fractions at $z = 0.6$ in the $K$ and $B$ bands are compatible within the errors. \\[-0.3cm]

%5%
\indent 5. \emph{Merger fractions as a function of redshift for different merger types}.--- Mergers involving gas-rich (late-type) galaxies dominate at $z> 0.8$, whereas mergers involving gas-poor (early-type) galaxies are more frequent at recent epochs \citep{2005ApJ...627L..25T,2005AJ....130.2647V,2008ApJ...681..232L,2008ApJ...672..177L,2009A&A...498..379D,2009ApJ...697.1369B}.  Percentages of wet, mixed, and dry mergers have been estimated up to $z\sim 1$ by \citet[][hereandafter L08]{2008ApJ...681..232L}. We will consider their relative fractions in order to estimate the net number of wet, mixed, and dry mergers taking place at each redshift. Figure \ref{fig:drywetmixedmergers} shows the linear extrapolations in redshift derived from L08 values at $z\sim 0.1$ and $z\sim 1$ that have been considered in the model.\\[-0.3cm]

%6%
\indent 6. \emph{Merger fractions by LSJ09 are only sensitive to the merging-nuclei stage of gas-rich mergers}.--- Observations have stated that CAS methods are not efficient detecting dry mergers \citep{2005AJ....130.2647V,2006ApJ...652..270B,2006AJ....132...71H,2007AJ....134.1118D,2008ApJ...684.1062F}. Therefore, we will assume that the merger fractions by LSJ09 only account for wet and mixed processes. Under this assumption, eq.\,\ref{eq:mergerratevol2} becomes:

\vspace{-0.2cm}\begin{equation}\label{eq:mergergas}
\phi_\mathrm{m,W} (z) + \phi_\mathrm{m,M}(z)= f_\mathrm{m} (z)\cdot \phi_\mathrm{total}(z) ,
\end{equation}

\noindent where $f_\mathrm{m}$ represents the merger fraction by LSJ09 and $\phi_\mathrm{m,W}$ and $\phi_\mathrm{m,M}$ the number density of wet and mixed mergers detected at $z$. The relative fractions of wet, mixed, and dry mergers at $z$ are estimated as the number of mergers of each type detected at that $z$, divided by the total number of mergers at that $z$ (L08). Therefore, the relative fractions of dry, mixed, and wet major mergers can also be computed as:

\vspace{-0.2cm}\begin{equation}\label{eq:mixedwetdry2b}
\begin{array}{rcl}
f_\mathrm{D} (z) &=& \phi_\mathrm{m,D}(z)/\phi_\mathrm{m}(z) ,\\
f_\mathrm{M} (z) &=& \phi_\mathrm{m,M}(z)/\phi_\mathrm{m}(z) ,\\
f_\mathrm{W} (z) &=& \phi_\mathrm{m,W}(z)/\phi_\mathrm{m}(z) ,
\end{array}
\end{equation}

\noindent where $\phi_\mathrm{m}(z)$ and $\phi_\mathrm{m,D}(z)$ represent the number density of all and dry major mergers at $z$, respectively. The total number of merger events detected by L08 also contain dry mergers, and thus:

\vspace{-0.2cm}\begin{equation}\label{eq:totalmerger}
\begin{array}{rcl}
\phi_\mathrm{m}(z) & =& \phi_\mathrm{m,D}(z) + \phi_\mathrm{m,W+M}(z),
\end{array}
\end{equation}

\noindent with $\phi_\mathrm{m,W+M}$ accounting for the sum of $\phi_\mathrm{m,W}$ and $\phi_\mathrm{m,M}$. Therefore, dividing eq.\,\ref{eq:mergergas} by $\phi_\mathrm{m}(z)$ and replacing the definitions given in eqs.\,\ref{eq:mixedwetdry2b}-\ref{eq:totalmerger}, we can estimate the total number density of major mergers at redshift $z$ through:

\vspace{-0.2cm}\begin{equation}\label{eq:mergercomputations1}
\phi_\mathrm{m}  (z)= f_\mathrm{m}(z)  \cdot \phi_\mathrm{total} (z)/[ f_\mathrm{M}(z)+f_\mathrm{W}(z)] , 
\end{equation}

\noindent as well as the number density of dry, mixed, and wet mergers at $z$ through eqs.\,\ref{eq:mixedwetdry2b}. 
 
CAS methods detect mergers in the immediate pre-merger and merging-nuclei phases \citep[][]{2000MNRAS.311..565L,2002ApJ...565..208P,2003AJ....126.1183C,2006ApJ...652..270B,2006MNRAS.373.1389C,2007ApJ...666..212D,2009A&A...498..379D}. Nevertheless, the visual inspection of the merger samples by LSJ09 indicates that the mergers correspond basically to merging-nuclei stages of gas-rich encounters (probably due to the image degradation performed by these authors). So, we have assumed that all the mergers included in LSJ09 merger fractions have been detected at this evolutionary stage.\\[-0.3cm]

%7%
\indent 7. \emph{Merger fractions depend on the stellar mass}.--- The merger rate increases as the galaxy mass increases at $z<0.6$, whereas at higher redshifts the merger rate does not depend on mass for galaxies with $\mathcal{M}_*>10^{11}\Msun$ \citep{1999RSPTA.357..167C,2000ApJ...532L...1C,2008MNRAS.383..557P,2009ApJ...701.2002G,2009ApJ...697.1369B,2009MNRAS.394.1956C,2009A&A...498..379D,2009ApJ...694..643L,2009arXiv0901.4545D}. As the model is limited to bright galaxies ($L\gtrsim L^\ast$), we will distribute the computed number of merger events at a given redshift according to these trends. This means that the fractional increment in stellar mass at each mass bin is the same for all masses in the model at $z>0.6$ by assumption ($\Delta \mathcal{M}_* /\mathcal{M}_* = \mathrm{constant,}\quad\forall\mathcal{M}_*$), in agreement with observations and semianalytical models \citep[][]{2007MNRAS.375....2D,2009MNRAS.397.1776F}.\\[-0.3cm]

%8%
\indent 8. \emph{Starbursts in gas-rich major mergers}.--- Observations and simulations have posed that mergers between gas-rich galaxies induce violent starbursts in the remnant, widely-spread along the whole galaxy \citep{1985MNRAS.214...87J,1996ApJ...464..641M,2003ApJ...594L..31Z,2004AJ....127.1371K,2007ApJ...659..931B,2009ApJ...697.1971J,2009ApJ...694L.123K,2010MNRAS.401.1552D}. The efficiency of mergers in triggering strong starbursts is high enough as to consume $\sim 50$-75\% of the original gas content \citep[][]{2004ApJ...607L..87C,2006MNRAS.373.1013C}. Therefore, we have assumed that: 1) each major merger involving a gas-rich progenitor must have converted a certain percentage of its original gas content into stars, and 2) the starburst affects the color of the whole galaxy. Assuming that the typical gas content in disks is $\sim 10$\% of their baryonic masses \citep{1975dgs..conf..113R,1982ApJ...260L..41Y,1983ApJ...265..148S}, we have tested how the efficiency of the merger-induced starbursts can affect to the predicted LFs (see \S\ref{sec:modelparametersinfluence}). We have considered that the increment in stellar mass induced by the merger is concentrated at the pre-merger phase. Assuming that a fraction $f$ ($0\leq f\leq 0.1$) of the original gas content of each disk progenitor has transformed into stellar mass in the remnant ETG, a stellar mass $\mathcal{M}_*\, (1-f)$ (instead of a mass $\mathcal{M}_*$) must be distributed into the original gas-rich progenitors, affecting to eq.\,\ref{eq:masses3}. \\[-0.3cm]

%9%
\indent 9. \emph{DSF phases in gas-rich major mergers}.--- The strong starbursts induced by major mergers are frequently accompanied by large amounts of dust, that extinct and redden noticeably the galaxy spectrum \citep{1995AJ....110..129K,2001ApJ...550..212B,2003MNRAS.341...33K,2006MNRAS.370...74R,2009ApJ...693..112P}. In the local Universe, gas-rich major mergers exhibit higher dust contents than normal star-forming disks \citep[][]{2000A&A...362...97B,2007ApJ...660..288P,2009ApJ...694..751I}. In fact, colors of local ULIRGs (which are basically gas-rich major mergers) are as red as those of local ETGs \citep{2000AJ....120..604S,2000ApJ...529..170S,2004AJ....127.2522A,2006A&A...453..809I,2010MNRAS.401.1552D}. At high redshift, major mergers exhibit even higher amounts of SF and dust extinction than their local counterparts \citep{1999Ap&SS.266..243C,2004ApJS..154..155A,2006ApJ...637..242C,2009ApJ...695.1537I}. Consequently, ETGs and DSFs at high redshift exhibit similar SEDs, except for some absorption and emission lines present in star-forming systems \citep{2000AJ....119.2556L,2001A&A...372L..45P,2001AJ....122.1861S,2002A&A...381L..68C,2005ApJ...620..595W,2006ApJ...638L..59V,2009MNRAS.394.2001S}. In fact, \citeauthor{2007MNRAS.382.1415S} (\citeyear{2007MNRAS.382.1415S,2010ApJ...714L.108S}) have proved observationally the existence of an evolutionary sequence in local ETGs from a merger-driven starburst phase to quiescence, that lasts $\sim 1$\,Gyr. The transition between both phases is driven by an AGN phase that peaks at $\sim 0.5$\,Gyr after the starting of the starburst phase, responsible of the suppression of the SF in these systems. These authors report colors of ETGs in the AGN phase quite similar to those exhibited by quiescent ETGs on the red sequence. Therefore, we will assume that gas-rich mergers undergoing transitory DSF phases (during the pre-merger and merging-nuclei phases) exhibit the same SED as ETGs at the same redshift. This assumption is reasonable in global terms at least for rest-frame wavelengths longer than the 4000\,\AA\ break feature. Moreover, we will show that DSFs become numerically-relevant at $z\gtrsim 0.7$ in the model (see \S\ref{sec:lf}), epochs at which real DSFs exhibit the reddest colors, making our assumption more accurate. \\[-0.3cm]

%10%
\indent 10. \emph{The gas-rich progenitors in a wet or mixed major merger are usually spirals of very late Hubble type}.--- This assumption is supported by observations posing that gas-rich galaxies involved in wet or mixed mergers tend to be of Sc-Irr type \citep{2005AJ....130.1516D}, and by the fact that the relative fraction of early-type galaxies in high-density environments seems to have increased since $z\sim 1$ at the expense of that of very late-type galaxies (Sc-Irr's), instead of intermediate spirals \citep{2009A&A...503..379T}. So, we assume that, if an ETG was built up through a mixed merger, its gas rich progenitor was a Sc-Irr galaxy (the another one was a gas-poor galaxy), and if the ETG was assembled through a wet merger, both gas-rich progenitors were Sc-Irr's.

\subsection{Computing the LF evolution}
\label{sec:computingEvolutionOnLF}

In our model, the backwards-in-time evolution of the local LFs is computed assuming that only ETGs are susceptible of coming from a major merger. In order to derive the evolved LFs at a given redshift $z$, we carry out a procedure analogous to the original G98 method (\S\ref{sec:originalcode}), but using the correct counting method (\S\ref{sec:countingmethod}) and all the improvements commented in \S\ref{sec:improvementstothemodel}. 

We assume that the number density of E-S0a's that are formed at each redshift $z$ through major mergers is equal to the number density of major mergers that are observed at $z$ (assumption \#2 in \S\ref{sec:assumptions}). The number density of the assembled ETGs at each redshift is estimated using LSJ09 merger fractions (eqs.\,\ref{eq:LSJ09} and \ref{eq:mergercomputations1}). The corresponding number densities of ETGs formed through wet, mixed, and dry major mergers at each redshift are estimated through eq.\ref{eq:mixedwetdry2b}. The number of ETGs that must be removed from their LF at each luminosity bin (and at each redshift) is estimated assuming the distribution of mergers with luminosity stated in assumption \#7 of \S\ref{sec:assumptions}. As the merger fractions by LSJ09 are detecting remnants before their post-merger phases, we can override this phase in the model (assumptions \#3-6 in \S\ref{sec:assumptions}). 

We have decomposed each ETG of luminosity $L$ coming from a major merger into two progenitors, with luminosities according to equations derived in \S\ref{sec:assumptions} (assumptions \#1 and \#8). In reversed merging, these recently-assembled ETGs must disappear from the cosmic scenario backwards-in-time, but firstly they come into the nuclei-merging phase during $\tau_2$ Gyrs, split into their progenitors in the pre-merger phase for $\tau_1$ Gyrs, and then become two typical progenitor galaxies (ETGs or blue disks) after this phase (as described in \S\ref{sec:evolutionphases}). This procedure has into account that the conversion of two merging progenitors into an ETG is not instantaneous, and the colors of galaxies at each phase depending on the merger type. The product of a reversed merger in our back-in-time technique results in the decomposition of the mETG assembled through this merger into their two progenitor galaxies (ETGs or gas-rich disks, depending on the merger type), after a period of $\tau_2 +\tau_1$\,Gyr. Gas-rich progenitors have been considered to be Sc-Irr galaxies (assumption \#10 in \S\ref{sec:assumptions}). Galaxies undergoing a DSF phase during a gas-rich major merger (in the pre-merger and in the merging-nuclei phases) are considered to exhibit the same SED as ETGs at each redshift (non-corrected from dust extinction, see assumption \#9 in \S\ref{sec:assumptions}). Notice that, as the model traces the evolution backwards-in-time, the fate of the most massive galaxies is not affected by the evolution of the less massive systems, as all galaxies are continuously losing stellar mass backwards in time. 

The timescales adopted for the DSF phases of wet and mixed major mergers control the ``life period`` of the transitory DSFs before becoming typical late-type disks, and hence, affect to the LF of these systems. However, it is worth noting that these timescales ($\tau_1\sim 0.7$\,Gyr for the pre-merger stage and $\tau_2\sim 1$\,Gyr for the merging-nuclei phase) do not affect the net assembly rate of mETGs at each redshift, but they just delay more or less the transformation of the progenitor galaxies in transitory merging stages into "typical" systems. The number of mETGs dissapearing at each redshift depends only on the merger fractions, the relative percentages of each merger type, and the total galaxy LF at each redshift, but not on any timescale (see eq.\,\ref{eq:mergercomputations1}). As observational merger fractions are independent on author's hypotheses on the timescale of a merger \citep[contrary to merger rates, see][LSJ09]{2000MNRAS.311..565L,2006ApJ...652..270B,2009ApJ...697.1971J}, the number of mETGs assembled through major mergers estimated by our procedure is independent on the selected timescales of the pre-merger and merging-nuclei phases (at which the built up ETG does not exist, but their merging progenitors).

\subsection{Model uncertainties and error assessment}
\label{sec:modelparametersinfluence}

The values indicated in \S\S\ref{sec:originalcode}-\ref{sec:merging} for parameters such as the SFH per morphological type, the merger fractions, or the merging timescales are average values derived from observations and simulations. We have tested how errors in the nominal model parameters affect the model results. \\[-0.2cm]

\indent \emph{i- SFHs for each morphological type (\S\ref{sec:sfh})}.--- \citet{2006ApJ...639..644E} showed that, although galaxy number counts predicted with \ncmod\/ are not sensitive to the SFH selected for each galaxy type, the predicted color distributions depend noticeably on them. We have tested how the LFs predicted by our model change when the standard parametrizations assumed for the SFH of each galaxy type are modified. Our tests confirm that the resulting LFs are quite robust against small changes of these parameters, provided that the input values are nearly realistic. Moreover, the morphological types used in the present model group galaxies with more similar spectral properties than in \citet{2006ApJ...639..644E}, making more realistic the representation of all galaxies within a given type with the same SFH (see \S\ref{sec:lfbymorphologicaltype} and Appendix\,\ref{append:LFdistribution}). 

The SFH assigned to the Sa-Sab's gives place to slightly bluer colors for these galaxies than those exhibited by this population (see \S\ref{sec:lfmorphological}). If we consider a shorter timescale for its SFR, in order to make them redder, the fitting in NIR bands of their LFs (and even of galaxy number counts, see the forthcoming paper III of this series) improves, but the LFs of the total galactic population are a bit overestimated. This galaxy population has evolved probably through secular evolutionary process from late-type disks \citep{2008mgng.conf...47A,2008ASPC..396..325C,2008ASPC..396..297K}, processes that are not being considered in the present model. So we have fixed their SFH to the typical values reported by observations for this galaxy type. \\[-0.3cm]

\indent \emph{ii- Dust extinction for each type (\S\ref{sec:dust})}.--- An accurate description of extinction is critical for comparing model predictions with data for bands from $U$ through $K$. We found G98's assumption on dust extinction ($\tau_\mathrm{dust}(B) = 0.2$ for all galaxy types) too conservative attending to studies of local disk galaxies (see \S\ref{sec:dust}). We have tested that, for typical observational uncertainties of the optical depth ($\Delta\tau_\mathrm{dust,B}\sim \pm 0.2$ in the values of Table\,\ref{tab:sfh}), results do not change appreciably. Higher dust extinctions in late-type galaxies lead to an underestimation of the LF of blue galaxies at all redshifts up to $z\sim 1$. However, results in ETGs are less sensitive to dust extinction assumptions. In fact, they could tolerate dust extinction values beyond the above $\Delta\tau_\mathrm{dust,B}\sim \pm 0.2$, although the LF at $0.8<z<1$ is worst-reproduced in this case.\\[-0.3cm]

\indent \emph{iii- Assumed colors for DSFs (\S\ref{sec:assumptions})}.--- We have assumed the same SED for representing DSFs and ETGs at each redshift, in order to simplify the model (see assumption \#9 in \S\ref{sec:assumptions}). This assumption implies that we are underestimating the UV emission of the DSF spectrum. So, we have not considered bands bluer than rest-frame $B$-band in our results.\\[-0.3cm]

\begin{figure*}[t]
\begin{center}
\includegraphics*[width=\textwidth,angle=0]{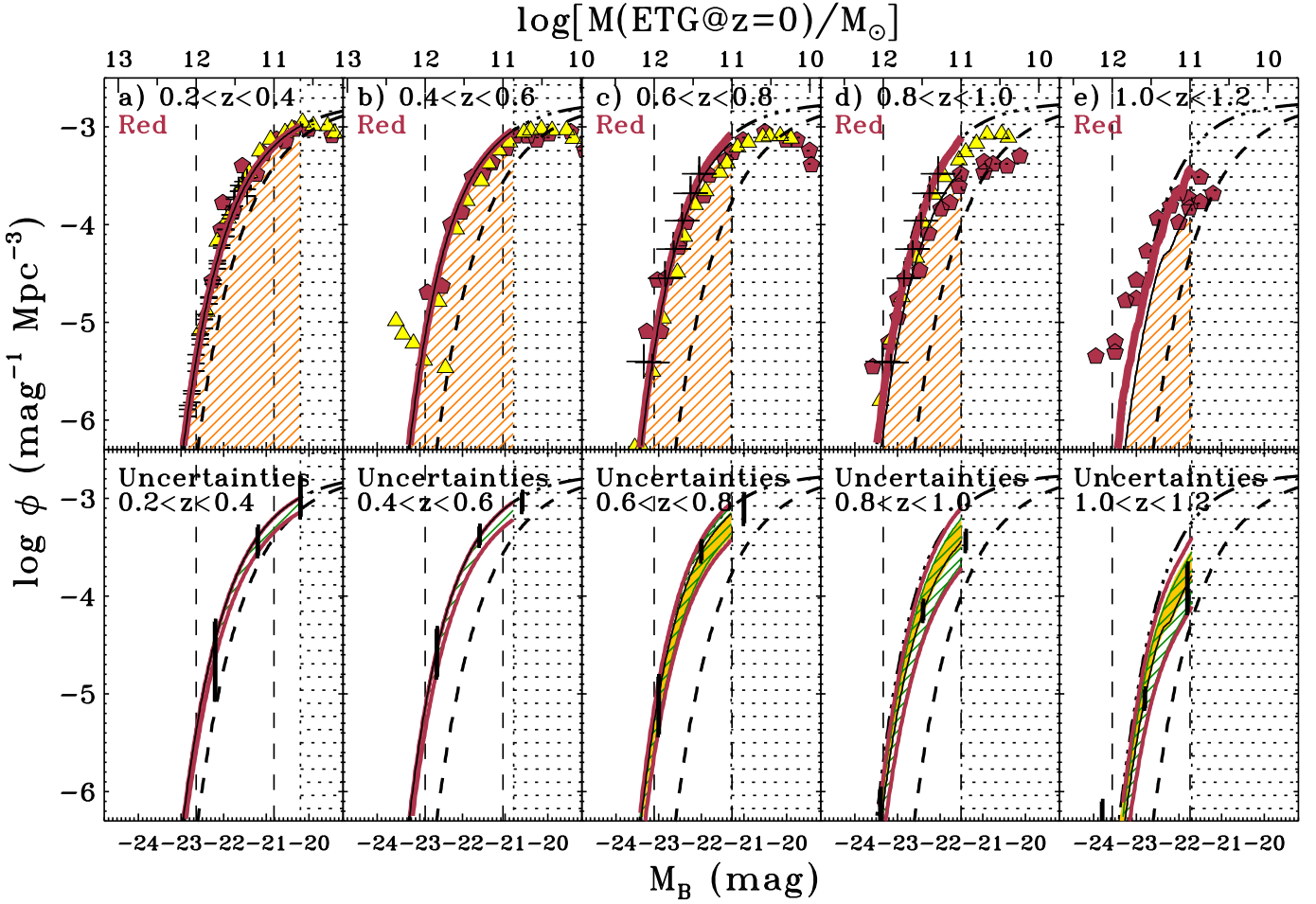}
\caption{Evolution of the $B$-band LFs with redshift of red galaxies up to $z\sim 1.2$, in redshift bins of $\Delta z=0.2$. \emph{Upper panels}: Comparison with observational $B$-band LFs. \emph{Triangles}: Data by \citet{2004ApJ...608..752B}. \emph{Crosses}: Data by \citet{2008ApJ...682..919C}. \emph{Pentagons}: Data by F07. \emph{Lines}: Model predictions. \emph{Thin, solid lines}: Predictions considering that only ETGs are classified as red galaxies. The area below them in each panel has been lined in benefit of clarity. \emph{Dashed line}: LFs at $z=0$-0.2, according to the model. \emph{Dashed-dotted lines}: Evolution of the LFs at each redshift in case of PLE. \emph{Dotted shaded region at the right of each panel}: Magnitude range where the model is not valid, because galaxies have $M(B)>M_\mathrm{lim}(B)$. \emph{Thick, solid lines}: Predictions considering that also DSFs are classified as red galaxies. \emph{Lower panels}: Model uncertainties, derived from the merger timescales (\emph{shaded area}) and from errors in LSJ09 merger fractions (\emph{lined area}). Typical error bars of observational data have been overplotted at three different magnitudes, covering the magnitude range of the model. \emph{Solid lines}: Limits of the total uncertainty region. [\emph{A color version of this plot is available at the electronic edition}].
}\label{fig:lfcolor1}
\end{center}
\end{figure*}

\begin{figure*}[t]
\begin{center}
\includegraphics*[width=\textwidth,angle=0]{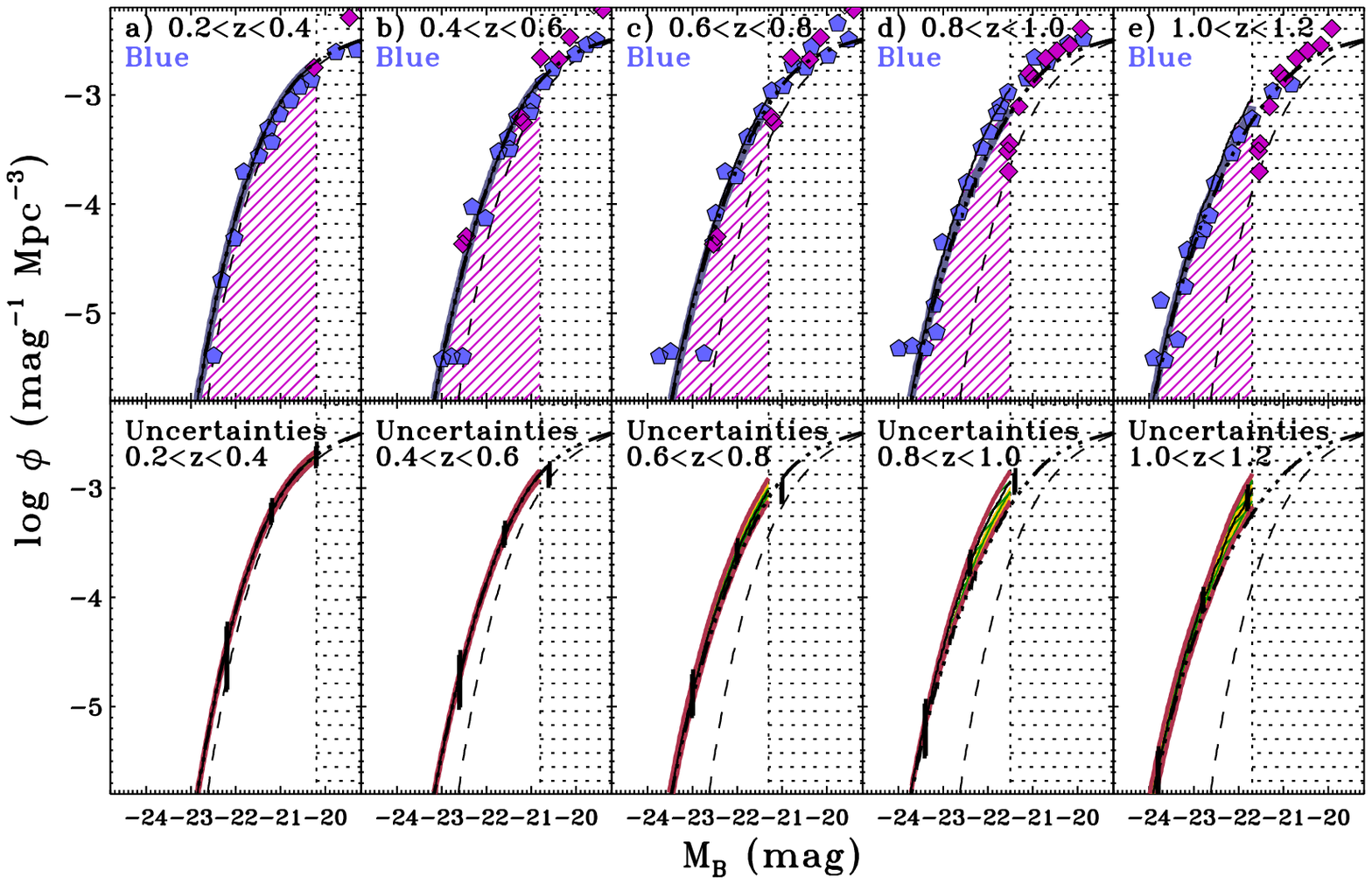}
\caption{Evolution of the $B$-band LFs with redshift for blue galaxies up to $z\sim 1.2$, in redshift bins of $\Delta z=0.2$. \emph{Upper panels}: Comparison with observational $B$-band LFs. \emph{Pentagons}: Data by F07. \emph{Diamonds}: Data by \citet{2006A&A...453..809I}. \emph{Lower panels}: Model uncertainties, derived from the merger timescales and from errors in LSJ09 merger fractions. The legend is analogous to that of Fig.\,\ref{fig:lfcolor1}. [\emph{A color version of this plot is available at the electronic edition}].
}\label{fig:lfcolor2}
\end{center}
\end{figure*}

\begin{figure*}[t]
\begin{center}
\includegraphics*[width=\textwidth,angle=0]{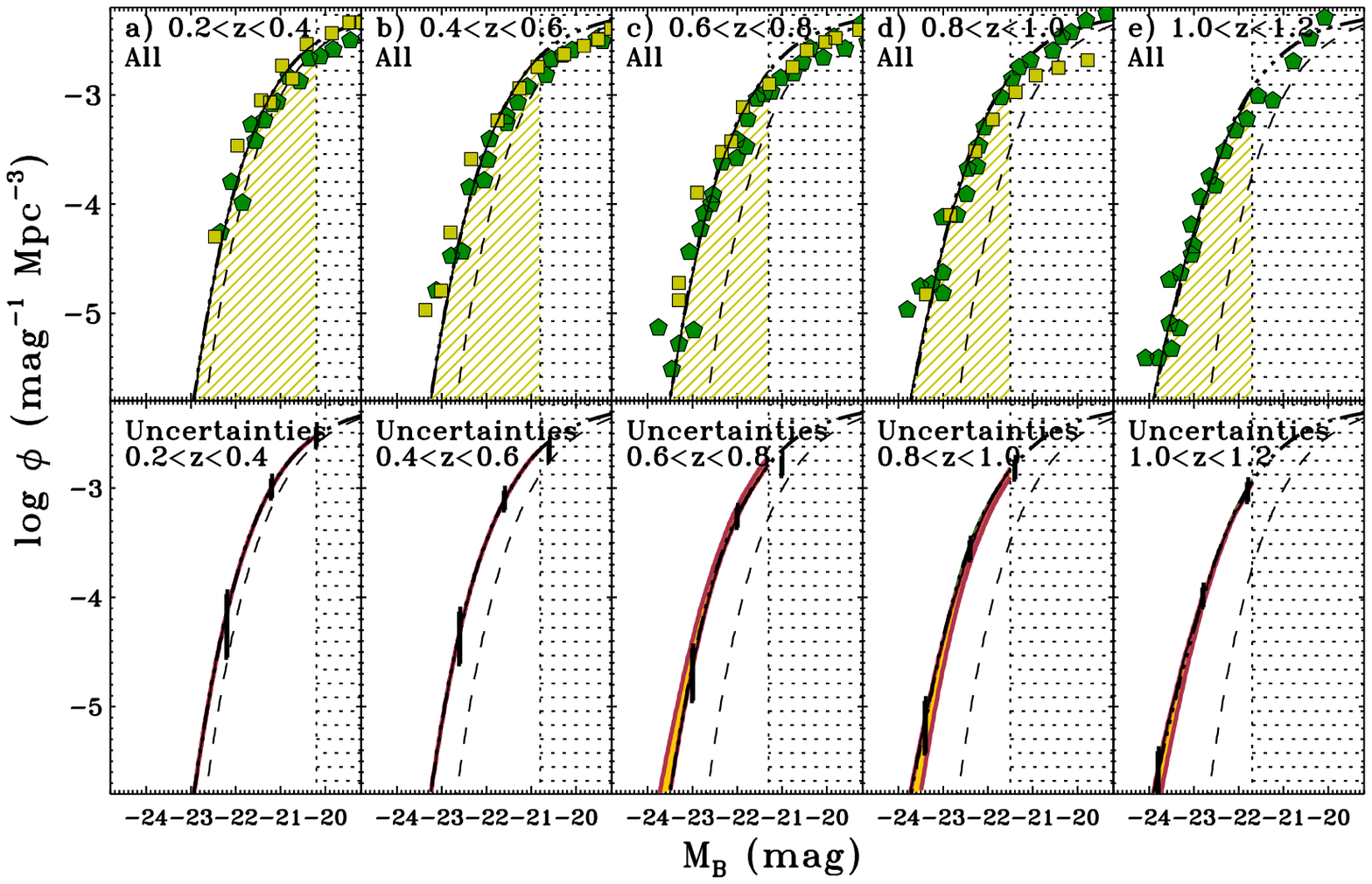}
\caption{Evolution of the $B$-band LFs with redshift for all galaxies up to $z\sim 1.2$, in redshift bins of $\Delta z=0.2$. \emph{Upper panels}: Comparison with observational $B$-band LFs. \emph{Pentagons}: Data by F07. \emph{Squares}: Data by \citet{2007ApJS..172..406S}. \emph{Lower panels}: Model uncertainties, derived from the merger timescales and from errors in LSJ09 merger fractions. The legend is analogous to that of Fig.\,\ref{fig:lfcolor1}. [\emph{A color version of this plot is available at the electronic edition}].
}\label{fig:lfcolor3}
\end{center}
\end{figure*}

\indent \emph{iv- Merger fractions (\S\ref{sec:assumptions})}.--- These uncertainties have turned out to be the most relevant source of error in the model, so we show them together with results in \S\ref{sec:lf}. We have also implemented different the parametrizations of merger fractions apart from the one by LSJ09. In particular, we have tried the parametrization derived by \citet{2006ApJ...638..686C}. This function provides a slightly better fit to LFs at $1<z<2$ (by $\sim 10$-20\%). At $z\sim 0.8$, both parametrizations imply similar merger fractions. However, the scenario proposed by our model overestimates largely the number evolution of ETGs at $z>1$, considering the major mergers reported by observations at $z>1$. In fact, we would require less than 20\% of the major mergers derived from the observational merger fractions at $z>1$ to reproduce the galaxy number counts by redshift bins and the LFs of red and blue galaxies at $1<z<1.5$. Nevertheless, it seems that the properties of major mergers start to differ appreciably from their counterparts at lower redshifts at $z>1$, and hence, assumptions adopted in the model are not fulfilled by real galaxies (see comments in \S\ref{sec:limitations}). Results are extremely sensitive to the adopted merger fractions (see \S\ref{sec:lf} and the model uncertainty regions plotted in Figs.\,\ref{fig:lfcolor1}-\ref{fig:lfsmorph}).\\[-0.3cm]

\indent \emph{v- Merger timescales (\S\ref{sec:evolutionphases})}.--- In order to account for the uncertainties in these parameters, we have considered the whole range of values reported by N-body simulations \citep[$\tau_1= 0.05$-1.7\,Gyr and $\tau_2= 0.003$-2\,Gyr, see][]{2008MNRAS.384..386C}. Although non-physical, we have also considered the possibility of "instantaneous" mergers ($\tau_1=\tau_2=0$), in order to compare our results with those that would be obtained in this case, as assumed by several semi-analytical codes (see references in \S\ref{sec:countingmethod}). We comment these uncertainties together with results in \S\ref{sec:lf} (see also the model uncertainty regions plotted in Figs.\,\ref{fig:lfcolor1}-\ref{fig:lfsmorph}). \\[-0.3cm]

\indent \emph{vi- Merger-induced SF (\S\ref{sec:assumptions})}.--- We have performed several tests changing the fraction of ETG stellar mass formed during the major merger that built it up. The typical increments of stellar mass stated by observations \citep[between 0-10\% of the total stellar mass, see][]{2004ApJ...607L..87C,2006MNRAS.373.1013C} affects negligibly the final results. However, we have also tested how higher merger-induced star formation can affect to the results, finding that increments of stellar mass above $\sim 40$\% of the total mass of the galaxies could affect the shape of the LFs of late-type galaxies at high redshift (see also item iii of \S\ref{sec:limitations}).

\section{Results}
\label{sec:lf}

In this section, we compare the evolution of the LFs predicted by the model for each galaxy population with observational data in different bands and according to different selection criteria (on color, on morphology). 

\subsection{LF evolution of red and blue galaxies at $z\lesssim 1$}
\label{sec:lfredblue}

As pointed out by F07, studies focusing on red galaxies are implicitly assuming that rest-frame color is a good way of finding E-S0 types at all redshifts. These authors claim that the identification red galaxy-ETG ''works well at low redshifts, where the contamination of the red sequence by DSFs with Hubble types later than S0 is negligible \citep[$\sim 10$-15\%, see][]{2004ApJ...600L..11B,2005ApJ...620..595W}. However, contamination by non-spheroidal DSFs is larger at higher redshifts, amounting to 30\% at $z\sim 0.75$, and may increase beyond that \citep{2002A&A...381L..68C,2003ApJ...586..765Y,2003MNRAS.346.1125G,2004ApJ...600L.131M}''. In fact, most of the authors selecting red galaxies admit that the similarity of the rest-frame colors of ETGs and DSFs makes difficult to disentangle both galaxy populations at high redshifts  \citep[][]{2001AJ....122.1861S,2006A&A...453..809I,2007MNRAS.380..585C,2007A&A...465..711F}. As the level of contamination by DSFs of a red galaxy sample is going to depend strongly on the color criteria used by each author, we are going to consider the two extreme situations in our model in order to compare with observational data: 1) the case that the red galaxy sample includes exclusively ETGs, and 2) the case that all DSFs are also included into the red sample.

\subsubsection{$B$-band LFs for red and blue galaxies up to $z\sim 1$}
\label{sec:lfredblueB}

In Figs.\,\ref{fig:lfcolor1}-\ref{fig:lfcolor3}, we have compared the evolution of the rest-frame LFs in the $B$-band as predicted by the model (using our nominal parameters) with real data on red, blue, and total galaxy populations. Symbols represent the observational LFs derived by different authors up to $z\sim 1.2$ (upper panels of each figure). We have overplotted a PLE model in all the panels (dashed-dotted lines). The abscissas axis (absolute magnitude axis) in Fig.\,\ref{fig:lfcolor1} has been transformed into stellar masses of mETGs at $z=0$ using the expression derived for local ETGs by \citet{2006A&A...453L..29C} and considering the modelled L-evolution of mETGs (axis at the top of the figure). The vertical dashed lines in panels a-e of Fig.\,\ref{fig:lfcolor1} indicate the magnitudes that correspond to ETGs with $\log(\mathcal{M}_*/\Msun)=11$ and 12 at $z=0$. So, our results concern to ETGs with $\mathcal{M}_*> 10^{11}\Msun$ at $z=0$, and roughly to $L\gtrsim L^*$ galaxies at $z\lesssim 1$.  

Attending to the LFs of red galaxies at $z<0.8$ (panels a to c of Fig.\,\ref{fig:lfcolor1}), we corroborate the negligible number evolution that all the observational studies report up to this redshift for the luminous red galaxies: notice that the PLE model for ETGs lies exactly on top of the observational data. However, observational studies disagree on the number evolution experienced by red galaxies at $0.8<z<1$ (see panel d in the figure). Depending on the author, the number density of ETGs is found to decrease by a factor of $\sim 2$-3 or negligibly. The question is why authors derive such a different number evolution for red galaxies at $0.8<z<1$.

The predictions of our model assuming that only ETGs are classified as red galaxies are also shown in Fig.\,\ref{fig:lfcolor1} (thin solid lines in top panels with lined area below them). The model is capable of reproducing the negligible number evolution reported by observations for massive, red galaxies up to $z\sim 0.8$ (panels a-c in the figure). However, it predicts a significant drop in the number density of mETGs at $0.8<z<1$ (panel d), reproducing the data for red galaxies obtained by F07 at that redshift bin. \emph{So, the model shows that it is completely feasible to build up $\sim 50$-60\% of the present-day number density of mETGs through the major mergers reported by observations since $z\sim 1$}. Another striking result of the model is that \emph{it predicts that the bulk of this recent mETGs assembly takes place basically during a time period of $\sim 1$\,Gyr elapsed at $0.8<z<1$}, in agreement with observations by F07, but in contrast with other studies \citep{2004ApJ...608..752B,2008ApJ...682..919C}. However, the contradiction between these observations and our model predictions is just apparent, as shown below.

Let us now consider that DSFs are also identified as red objects. In Fig.\,\ref{fig:lfcolor1}, we have also overplotted the LFs of ETGs$+$DSFs in the panels corresponding to the red galaxy population (thick, solid lines). Notice that, if DSFs are included into the red galaxy sample, the model now reproduces the data reporting an apparent negligible number evolution of massive, red galaxies up to $z\sim 1$ \citep{2004ApJ...608..752B,2008ApJ...682..919C}. \emph{This result suggests that the discrepancy on the number evolution reported by different studies for bright, red galaxies up to $z\sim 1$ could be due to the inclusion of a significant amount of DSFs into the red galaxy sample, depending on the color criteria used by each author}. 

This possibility has also been suggested by other recent studies \citep{2009ApJ...706L.173B,2009MNRAS.tmp.1955B}. \citet{2009A&A...503..445K} reports that, accounting also for the late-type progenitors of mETGs in a $\Lambda$CDM scenario (and not just those of early-types), more than 50\% of the stellar mass which ends up in mETGs today is in late-type progenitors at $z\sim 1$ (in agreement with our results). Our model corroborates their conclusion on the relevance of accounting for the progenitor bias in observational studies, in order to derive accurate conclusions regarding the evolution of mETGs in the last $\sim 8$\,Gyr.

Note that the model predicts a contamination of the galaxy red sequence by dusty galaxies amounting to $\sim 50$\% at $z\sim 1$ for our magnitude range, in agreement with observations \citep{2002A&A...381L..68C,2003ApJ...586..765Y,2003MNRAS.346.1125G,2004ApJ...600L.131M}. \emph{This means that the existence of this DSF population at $z\sim 1$ can be explained just accounting for the gas-rich major mergers strictly reported by observations}.

The model starts failing to reproduce observational data at $z>1.2$, independently on if DSFs are classified as red galaxies or not. In \S\ref{sec:limitations}, we discuss in detail the model limitations, especially at redshifts $z\gtrsim 1$. 

Our scenario also works for disk galaxies. Attending to the panels a-e of Fig.\,\ref{fig:lfcolor2}, we can see that the model predictions for blue galaxies match their observational evolution at $z\lesssim 1.2$ as reported by F07, better if DSFs are considered as blue galaxies together with Sa-Irr's (thin, solid lines in the panels) than if not (thick, solid lines). \emph{The model predicts naturally the decrease by $\sim 30$-40\% of the number density of blue galaxies since $z\sim 1$ to $z\sim 0$ reported by observations \citep[][F07]{2006A&A...453..809I}, just considering the transformation of disks into ETGs driven by the major mergers at $z\lesssim 1$}. Moreover, the majority of these gas-rich progenitors are predicted to be undergoing a transitory DSF phase of the major merger at $0.8<z<1$ (compare the thick, solid line and the thin, solid line in panels i-j of the figure). Obviously, as the model reproduces the observational evolution of the LFs of blue and red galaxies up to $z\sim 1$, it also fits that of the total LFs (see panels a-e in Fig.\,\ref{fig:lfcolor3}). 

In the lower panels corresponding to each redshift bin of Figs.\,\ref{fig:lfcolor1}-\ref{fig:lfcolor3}, we show the model uncertainties due to the merger timescales and to the errors in the observational merger fractions (see \S\ref{sec:modelparametersinfluence}). These uncertainties are relevant enough to make difficult to draw a conclusion on the production rate of ETGs, as they could imply from no number evolution of mETGs to a decrement of their number density by a factor of $\sim 4$. However, we must keep in mind that the nominal values used in the model are robust and realistic, derived as averages of the possible observational values. By the contrary, the values used here for uncertainty estimates are quite extreme. Nevertheless, we will show that these uncertainties affect more weakly to the results in the $K$-band (\S\ref{sec:lfredblueK}), allowing us to rule out the PLE scenario not only for red galaxies at $0.8<z<1$, but also for the blue galaxy population. This supports the robustness of the results derived for the $B$-band in this section. 

Summarizing, the model offers a framework in which the studies deriving a negligible evolution of red galaxies up to $z\sim 1$ and those finding a significant evolution for them can be reconciled, just having into account that red galaxy samples can be highly contaminated by DSFs, probably intermediate stages of gas-rich major mergers evolving into ETGs. The model suggests that the follow up of populations selected by color (red vs.\,blue) backwards-in-time, as the majority of studies do since the pioneering study by \citet{2001AJ....122.1861S}, might be providing a biassed view of the galaxy assembly at $z\lesssim 1$, as already noted by \citet{2009A&A...503..445K}.

\begin{figure*}[t]
\begin{center}
\includegraphics*[width=0.85\textwidth,angle=0]{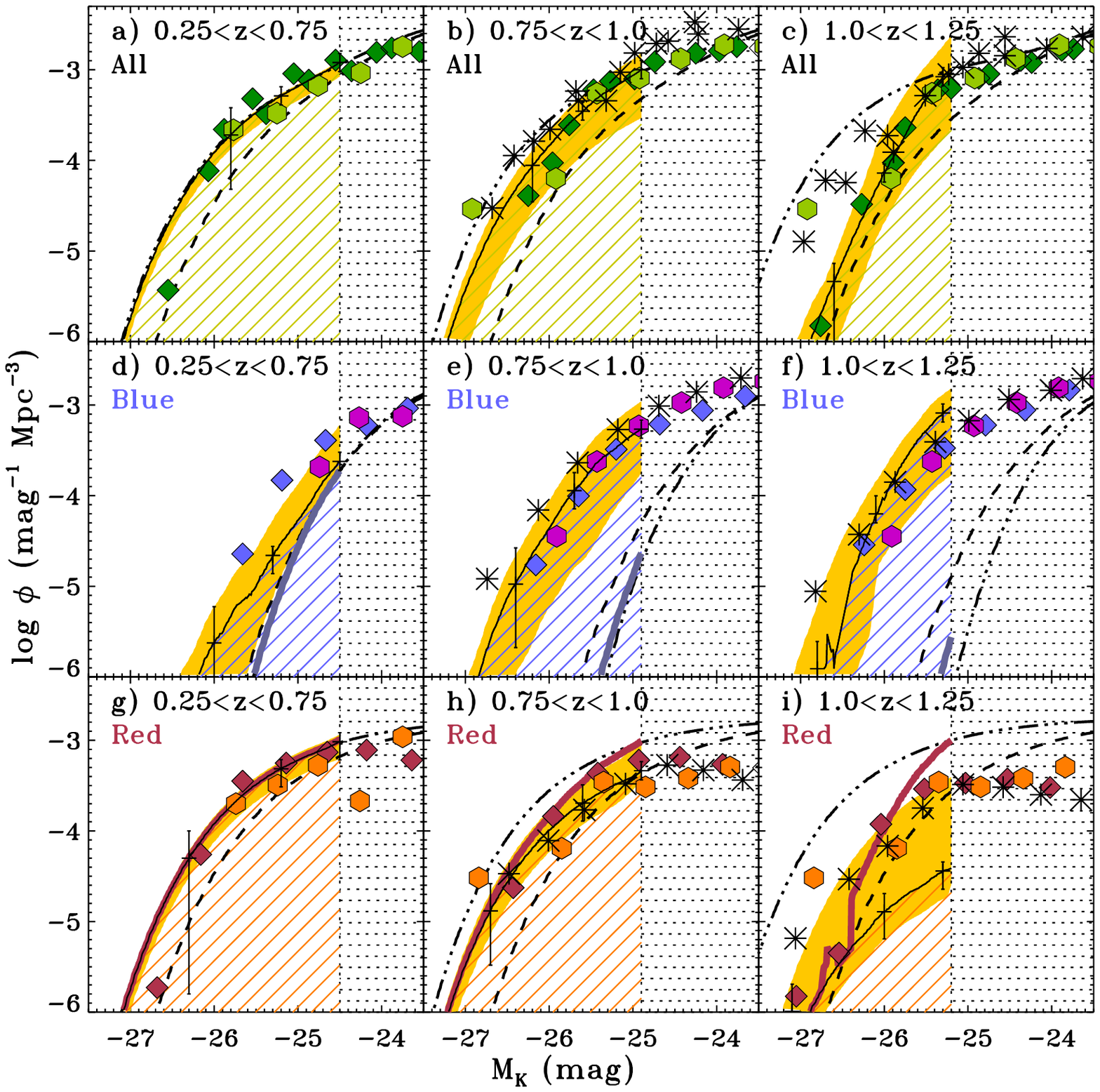}
\caption{Evolution of the $K$-band LFs with redshift of red galaxies (panels g-i), blue galaxies (panels d-f) and all galaxies (panels a-c), as predicted by the model, up to $z\sim 1.25$. \emph{Symbols}: Observational LF data in the $K$-band. \emph{Diamonds}: Data by \citet{2007MNRAS.380..585C}. \emph{Hexagons}: Data by \citet{2003A&A...402..837P} for $0.2<z<0.65$ and  $0.75<z<1.3$. \emph{Asterisks}: Data by \citet{2007A&A...476..137A} for $0.8<z<1.0$ and $1.0<z<1.2$. \emph{Lines}: Model predictions. The legend for lines is the same as in Fig.\,\ref{fig:lfcolor1}, but now for the $K$-band. \emph{Shaded area}: Total uncertainty region due to errors in the nominal values used in the model for the merger timescales and the merger fractions. Typical error bars of observational data have been drawn along the whole magnitude range in all the panels, just as reference. [\emph{A color version of this plot is available at the electronic edition}].}\label{fig:lfk}
\end{center}
\end{figure*}

\subsubsection{$K$-band LFs for red and blue galaxies up to $z\sim 1$}
\label{sec:lfredblueK}

In Fig.\,\ref{fig:lfk}, we show the model predictions on the evolution of the LFs for red, blue, and all galaxies for the $K$-band. The legend is the analogous to the one in Fig.\,\ref{fig:lfcolor1}. We have overplotted in the figure the model uncertainty region due to errors in the merger timescales and in the observational merger fractions by LSJ09.

Comparing the observational data of red galaxies with the PLE model (panels g-i in the figure), we can see that red galaxies do not show any relevant number evolution at $0.25<z<0.75$ (panel g). However, their number density decreases suddenly by a factor of $\sim 2$-3 at $0.75<z<1$, laying quite below the PLE model (panel h). So, $K$-band data are ruling out directly a PLE scenario for bright, red galaxies at $0.75<z<1$.

We have overplotted the $K$-band LFs predicted by the model in the same panels. Assuming that exclusively ETGs are identified as red galaxies, the model predicts the negligible number-evolution observed for this galaxy population at $0.25<z<0.75$, as well as the significant drop in their number density observed at $0.75<z<1$ (see panels g and h in Fig.\,\ref{fig:lfk}). The good agreement of the model predictions with $K$-band observational data even considering the model uncertainties stresses the feasibility of building up $\sim 50$-60\% of the present-day number density of mETGs accounting for the major mergers reported by observations at $z\lesssim 1$. Again, the model predicts naturally the observational fact that the bulk of this assembly occurs during the period of $\sim 1$\,Gyr elapsed at $0.75<z<1$, as in the $B$-band (see \S\ref{sec:lfredblueB}). Moreover, considering that all DSFs are classified as red galaxies (thick, solid lines in panels g-i in Fig.\,\ref{fig:lfk}), the model still reproduces better the observational results on red galaxies than the PLE model.

Panels d-f in Fig.\,\ref{fig:lfk} represent the observed LF evolution for blue galaxies in the $K$-band up to $z\sim 1.25$. The figure shows that the PLE model (dotted-dashed lines) lies quite far away from the observational data not just for red galaxies, but also for the blue ones. So, the PLE scenario for blue galaxies is also discarded just considering observational $K$-band data. 

The predictions of our nominal model on blue galaxies are overplotted in the panels. If DSFs are not included into the blue galaxy population (thick, solid line in panel d), the model for blue galaxies (Sa-Irr's) overlaps with the PLE model at $0.25<z<0.75$, not fitting data. If the DSFs predicted by the model are considered as blue galaxies (thin, solid line with lined area below it), the model reproduces the data better than before, although it still underestimates the number density of blue galaxies at this redshift bin. However, \emph{DSFs become essential to reproduce the huge increase in the number density of blue galaxies reported by observations at $0.75<z<1$, with respect to their present-day value} (by a factor of $\sim 10$, compare our model with the PLE model in panel e of Fig.\,\ref{fig:lfk}). \emph{So, the model predicts two observational excesses of blue galaxies with respect to PLE models simultaneously: 1) that of bright blue galaxies registered in the $B$-band, and 2) that of massive blue galaxies detected in the $K$-band}. And both are explained just considering the gas-rich progenitors of the mETGs that have been assembled through major mergers at $z\lesssim 1$. As in the $B$-band, \emph{the majority of these gas-rich progenitors are predicted to be undergoing a major merger at $0.8<z<1$}. The model reproduces the evolution of total $K$-band LFs up to $z\sim 1.25$ better than a PLE model, even accounting for the model uncertainties (see panels a-c  in the figure). 

\subsubsection{$I$-band LFs for red and blue galaxies up to $z\sim 1$}
\label{sec:lfredblueI}

\begin{figure*}[t]
\begin{center}
\includegraphics*[width=\textwidth,angle=0]{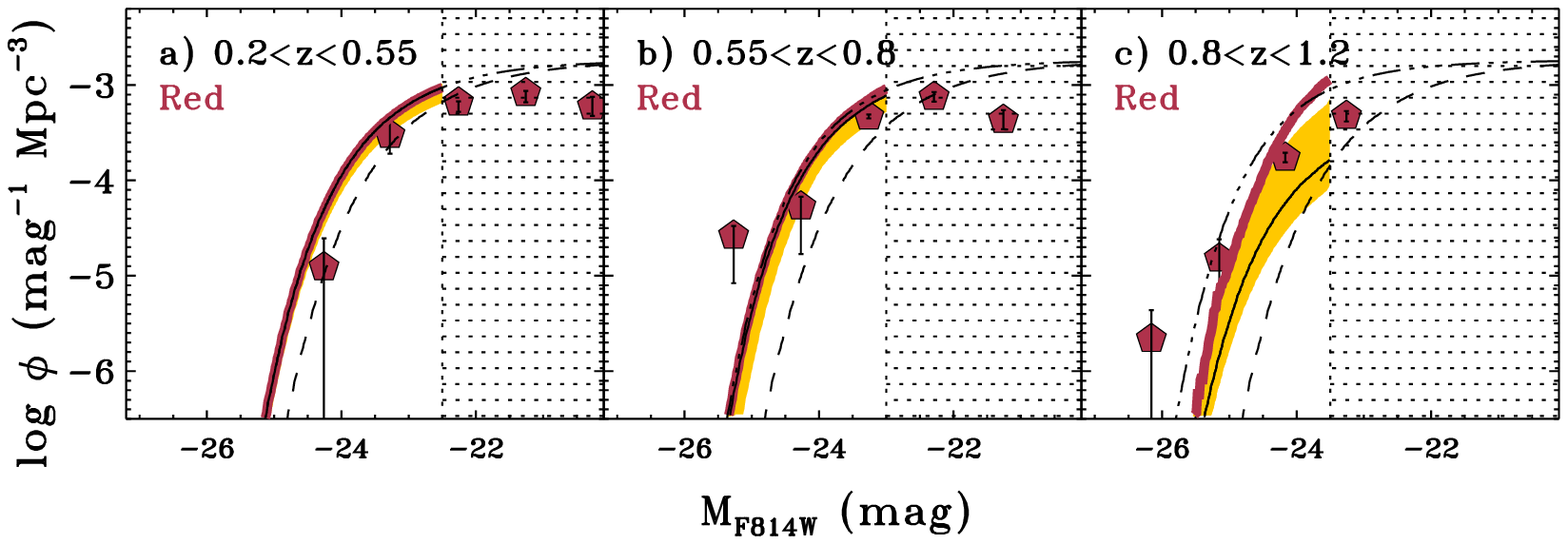}
\caption{Evolution of the $I$-band LFs for red galaxies with redshift, as predicted by the model. \emph{Pentagons}: Observational $I$-band LFs for red galaxies derived by \citet{1996ApJ...461L..79I}. \emph{Lines}: Model predictions. The legend for lines is the same as in Fig.\,\ref{fig:lfcolor1}, but for the $I$-band. \emph{Shaded area}: Total uncertainty region due to errors in the nominal values used in the model for the merger timescales and the merger fractions. [\emph{A color version of this plot is available at the electronic edition}].}\label{fig:lfi}
\end{center}
\end{figure*}

In Fig.\,\ref{fig:lfi}, we compare the LFs for red galaxies predicted by the model in the $F814W$-band ($\sim I$-band) with the observational results by \citet{1996ApJ...461L..79I}. The model reproduces quite properly the data at $z<0.8$. However, notice that data at $0.8<z<1.2$ are better reproduced if DSFs are being identified as red galaxies.

\subsection{LF evolution per morphological types up at $z\lesssim 1$}
\label{sec:lfmorphological}

In Fig.\,\ref{fig:lfsmorph}, we plot the predicted evolution with redshift of the $B$-band LFs of different galaxy morphological types. DSFs have been included into the LFs of the Sc-Irr$+$M type, as they represent stages of major mergers with strong morphological distortions. Observational data by different authors are also shown for comparison  \citep[][]{2006A&A...453..809I,2006A&A...453..397F,2006A&A...455..879Z}. The legend for lines is analogous to that of Figs.\,\ref{fig:lfcolor1}-\ref{fig:lfk}, but now refers to morphological types, not to color classes.

Figure\,\ref{fig:lfsmorph} shows that the model can reproduce the observed evolution of the LFs for different morphological types at $z\lesssim 1$: E-S0a's (panels a-e), Sa-Sab's (panels f-k), Sb-Sbc's (panels l-r), and Sc-Irr's (panels s-y). The figure suggests again that the major mergers that are observed at $z\lesssim 1$ provide a natural explanation for the observed numerical growth of mETGs since $z\sim 1$, and that the gas-rich progenitors of these recently-formed mETGs can explain the observational excess by a factor of $\sim 4$-5 of late-type galaxies at $0.8<z<1$ with respect to a PLE model (see panel w in the figure). Intermediate types (Sa-Sab's and Sb-Sbc's) are consistent with a PLE evolution up to $z\sim 1$, according to their typical SFHs.

Figure\,\ref{fig:lfcolor2} shows that the decrease in the number density of the spiral (\ie, blue) galaxies from $z\sim 1$ to $z\sim 0$ is $\sim 30$-40\% of its present-day value. However, the same numerical decrease in terms of the present-day Sc-Irr$+$DSF population means that this population was higher by a factor of $\sim 4$-5 at $z\sim 1$ than now \citep{2006A&A...453..809I,2006A&A...453..397F,2006A&A...455..879Z,2007ApJ...654..858B,2007ApJS..172..494S,2008ApJ...682..919C}. \emph{So, the scenario proposed in the model for describing the evolution of the galaxy LFs selected by color up to $z\sim 1$ also explains the observed evolution of the LFs selected by morphology}.

\section{Discussion}
\label{sec:discussion}

\subsection{Mass-downsizing and the hierarchical assembly of mETGs at $z\lesssim 1$}
\label{sec:reconciling}

The most striking result of the model is that it reproduces the observational evolution of the LFs at $z\lesssim 1$, predicting \emph{at the same time} a late hierarchical assembly for at least half of the present-day mETGs (at $z\lesssim 1$). Recent semianalytical model within the standard $\Lambda$CDM framework reproduce the observed increase of the number density of mETGs since $z\sim 1$ \citep{2008ApJS..175..390H,2009A&A...503..445K,2009arXiv0911.1126O,2010arXiv1002.3257C}. As our model proves that the major mergers strictly reported by current observations are capable of explaining the mETGs assembly since $z\sim 1$, our approach is directly linking the observed number of major mergers and the predicted one by hierarchical models, demonstrating that the hierarchical scenario and observations are quite compatible afterwards.

Studies supporting galaxy mass-downsizing report that only $\sim 40$-50\% of the galaxies with $\mathcal{M}_*> 10^{11}\Msun$ were completely assembled at $z\sim 1$ \citep{2008ApJ...675..234P,2010ApJ...709..644I}. Having into account that ETGs dominate the mass function at this mass range, we can conclude that the main result of our model is in agreement with the global trends of mass-downsizing. Moreover, the model predicts that mETGs have acquired their actual volume density at $z\sim 0.8$, again in agreement with the previous studies. In Paper II of this series, we will show that the assembly epoch derived by the model for field mETGs is in agreement with that derived by semianalytical models and studies on the SPs of mETGs \citep[see][]{2006MNRAS.366..499D,2006MNRAS.372..933N}. Moreover, the model predicts that $\sim 50$\% of the present-day number density of mETGs already exist at $z\sim 1$, consistently with the detection of numerous "read\&dead" mETGs in high-density environments at higher redshifts (see references in \S\ref{sec:introduction}). 

The model assumes that the numerical growth for each mass bin at $z>0.6$ is similar, in agreement with several observations (see assumption \#7 of \S\ref{sec:assumptions}). This means that the mETGs at different masses are assembled at the same numerical rate at $0.6<z<1$ in the model. Obviously, this growth rate is in contradiction with mass-downsizing, that derives that the most-massive galaxies must have experienced a lower number evolution at $z\lesssim 1$ than galaxies with lower masses \citep[][]{2007MNRAS.380..585C,2007ApJS..172..406S,2007ApJS..172..494S}. In fact, ETGs with masses $\log(\mathcal{M}_*/\Msun) > 11.5$ seem to be already formed at $z>1.7$ \citep{2006A&A...453..397F}. This does not mean that the model and mass-downsizing can not be reconciled. Firstly, because it can not be ruled out that mass-downsizing can be an artifact of large observational errors at the bright-end of the derived galaxy mass functions, as claimed by \citet{2009MNRAS.397.1776F}. And secondly, because even if galaxies more massive than $10^{12}\Msun$ were already in place at $z>2$, the model could reproduce it just considering a slower numerical growth rate at the high-mass end than at lower masses, without affecting its main results (these galaxies represent $<5$\% of the total mETGs population considered by the model). However, this would require the prior knowledge of the distribution of major mergers with galaxy masses up to $z\sim 1$, a question unsettled by the moment.

In conclusion, the model shows that the observational phenomenon of mass-downsizing can be compatible with the hierarchical assembly of nearly half of the present-day number density of mETGs at $z\lesssim 1$. In fact, other recent studies confirm upsizing in galaxies at $z<1.5$ \citep[see][]{2007ApJ...668..839R,2008ApJ...688..770F}. However, a complete reconciliation of hierarchical models and mass-downsizing might require a better implementation of the physical processes affecting to massive systems in the simulations, and a more exhaustive analysis of the biasses and uncertainties of observational data.

\subsection{Comparison with other observational results}
\label{sec:comparison}

The model offers a framework for galaxy evolution that can explain or agrees with several recent observational results:\\[-0.2cm]

\indent 1. \emph{The bright-end of the LF in H$\alpha$ at $z\sim 0.84$ is dominated by Irr and merging systems \citep[][]{2008ApJ...677..169V,2009MNRAS.398...75S}}.--- The model predictions are coherent with this fact, as the gas-rich progenitors at $0.8<z<1$ of the recently-assembled mETGs represent $\sim 30$-40\% of the present-day number density of all spirals, and are Sc-Irr systems. Therefore, the model predicts that $\sim 50$\% of the star-forming galaxies at $z\sim 0.8$ are Sc-Irr or DSF merging systems. As Sc-Irr's have enhanced H$\alpha$ emissions as compared to normal disks \citep[and much more if they are involved in a major merger, see][]{2010ApJ...708..534W}, the model predicts that they must control the H$\alpha$ luminosity at $z\sim 0.8$, as observed.  \\[-0.3cm]

\indent 2. \emph{Nearly 75-85\% of quiescent galaxies with $\log(\mathcal{M}_*/\Msun)>10^{11}$ at $z<0.8$ have an elliptical morphology \citep{2010ApJ...709..644I}}. --- The model predicts that the major mergers start affecting the LF of ETG appreciably at $z>0.8$. At lower redshifts, the mETGs are nearly in place in the model (some evolution is still ongoing, but it is negligible, see figures in \S\ref{sec:lf}), as observed. \\[-0.3cm]

\indent 3. \emph{An increased fraction of the SF takes place in ETGs at $z>0.8$ \citep{2007ApJ...654..172D}}.--- The model predicts that the bulk of the number evolution of ETGs through major mergers at $z\lesssim 1$ basically takes place during a period of $\sim 1$\,Gyr at $0.8<z<1$. As these major mergers are predominantly wet, their remnant mETGs must have undergone strong starbursts during the merging-nuclei phase of the encounter, meaning that a relevant fraction of the SF at $z>0.8$ must be occurring in spheroidal remnants that are evolving into typical ETGs, in agreement with the previous result. \\[-0.3cm]

\indent 4. \emph{Redshift $z\sim 0.8$-1 is a transition epoch in the formation history of ETGs \citep[][]{2006A&A...453L..29C,2006A&A...453..397F,2010ApJ...709..644I}}.--- 
The number evolution of mETGs is nearly frozen at $z<0.8$ in the model, in agreement with numerous observational studies \citep{2007ApJ...670..206V,2010MNRAS.tmp...24B,2009A&A...505...83H,2009ApJ...700.1559R}. The slow down of the generation of ETGs at $z\lesssim 0.8$ is due to two facts: basically, to the observational decrease of the merger fractions at these redshifts and, secondarily, to the more relevant role played by dry and mixed mergers at $z<0.8$ in comparison to wet events (see Fig.\,\ref{fig:drywetmixedmergers}). Our model proves that the declining of the wet mergers rate at $z\lesssim 1$ can explain the observed slow down in the assembly of massive ellipticals since that epoch, as proposed by \citet{2010ApJ...709..644I}. For more details, consult Paper II.\\[-0.3cm]

\indent 5. \emph{The evolution of the global LF is mainly driven by the SF associated to disk galaxies, instead of by the number evolution of galaxies \citep{2008A&A...491..713W}}.--- If we compare the LFs of the total galaxy populations with the LFs of disks and ETGs in Figs.\,\ref{fig:lfcolor1}-\ref{fig:lfcolor3}, we can conclude that the evolution of disk galaxies (which is basically due to their SFH) determines the global shape and behaviour of the LF of all galaxies at $z\lesssim 1$, in agreement with this result by \citeauthor{2008A&A...491..713W} \\[-0.3cm]

\indent 6. \emph{The SFRs of star-forming galaxies seem to be largely unaffected by the local processes that truncate SF at $z\sim 0.8$ in high density regions \citep{2009ApJ...705L..67P}}.--- In the model, the SF quenching takes place during the post-merger phase of major mergers. Normal disks not involved in major mergers continue with their exponentially-decaying SFR, without experiencing SF quenching. If we also consider that the enhancement of the SFR due to major mergers is negligible as compared to the typical SFR undergone by normal disks even at $z\sim 0.8$-1 \citep[][]{2009MNRAS.398...75S,2009ApJ...704..324R,2010ApJ...710.1170L}, the model implies that the global SFR of typical disks is not affected by the processes that shut down the SF at $z\sim 0.8$, in agreement with the result stated above.\\[-0.3cm]

\indent 7. \emph{Observational major merger rates must account for $\sim 50$\% of the mass build-up at the massive end at $z\lesssim1$ \citep{2008A&A...491..713W}}.--- At the brightest magnitudes, where the LF of mETGs dominate over the rest of galaxy types in the global LF, the number evolution of mETGs determines the evolution of the global LF, and explains the buildup of $\sim 50$\% of the systems in this range (see Figs.\,\ref{fig:lfcolor1}-\ref{fig:lfk}). The late buildup of mETGs is being confirmed by recent spectral studies that conclude that $\sim 50$\% of the total stellar mass density of mETGs has appeared at $z<1$ \citep[][]{2005ApJ...621..673T,2006A&A...453..397F,2008ApJ...682..919C,2010ApJ...709..644I}. Our model proves the feasibility of the conclusion derived by \citeauthor{2008A&A...491..713W}\\[-0.3cm]

\indent 8. \emph{A significant DSF population appears at $0.8<z<1$ \citep{2002A&A...381L..68C,2003ApJ...586..765Y,2003MNRAS.346.1125G,2004ApJ...600L.131M,2005ApJ...620..595W}}.--- The contribution of gas-rich major mergers is negligible at $z<0.7$, and thus DSFs (understood as transitory stages of major mergers by the model) are not relevant numerically at these redshifts, in agreement with observations \citep{2004ApJ...600L..11B}. However, the model predicts that large amounts of DSFs must appear at $z>0.8$, associated to the rise of gas-rich major mergers. According to the model predictions, $\sim 50$\% of the galaxies in the massive-end of the red sequence at $z\sim 1$ must be DSFs, in agreement with observations \citep{2002A&A...381L..68C,2003ApJ...586..765Y,2003MNRAS.346.1125G,2004ApJ...600L.131M,2009MNRAS.395..144W,2009arXiv0910.1598W}. \\[-0.3cm]

\indent 9. \emph{The build up of ETGs at $z\lesssim 1.2$ seems to have been confined only to low-density environments, and not to galaxy clusters \citep{2009ApJ...697L.137P}}.--- The model is also coherent with this observational fact. Notice that it does not consider the effects of the environmental density on galaxy evolution, but a global, numerical evolution of field galaxies. In fact, the observational merger fractions and local LFs that we use are derived from samples of field galaxies. So, the model is tracing the average build up of mETGs in the field since $z\sim 1$. \\[-0.3cm]

\indent 10. \emph{Above $\mathcal{M}_* \gtrsim 10^ {10.8}\Msun$, blue E-S0's at $z<1.4$ resemble to merger remnants probably migrating to the red-sequence \citep{2010arXiv1002.3076H}}.--- Our model is coherent with this result, and proves the feasibility of reproducing the observed mass migration from the blue galaxy cloud to the red sequence at $z\lesssim 1$ for masses $\mathcal{M}_* \gtrsim 10^ {11}\Msun$, just accounting for the major mergers reported by observations.\\[-0.3cm]

\indent 11. \emph{The evolution of luminous red galaxies since $z\sim 0.5$ is consistent with that expected from passive evolution \citep{2010arXiv1001.2015T}}.--- Our model is in agreement with this result, as the buildup of mETGs is nearly frozen at $z\lesssim 0.7$ (see paper II of this series for more details).

\subsection{Model limitations}
\label{sec:limitations}

The model reproduces the evolution of the number density of all galaxy types at $z\lesssim 1$, but fails at $z\gtrsim 1$ (see Figs.\,\ref{fig:lfcolor1}-\ref{fig:lfsmorph}). Several aspects on galaxy evolution at $z>1$ can be limiting the validity of the model at higher redshifts:\\[-0.2cm]

\indent \emph{i. Validity of the Hubble classification scheme at $z>1$}.--- It is known that the fraction of galaxies with highly-distorted morphologies increases at $z>1$. In fact, the emergence of the Hubble Sequence seems to occur around $z\sim 1.4$  \citep{1996ApJ...472L..13O,2001defi.conf..170W,2005ApJ...631..101P,2006ApJ...652..963R}. This means that the correspondence between SFH and morphological type (as assumed in the model) lose its meaning at $z>1$. \\[-0.3cm]

\indent \emph{ii. Significant contribution of other evolutionary processes}.--- Besides major merging, other processes are capable of transforming spirals into S0-S0a galaxies, as the effects of the intracluster medium on the infallen star-forming galaxies or the minor mergers \citep{2006A&A...457...91E,2008IAUS..245..285A,2008A&A...483L..39C,2009ApJ...693..112P,2009MNRAS.395L..62G,2009ApJ...695....1T,2009MNRAS.393.1302W}. However, it seems that the effects of these processes are more relevant in the formation of intermediate-mass disks of types later than Sa than in the buildup of ETGs at $z\sim 1$ \citep{2009A&A...501..119A,2009ApJ...696..411W,2010ApJ...710.1170L}. At higher redshifts, the relative relevance of these processes might be quite different. In fact, observations at high redshift and cosmological simulations favour a scenario in which galaxies at $z>>1$ formed through steady, narrow, cold gas streams that penetrate the shock-heated media of massive dark matter haloes \citep{2008ApJ...680..246T,2009Natur.457..451D,2009ApJ...703..785D,2009ApJ...706.1364F,2010MNRAS.tmp..440C}.\\[-0.3cm]

\indent \emph{iii. Higher gas amounts in disks at $z>1$}.--- Major mergers at $z>1$ basically involve gas-rich galaxies with higher gas contents than their local counterparts \citep{2005ApJ...631..101P,2008ApJ...687...59G,2008ApJ...680..246T,2009ApJ...706.1364F,2009ApJ...697.2057L}. Recent observations and simulations show that major mergers in extremely gas-rich media can give place to galaxies of later types than S0, as gas infall can rebuild the disk in $\sim 2$-3\,Gyr \citep[see][]{2007AJ....133.2132S,2009MNRAS.398..312G,2009A&A...496..381H}. So, as this situation seems to be common for major mergers at $z>1$, the main model hypothesis may be not fulfilled, invalidating model results at $z>1$. Moreover, numerical simulations suggest that extremely gas-rich disks tend to form massive clumps \citep{2008A&A...492...31D,2009arXiv0903.1937E}. These clumps could lead to an overestimation of the real fraction of major mergers at $z>1$ through the classical CAS techniques, making the model predictions unrealistic. Higher gas reservoirs in disks at $z>1$ can also imply more massive starbursts induced by mergers. In this regard, CAS techniques could identify a minor merger event as a major one, overestimating again the number of major mergers at $z>1$.\\[-0.3cm]

\indent \emph{iv. Observational merger fractions at $z>1$}.--- The merger fractions used in the model are robust at $z<1$, where they were derived (see \S\ref{sec:assumptions}). Results for higher redshifts use the extrapolation of the trend at $z<1$ to earlier epochs.  Although this extrapolation is very similar at $1<z<1.5$ to the merger fractions derived by \citet{2008MNRAS.386..909C}, they may be overestimated by a factor of $\sim 2$-3 \citep{2009ApJ...697.1971J}, contributing to the uncertainties of the model results at $z>1$.

\section{Summary and conclusions}
\label{sec:conclusions}

We present a model that predicts the backwards-in-time evolution of the local mETG population (with $\log(\mathcal{M}_*/\Msun)>11$ at $z=0$), under the hypothesis that each observed major merger gives place to an ETG. The model traces back-in-time the evolution of the local galaxy populations considering the number evolution derived from observational merger fractions and the L-evolution of each galaxy type due to typical SFHs. The main novelty of the model is the realistic treatment of the effects of major mergers on the LFs. In particular, we have had into account the different timescales of a major merger, the colors of merging galaxies at each merger phase, and the different morphological types of the progenitors depending on whether the merger is wet, mixed, or dry. All the model parameters and evolutionary processes considered in the model are based on robust results of observational and computational studies. 

The model is capable of reproducing the observational evolution of the bright-end of the galaxy LFs at $z\lesssim 1$, simultaneously for different rest-frame bands ($B$, $I$, $K$) and selection criteria (on color, on morphology). Moreover, it can explain several recent observational results, such as the morphology of the star-forming and quiescent populations at $z\sim 0.8$, the global evolution of the SFR at $z<1$, and the fraction of red sequence contamination by DSFs up to $z\sim 1$. 

The model predicts the appearance of $\sim 50$-60\% of the present-day number density of mETGs since $z\sim 1$ (in agreement with some observational studies), just accounting for the major mergers strictly reported by observations during this epoch. The bulk of this evolution has taken place during a time period of $\sim 1$\,Gyr elapsed at $0.8<z<1$, also in agreement with observations. So, the model reproduces the progressive migration of stellar mass from the bright-end of the blue galaxy cloud to that of the red sequence observed during the last $\sim 8$\,Gyr through mixed and wet major mergers, and from the lower masses to the higher masses of the red sequence through dry and mixed mergers. In this sense, the model corroborates the feasibility of the mixed evolutionary scenario proposed by F07, al least for the bright-end of the LFs at $z\lesssim 1$. 

The framework proposed by the model can also reconcile conflicting observational results on the amount of number evolution experienced by massive red galaxies, just considering the usual contamination by DSFs exhibited by red galaxy samples at $z>0.7$. Moreover, the number evolution of massive galaxies predicted by the model at $z\lesssim 1$ is compatible with global trends of mass-downsizing, a fact that suggests that the hierarchical scenario and this observational phenomenon can be made compatible. 

The model provides several secondary results on the evolution of galaxy populations, summarized below:
\begin{itemize}

\item The model reproduces naturally the excess of bright late-type galaxies (by a factor $\sim 4$-5) observed at $0.8<z<1$ with respect to PLE models (or, equivalently, of $\sim 30$-40\% of bright blue galaxies), just considering the gas-rich progenitors of these recently-assembled mETGs. 

\item It also explains the observational fact that $z\sim 0.8$ is a transition redshift in the recent assembly history of mETGs. The combination of the decrease of major merger fractions at $z<1$ with the dominant role played by mixed and dry mergers since $z\sim 0.8$ over wet mergers makes the assembly of mETGs to be nearly frozen at $z< 0.8$ (in terms of net growth). 

\item The model predicts the appearance of a significant population of DSFs at $z\gtrsim 0.8$, associated to the rise of gas-rich major mergers at these epochs. The predicted number density of DSFs at $z\sim 1$ (considered as transitory stages of gas-rich major mergers) can explain the $\sim 50$\% of contamination of the red sequence by dusty galaxies reported by observations at $z\sim 1$. This DSF population is essential for reproducing the noticeable observational number evolution experienced by bright blue galaxies in the $K$-band since $z\sim 1$. 

\end{itemize}

The role of major mergers in the assembly of mETGs has been supported by several observational studies, but this is the first time that it is proved the feasibility of building up nearly half of the present-day number density of mETGs through the major mergers strictly reported by observations at $z\lesssim 1$. The model provides an evolutionary scenario that establishes naturally a link between the major mergers occurring at $z<1$, the appearance of a significant population of DSFs at $0.8<z<1$, and a significant hierarchical assembly of mETGs during the last $\sim 8$\,Gyr, in agreement with predictions of hierarchical models and with observations.

\begin{acknowledgements}
The authors want to thank the anonymous referee for useful suggestions that clearly improved the paper. We wish to acknowledge J.\,P.Gardner for making its \ncmod\/ code publicly available, and Antonio Ben\'{\i}tez for interesting discussions and comments. Supported by the Spanish Programa Nacional de Astronom\'{\i}a y Astrof\'{\i}sica, under projects AYA2006-02358, AYA2009-10368, and AYA2006–12955. MCEM acknowledges support from  the Madrid Regional Government through the ASTRID Project (S0505/ESP-0361), for development and exploitation of astronomical instrumentation (http://www.astrid-cm.org/). Partially funded by the Spanish MICINN under the Consolider-Ingenio 2010 Program grant CSD2006-00070: "First Science with the GTC" (http://www.iac.es/consolider-ingenio-gtc/).
 
\end{acknowledgements}

\begin{appendix}

\section{Decomposition of the LFs in $r^*$ by N03 in finer morphological type bins}
\label{append:LFdistribution}

\subsection{General description of the procedure}\label{sec:procedureintroduction}

As commented in \S\ref{sec:lfbymorphologicaltype}, we have decomposed the $r^*$-band LFs derived by N03 into thinner classes, using the information of the LFs in the $B$-band by C03. The morphological types defined by C03 are thin enough as to reproduce each of the coarser morphological classes defined by N03 as a sum of "thin" types. As the morphological classifications of a galaxy in the $B$ and $r^*$ bands are similar for local galaxies \citep[][]{2000ApJS..131..441K}, we can expect that a galaxy at $z\sim 0$ exhibits nearly the same morphological classification at different bands. If a univocal relation between the absolute magnitudes in the $B$ and $r^*$ bands exists for each galaxy type at $z\sim 0$, the number of galaxies of a certain type with magnitude $M(r^*)$ would be the same as the number of galaxies of the same type exhibiting the corresponding $M(B)$\,magnitude. Let us assume that this univocal relation exists at $z\sim 0$, at least fulfilled by all galaxies within the same morphological type (we will justify it in \S\ref{sec:proportionality}). This means that, for all the galaxies within a given morphological type:

\begin{equation}\label{eq:assumption}
M(B) = a_0 + a_1 M(r^*),
\end{equation}
\noindent being $a_0$ and $a_1$ constants that depend on the morphological type.

\noindent We can also assume that all the galaxies detected in $B$ are going to be also detected in $r^*$ (this is true at least for bright magnitudes). Therefore, the number of galaxies with $M(r^*)$ and with $M^\mathrm{eq}(B)$ will be the same for each morphological type\footnote{We will note the $M(B)$ magnitude corresponding to a given value of $M(r^*)$ through eq.\,\ref{eq:assumption} as $M^\mathrm{eq}(B)$.}. In particular, this occurs for the E's and S0's, and hence: $\phi[M(r^*),\mathrm{"E"}] = \phi[M_\mathrm{E}^\mathrm{eq}(B),\mathrm{"E"}]$ and $\phi [M(r^*),\mathrm{"S0"}] = \phi[M_\mathrm{S0}^\mathrm{eq}(B),\mathrm{"S0"}]$. Notice that the $M(B)$\,magnitude corresponding to the same $M(r^*)$ for different morphological types will be different a priori (that is why we note them by the subindices "E" and "S0"). The previous expressions imply that: 
{\small
\begin{equation}\label{eq:coc1}
\phi[M(r^*),\mathrm{"E"}]\,/\, \phi[M(r^*),\mathrm{"S0"}]= \phi[M_\mathrm{E}^\mathrm{eq}(B),\mathrm{"E"}]\, \,/ \phi[M_\mathrm{S0}^\mathrm{eq}(B),\mathrm{"S0"}].
\end{equation}
}

\noindent This equation would be valid for all $M(r^*)$. The last term of the previous equation is given by the LFs by C03, provided we know the $M(r^*)$-$M(B)$ relation for each morphological type. Obviously, we can consider that:

\begin{equation}\label{eq:coc2}
\phi[M(r^*),\mathrm{"E-S0"}] = \phi[M(r^*),\mathrm{"E"}] + \phi[M(r^*),\mathrm{"S0"}] ,
\end{equation}

\noindent where the left term is given by the LF derived by N03. Therefore, if the univocal relation $M(r^*)$-$M^\mathrm{eq}(B)$ for each galaxy type exists and it is known, we can decompose the local LF in $r^*$ for E-S0's into thinner classes ($\phi[M(r^*),\mathrm{"E"}]$ and $\phi[M(r^*),\mathrm{"S0"}]$), using eqs.\,\ref{eq:coc1} and \ref{eq:coc2}. 

As this procedure can be performed also with the rest of morphological classes, we could split the four LFs in the $r^*$-band by N03 into the eleven thin types defined by C03, as required. Nevertheless, the introduction of so many types in the model does not insert any additional advantage, because the characteristic properties of adjacent, thin morphological classes are quite similar. However, C03 also derived the local $B$-band LFs for six coarser morphological classes (E, S0-S0a, Sa-Sab, Sb-Sbc, Sc-Scd, and Sd-Irr). Note that the four classes by N03 are not strictly compatible with these six coarser types, in the sense that they can not be reproduced as simple sums of these six types, except for certain classes. Therefore, we have combined the information by N03 with the information for thinner and coarser types by C03 to derive the LFs in the $r^*$-band for the following morphological types: E-S0a, Sa-Sab, Sb-Sbc, Sc-Scd, and Sd-Irr. Although we are just considering only one morphological type more than N03, the new distribution of morphological classes groups galaxies with more similar spectrophotometric and morphological properties than the classes defined by N03. The LFs in these five classes can be easily obtained through a procedure analogous to that used to derive eqs.\,\ref{eq:coc1} and \ref{eq:coc2}, combining the LFs by N03 and those by C03 in coarse and thin types. 

The resulting LFs are plotted Fig.\,\ref{fig:lfs}. Their Schechter parametrizations are listed in Table\,\ref{tab:LFs4}. In order to obtain an adequate Schechter fit to the LF of ETGs obtained through this procedure, we have fixed the $\alpha$ slope of ETGs. This does not affect the results in the model, as we are limited to the bright-end of all LFs by the limiting absolute magnitude of the observational merger fractions by LSJ09. The residuals between the total LF derived by N03 and the resulting one from the decomposition are lower than $\sim 0.1$ in $M(r^ *)>-22$\,mag, peaking at the brightest magnitudes. Residuals between the LFs obtained for each morphological type are lower than 5\% in the whole magnitude range of interest, supporting the robustness of the procedure.

\subsection{Proportionality between $M(B)$ and $M(r^*)$ for local galaxies}
\label{sec:proportionality}

\citet{2006astro.ph.12034Z} indicate that magnitudes in the $B$-band can be transformed to the $r^*$-band, according to: $r^* = V - 0.451\,(B-V) + 0.082$. We have selected a local galaxy sample  ($z<0.1$) from the GOYA survey \citep{2007RMxAC..29..165A}, which basically contains spiral galaxies. We have checked that the $(B-V)$ color distribution of local galaxies is quite narrow ($<B-V> = 0.8529$, with $\sigma= 0.1218$). So, for local spirals, we can derive that: $r^* \sim  V - 0.303$, by replacing this median color value into \citet{2006astro.ph.12034Z} conversion equation. 

We have plotted the magnitude of these local galaxies in the $V$-band against their magnitudes in the $B$-band. A well-defined linear trend is found, with a Pearson linear correlation coefficient of $\rho \sim 0.96$.  Therefore, we can consider that for local galaxies:  $B = c_0 + c_1 V $, being $c_0=-1.3\pm 0.4$ and $c_1=1.02\pm 0.06$ valid for Sa-Irr galaxies. Replacing this relation into the previous $r^ *$-$V$ relation, we find that: $B \sim c_1 r* + (0.303c_1 + c_0)$. This implies that, for local spiral galaxies, we can assume that $ m(B) \propto m(r^*)$. Considering that the k- and e-corrections at $z<0.1$ are negligible and that global effects of dust-extinction can be neglected, the previous relation could be extrapolated to absolute magnitudes, and thus: $M(B) \propto M(r^*)$. Therefore, the assumption in eq.\,\ref{eq:assumption} is plausible for local disk galaxies, resulting for them:

\vspace{-0.2cm}\begin{equation}\label{eq:b2r2}
 M(B) 	\sim 0.98 M(r^*) + 1.6.
\end{equation} 

The existence of the narrow red sequence of galaxies in the local Universe implies that a similar relation must exist for E-S0's, as it is basically composed by E-S0's at $z\sim 0$ \citep{2006ApJ...648..268B}. As the k- and e-corrections and the dust extinction are also negligible in the local Universe for E-S0's, we can consider that $M(B) -M(r^*)\sim c_2$ also for E-S0a's. In order to obtain $c_2$, we have compared the LFs of ETGs in the $B$ and $r^*$ bands, estimating the displacement on the magnitude axis required in order to make them to overlap. For local E-S0a's, we can derive that:

\vspace{-0.2cm}\begin{equation}\label{eq:b2r3}
 M(B) \sim M(r^*) + 1.5.
\end{equation} 
 
\noindent Equation\,\ref{eq:b2r2} is consistent with the typical, observational colors derived for Sa-Sc galaxies ($<B-R>\sim 1.4$), while eq.\,\ref{eq:b2r3} is corroborated by typical colors of real E-S0 galaxies \citep[$<B-R>\sim 1.6$, see][]{2001MNRAS.326..745M}. Equations \ref{eq:b2r2} and \ref{eq:b2r3} prove the assumption adopted in eq.\,\ref{eq:assumption}. The residuals obtained by comparing the total LF derived by N03 in $r^*$ with the total LF by C03 transformed to the $r^*$-band (using eq.\,\ref{eq:b2r2}-\ref{eq:b2r3}) are lower than $\sim 0.01$, supporting the assumption adopted in eq.\,\ref{eq:assumption}.

\end{appendix}

\newpage
%\begin{sidewaysfigure}[b]
\begin{sidewaysfigure*}[t]
\vspace{17cm}

\begin{center}
\includegraphics*[width=\textwidth,angle=0]{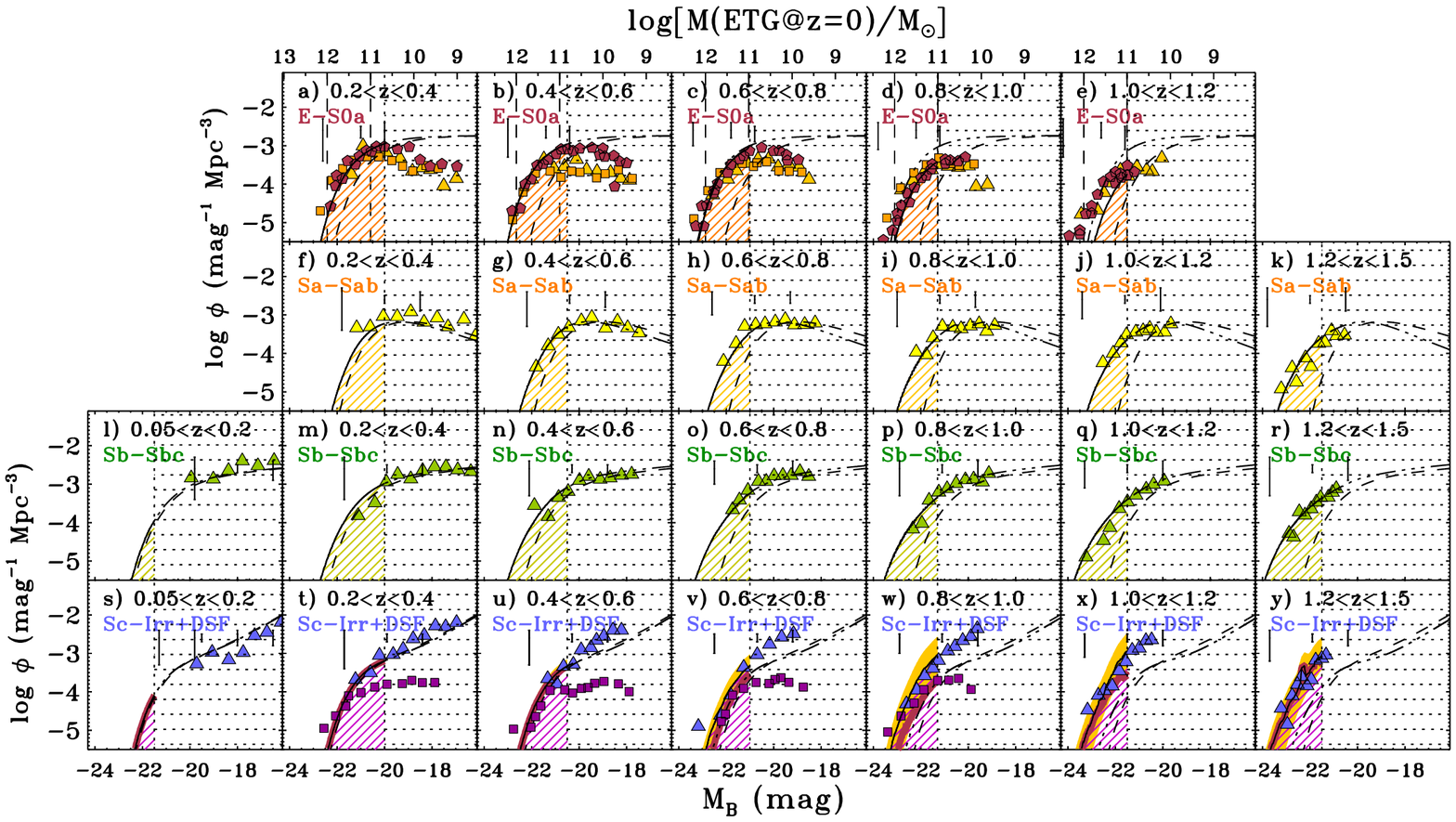}
\caption{Predicted evolution of the $B$-band LFs by morphological types (E-S0a, Sa-Sab, Sb-Sbc, Sc-Irr) with redshift, up to $z\sim 1.5$. Observational data have been overplotted for comparison. The abscissas axis (absolute magnitude axis) has been transformed into stellar masses using the expression derived for local ETGs by \citet{2006A&A...453L..29C} (axis at the top of the figure). \emph{Panels a-e}: Evolution of the LF for E-S0's. \emph{Panels f-k}: For Sa-Sab's. \emph{Panels l-r}: For Sb-Sbc's. \emph{Panels s-y}: For Sc-Irr$+$M's. {\bf Symbols\/}: Observational data in the $B$-band for each morphological type. Typical observational error bars in the whole magnitude range have been overplotted for comparison. \emph{Squares}: Data by \citet{2007ApJS..172..406S}. \emph{Triangles}: Data by \citet{2006A&A...455..879Z}. \emph{Pentagons}: Data for red galaxies by F07. {\bf Lines\/}: Model predictions. The legend is analogous to that of Fig.\,\ref{fig:lfcolor1}, but referred to morphological classes, instead of color classes. {\bf Shaded area\/}: Total uncertainty region due to errors in the nominal values used in the model for the merger timescales and the merger fractions. Notice that Sa-Sbc galaxies are not affected by major mergers in the model (and hence, they do not exhibit significant uncertainties). The uncertainties of the model predictions for ETGs are shown in Fig.\,\ref{fig:lfcolor1}. [\emph{A color version of this plot is available at the electronic edition}].}\label{fig:lfsmorph}
\end{center}

\end{sidewaysfigure*}
%\end{sidewaysfigure}
\newpage

\bibliographystyle{aa}{}
\bibliography{elic1_v1.bib}{}

\begin{thebibliography}{260}
\expandafter\ifx\csname natexlab\endcsname\relax\def\natexlab#1{#1}\fi

\bibitem[{{Abreu} {et~al.}(2007){Abreu}, {Balcells}, {Garc{\'{\i}}a-Dab{\'o}},
  {Prieto}, {Erwin}, \& {Eliche-Moral}}]{2007RMxAC..29..165A}
{Abreu}, D., {Balcells}, M., {Garc{\'{\i}}a-Dab{\'o}}, C.~E., {et~al.} 2007, in
  Revista Mexicana de Astronomia y Astrofisica Conference Series, Vol.~29,
  Revista Mexicana de Astronomia y Astrofisica Conference Series, ed.
  R.~{Guzm{\'a}n}, 165--165

\bibitem[{{Alonso-Herrero} {et~al.}(2004){Alonso-Herrero},
  {P{\'e}rez-Gonz{\'a}lez}, {Rigby}, {Rieke}, {Le Floc'h}, {Barmby}, {Page},
  {Papovich}, {Dole}, {Egami}, {Huang}, {Rigopoulou},
  {Crist{\'o}bal-Hornillos}, {Eliche-Moral}, {Balcells}, {Prieto}, {Erwin},
  {Engelbracht}, {Gordon}, {Werner}, {Willner}, {Fazio}, {Frayer}, {Hines},
  {Kelly}, {Latter}, {Misselt}, {Miyazaki}, {Morrison}, {Rieke}, \&
  {Wilson}}]{2004ApJS..154..155A}
{Alonso-Herrero}, A., {P{\'e}rez-Gonz{\'a}lez}, P.~G., {Rigby}, J., {et~al.}
  2004, \apjs, 154, 155

\bibitem[{{Arag{\'o}n-Salamanca}(2008)}]{2008IAUS..245..285A}
{Arag{\'o}n-Salamanca}, A. 2008, in IAU Symposium, Vol. 245, IAU Symposium, ed.
  M.~{Bureau}, E.~{Athanassoula}, \& B.~{Barbuy}, 285--288

\bibitem[{{Arnouts} {et~al.}(2007){Arnouts}, {Walcher}, {Le F{\`e}vre},
  {Zamorani}, {Ilbert}, {Le Brun}, {Pozzetti}, {Bardelli}, {Tresse}, {Zucca},
  {Charlot}, {Lamareille}, {McCracken}, {Bolzonella}, {Iovino}, {Lonsdale},
  {Polletta}, {Surace}, {Bottini}, {Garilli}, {Maccagni}, {Picat},
  {Scaramella}, {Scodeggio}, {Vettolani}, {Zanichelli}, {Adami}, {Cappi},
  {Ciliegi}, {Contini}, {de la Torre}, {Foucaud}, {Franzetti}, {Gavignaud},
  {Guzzo}, {Marano}, {Marinoni}, {Mazure}, {Meneux}, {Merighi}, {Paltani},
  {Pell{\`o}}, {Pollo}, {Radovich}, {Temporin}, \&
  {Vergani}}]{2007A&A...476..137A}
{Arnouts}, S., {Walcher}, C.~J., {Le F{\`e}vre}, O., {et~al.} 2007, \aap, 476,
  137

\bibitem[{{Arribas} {et~al.}(2004){Arribas}, {Bushouse}, {Lucas}, {Colina}, \&
  {Borne}}]{2004AJ....127.2522A}
{Arribas}, S., {Bushouse}, H., {Lucas}, R.~A., {Colina}, L., \& {Borne}, K.~D.
  2004, \aj, 127, 2522

\bibitem[{{Athanassoula}(2008)}]{2008mgng.conf...47A}
{Athanassoula}, E. 2008, in Mapping the Galaxy and Nearby Galaxies, ed.
  K.~{Wada} \& F.~{Combes}, 47

\bibitem[{{Azzollini} {et~al.}(2009){Azzollini}, {Beckman}, \&
  {Trujillo}}]{2009A&A...501..119A}
{Azzollini}, R., {Beckman}, J.~E., \& {Trujillo}, I. 2009, \aap, 501, 119

\bibitem[{{Banerji} {et~al.}(2010){Banerji}, {Ferreras}, {Abdalla}, {Hewett},
  \& {Lahav}}]{2010MNRAS.tmp...24B}
{Banerji}, M., {Ferreras}, I., {Abdalla}, F.~B., {Hewett}, P., \& {Lahav}, O.
  2010, \mnras, 24

\bibitem[{{Barnes} \& {Hernquist}(1996)}]{1996ApJ...471..115B}
{Barnes}, J.~E. \& {Hernquist}, L. 1996, \apj, 471, 115

\bibitem[{{Barnes} \& {Hernquist}(1991)}]{1991ApJ...370L..65B}
{Barnes}, J.~E. \& {Hernquist}, L.~E. 1991, \apjl, 370, L65

\bibitem[{{Bekki} \& {Shioya}(2000)}]{2000A&A...362...97B}
{Bekki}, K. \& {Shioya}, Y. 2000, \aap, 362, 97

\bibitem[{{Bell} \& {de Jong}(2001)}]{2001ApJ...550..212B}
{Bell}, E.~F. \& {de Jong}, R.~S. 2001, \apj, 550, 212

\bibitem[{{Bell} {et~al.}(2004{\natexlab{a}}){Bell}, {McIntosh}, {Barden},
  {Wolf}, {Caldwell}, {Rix}, {Beckwith}, {Borch}, {H{\"a}ussler}, {Jahnke},
  {Jogee}, {Meisenheimer}, {Peng}, {Sanchez}, {Somerville}, \&
  {Wisotzki}}]{2004ApJ...600L..11B}
{Bell}, E.~F., {McIntosh}, D.~H., {Barden}, M., {et~al.} 2004{\natexlab{a}},
  \apjl, 600, L11

\bibitem[{{Bell} {et~al.}(2006{\natexlab{a}}){Bell}, {Naab}, {McIntosh},
  {Somerville}, {Caldwell}, {Barden}, {Wolf}, {Rix}, {Beckwith}, {Borch},
  {H{\"a}ussler}, {Heymans}, {Jahnke}, {Jogee}, {Koposov}, {Meisenheimer},
  {Peng}, {Sanchez}, \& {Wisotzki}}]{2006ApJ...640..241B}
{Bell}, E.~F., {Naab}, T., {McIntosh}, D.~H., {et~al.} 2006{\natexlab{a}},
  \apj, 640, 241

\bibitem[{{Bell} {et~al.}(2006{\natexlab{b}}){Bell}, {Phleps}, {Somerville},
  {Wolf}, {Borch}, \& {Meisenheimer}}]{2006ApJ...652..270B}
{Bell}, E.~F., {Phleps}, S., {Somerville}, R.~S., {et~al.} 2006{\natexlab{b}},
  \apj, 652, 270

\bibitem[{{Bell} {et~al.}(2004{\natexlab{b}}){Bell}, {Wolf}, {Meisenheimer},
  {Rix}, {Borch}, {Dye}, {Kleinheinrich}, {Wisotzki}, \&
  {McIntosh}}]{2004ApJ...608..752B}
{Bell}, E.~F., {Wolf}, C., {Meisenheimer}, K., {et~al.} 2004{\natexlab{b}},
  \apj, 608, 752

\bibitem[{{Bendo} \& {Barnes}(2000)}]{2000MNRAS.316..315B}
{Bendo}, G.~J. \& {Barnes}, J.~E. 2000, \mnras, 316, 315

\bibitem[{{Benson} \& {Devereux}(2009)}]{2009MNRAS.tmp.1955B}
{Benson}, A.~J. \& {Devereux}, N. 2009, \mnras, 1955

\bibitem[{{Blanton}(2006)}]{2006ApJ...648..268B}
{Blanton}, M.~R. 2006, \apj, 648, 268

\bibitem[{{Boselli} {et~al.}(2002){Boselli}, {Gavazzi}, {Lequeux}, \&
  {Pierini}}]{2002A&A...385..454B}
{Boselli}, A., {Gavazzi}, G., {Lequeux}, J., \& {Pierini}, D. 2002, \aap, 385,
  454

\bibitem[{{Bournaud} {et~al.}(2004){Bournaud}, {Combes}, \&
  {Jog}}]{2004A&A...418L..27B}
{Bournaud}, F., {Combes}, F., \& {Jog}, C.~J. 2004, \aap, 418, L27

\bibitem[{{Bournaud} {et~al.}(2005){Bournaud}, {Jog}, \&
  {Combes}}]{2005A&A...437...69B}
{Bournaud}, F., {Jog}, C.~J., \& {Combes}, F. 2005, \aap, 437, 69

\bibitem[{{Bournaud} {et~al.}(2007){Bournaud}, {Jog}, \&
  {Combes}}]{2007A&A...476.1179B}
{Bournaud}, F., {Jog}, C.~J., \& {Combes}, F. 2007, \aap, 476, 1179

\bibitem[{{Brammer} {et~al.}(2009){Brammer}, {Whitaker}, {van Dokkum},
  {Marchesini}, {Labb{\'e}}, {Franx}, {Kriek}, {Quadri}, {Illingworth}, {Lee},
  {Muzzin}, \& {Rudnick}}]{2009ApJ...706L.173B}
{Brammer}, G.~B., {Whitaker}, K.~E., {van Dokkum}, P.~G., {et~al.} 2009, \apjl,
  706, L173

\bibitem[{{Bridge} {et~al.}(2007){Bridge}, {Appleton}, {Conselice}, {Choi},
  {Armus}, {Fadda}, {Laine}, {Marleau}, {Carlberg}, {Helou}, \&
  {Yan}}]{2007ApJ...659..931B}
{Bridge}, C.~R., {Appleton}, P.~N., {Conselice}, C.~J., {et~al.} 2007, \apj,
  659, 931

\bibitem[{{Brown} {et~al.}(2007){Brown}, {Dey}, {Jannuzi}, {Brand}, {Benson},
  {Brodwin}, {Croton}, \& {Eisenhardt}}]{2007ApJ...654..858B}
{Brown}, M.~J.~I., {Dey}, A., {Jannuzi}, B.~T., {et~al.} 2007, \apj, 654, 858

\bibitem[{{Bruzual} \& {Charlot}(1993)}]{1993ApJ...405..538B}
{Bruzual}, G. \& {Charlot}, S. 1993, \apj, 405, 538

\bibitem[{{Bruzual} \& {Charlot}(2003)}]{2003MNRAS.344.1000B}
{Bruzual}, G. \& {Charlot}, S. 2003, \mnras, 344, 1000

\bibitem[{{Bruzual} {et~al.}(1988){Bruzual}, {Magris}, \&
  {Calvet}}]{1988ApJ...333..673B}
{Bruzual}, G., {Magris}, G., \& {Calvet}, N. 1988, \apj, 333, 673

\bibitem[{{Buat} {et~al.}(2002){Buat}, {Boselli}, {Gavazzi}, \&
  {Bonfanti}}]{2002A&A...383..801B}
{Buat}, V., {Boselli}, A., {Gavazzi}, G., \& {Bonfanti}, C. 2002, \aap, 383,
  801

\bibitem[{{Buat} \& {Xu}(1996)}]{1996A&A...306...61B}
{Buat}, V. \& {Xu}, C. 1996, \aap, 306, 61

\bibitem[{{Bundy} {et~al.}(2006){Bundy}, {Ellis}, {Conselice}, {Taylor},
  {Cooper}, {Willmer}, {Weiner}, {Coil}, {Noeske}, \&
  {Eisenhardt}}]{2006ApJ...651..120B}
{Bundy}, K., {Ellis}, R.~S., {Conselice}, C.~J., {et~al.} 2006, \apj, 651, 120

\bibitem[{{Bundy} {et~al.}(2004){Bundy}, {Fukugita}, {Ellis}, {Kodama}, \&
  {Conselice}}]{2004ApJ...601L.123B}
{Bundy}, K., {Fukugita}, M., {Ellis}, R.~S., {Kodama}, T., \& {Conselice},
  C.~J. 2004, \apjl, 601, L123

\bibitem[{{Bundy} {et~al.}(2009){Bundy}, {Fukugita}, {Ellis}, {Targett},
  {Belli}, \& {Kodama}}]{2009ApJ...697.1369B}
{Bundy}, K., {Fukugita}, M., {Ellis}, R.~S., {et~al.} 2009, \apj, 697, 1369

\bibitem[{{Bundy} {et~al.}(2007){Bundy}, {Treu}, \&
  {Ellis}}]{2007ApJ...665L...5B}
{Bundy}, K., {Treu}, T., \& {Ellis}, R.~S. 2007, \apjl, 665, L5

\bibitem[{{Calzetti}(1999)}]{1999Ap&SS.266..243C}
{Calzetti}, D. 1999, \apss, 266, 243

\bibitem[{{Carlberg} {et~al.}(2000){Carlberg}, {Cohen}, {Patton}, {Blandford},
  {Hogg}, {Yee}, {Morris}, {Lin}, {Hall}, {Sawicki}, {Wirth}, {Cowie}, {Hu}, \&
  {Songaila}}]{2000ApJ...532L...1C}
{Carlberg}, R.~G., {Cohen}, J.~G., {Patton}, D.~R., {et~al.} 2000, \apjl, 532,
  L1

\bibitem[{{Carlberg} {et~al.}(1999){Carlberg}, {Yee}, {Morris}, {Lin},
  {Sawicki}, {Wirth}, {Patton}, {Shepherd}, {Ellingson}, \&
  {Schade}}]{1999RSPTA.357..167C}
{Carlberg}, R.~G., {Yee}, H.~K.~C., {Morris}, S.~L., {et~al.} 1999, Royal
  Society of London Philosophical Transactions Series A, 357, 167

\bibitem[{{Cassata} {et~al.}(2008){Cassata}, {Cimatti}, {Kurk}, {Rodighiero},
  {Pozzetti}, {Bolzonella}, {Daddi}, {Mignoli}, {Berta}, {Dickinson},
  {Franceschini}, {Halliday}, {Renzini}, {Rosati}, \&
  {Zamorani}}]{2008A&A...483L..39C}
{Cassata}, P., {Cimatti}, A., {Kurk}, J., {et~al.} 2008, \aap, 483, L39

\bibitem[{{Cattaneo} {et~al.}(2010){Cattaneo}, {Mamon}, {Warnick}, \&
  {Knebe}}]{2010arXiv1002.3257C}
{Cattaneo}, A., {Mamon}, G.~A., {Warnick}, K., \& {Knebe}, A. 2010, ArXiv
  e-prints

\bibitem[{{Ceverino} {et~al.}(2010){Ceverino}, {Dekel}, \&
  {Bournaud}}]{2010MNRAS.tmp..440C}
{Ceverino}, D., {Dekel}, A., \& {Bournaud}, F. 2010, \mnras, 440

\bibitem[{{Chabrier}(2001)}]{2001ApJ...554.1274C}
{Chabrier}, G. 2001, \apj, 554, 1274

\bibitem[{{Chabrier}(2002)}]{2002ApJ...567..304C}
{Chabrier}, G. 2002, \apj, 567, 304

\bibitem[{{Chabrier}(2003)}]{2003PASP..115..763C}
{Chabrier}, G. 2003, \pasp, 115, 763

\bibitem[{{Cimatti} {et~al.}(2002){Cimatti}, {Daddi}, {Mignoli}, {Pozzetti},
  {Renzini}, {Zamorani}, {Broadhurst}, {Fontana}, {Saracco}, {Poli},
  {Cristiani}, {D'Odorico}, {Giallongo}, {Gilmozzi}, \&
  {Menci}}]{2002A&A...381L..68C}
{Cimatti}, A., {Daddi}, E., {Mignoli}, M., {et~al.} 2002, \aap, 381, L68

\bibitem[{{Cimatti} {et~al.}(2006){Cimatti}, {Daddi}, \&
  {Renzini}}]{2006A&A...453L..29C}
{Cimatti}, A., {Daddi}, E., \& {Renzini}, A. 2006, \aap, 453, L29

\bibitem[{{Cimatti} {et~al.}(2004){Cimatti}, {Daddi}, {Renzini}, {Cassata},
  {Vanzella}, {Pozzetti}, {Cristiani}, {Fontana}, {Rodighiero}, {Mignoli}, \&
  {Zamorani}}]{2004Natur.430..184C}
{Cimatti}, A., {Daddi}, E., {Renzini}, A., {et~al.} 2004, \nat, 430, 184

\bibitem[{{Cirasuolo} {et~al.}(2007){Cirasuolo}, {McLure}, {Dunlop}, {Almaini},
  {Foucaud}, {Smail}, {Sekiguchi}, {Simpson}, {Eales}, {Dye}, {Watson}, {Page},
  \& {Hirst}}]{2007MNRAS.380..585C}
{Cirasuolo}, M., {McLure}, R.~J., {Dunlop}, J.~S., {et~al.} 2007, \mnras, 380,
  585

\bibitem[{{Combes}(2008)}]{2008ASPC..396..325C}
{Combes}, F. 2008, in Astronomical Society of the Pacific Conference Series,
  Vol. 396, Astronomical Society of the Pacific Conference Series, ed. J.~G.
  {Funes} \& E.~M. {Corsini}, 325

\bibitem[{{Conselice}(2006{\natexlab{a}})}]{2006ApJ...638..686C}
{Conselice}, C.~J. 2006{\natexlab{a}}, \apj, 638, 686

\bibitem[{{Conselice}(2006{\natexlab{b}})}]{2006MNRAS.373.1389C}
{Conselice}, C.~J. 2006{\natexlab{b}}, \mnras, 373, 1389

\bibitem[{{Conselice}(2009)}]{2009MNRAS.399L..16C}
{Conselice}, C.~J. 2009, \mnras, 399, L16

\bibitem[{{Conselice} {et~al.}(2003){Conselice}, {Bershady}, {Dickinson}, \&
  {Papovich}}]{2003AJ....126.1183C}
{Conselice}, C.~J., {Bershady}, M.~A., {Dickinson}, M., \& {Papovich}, C. 2003,
  \aj, 126, 1183

\bibitem[{{Conselice} {et~al.}(2008){Conselice}, {Rajgor}, \&
  {Myers}}]{2008MNRAS.386..909C}
{Conselice}, C.~J., {Rajgor}, S., \& {Myers}, R. 2008, \mnras, 386, 909

\bibitem[{{Conselice} {et~al.}(2009){Conselice}, {Yang}, \&
  {Bluck}}]{2009MNRAS.394.1956C}
{Conselice}, C.~J., {Yang}, C., \& {Bluck}, A.~F.~L. 2009, \mnras, 394, 1956

\bibitem[{{Cool} {et~al.}(2008){Cool}, {Eisenstein}, {Fan}, {Fukugita},
  {Jiang}, {Maraston}, {Meiksin}, {Schneider}, \& {Wake}}]{2008ApJ...682..919C}
{Cool}, R.~J., {Eisenstein}, D.~J., {Fan}, X., {et~al.} 2008, \apj, 682, 919

\bibitem[{{Cortese} {et~al.}(2006){Cortese}, {Boselli}, {Buat}, {Gavazzi},
  {Boissier}, {Gil de Paz}, {Seibert}, {Madore}, \&
  {Martin}}]{2006ApJ...637..242C}
{Cortese}, L., {Boselli}, A., {Buat}, V., {et~al.} 2006, \apj, 637, 242

\bibitem[{{Cowie} {et~al.}(1996){Cowie}, {Songaila}, {Hu}, \&
  {Cohen}}]{1996AJ....112..839C}
{Cowie}, L.~L., {Songaila}, A., {Hu}, E.~M., \& {Cohen}, J.~G. 1996, \aj, 112,
  839

\bibitem[{{Cox} {et~al.}(2006){Cox}, {Jonsson}, {Primack}, \&
  {Somerville}}]{2006MNRAS.373.1013C}
{Cox}, T.~J., {Jonsson}, P., {Primack}, J.~R., \& {Somerville}, R.~S. 2006,
  \mnras, 373, 1013

\bibitem[{{Cox} {et~al.}(2008{\natexlab{a}}){Cox}, {Jonsson}, {Somerville},
  {Primack}, \& {Dekel}}]{2008MNRAS.384..386C}
{Cox}, T.~J., {Jonsson}, P., {Somerville}, R.~S., {Primack}, J.~R., \& {Dekel},
  A. 2008{\natexlab{a}}, \mnras, 384, 386

\bibitem[{{Cox} {et~al.}(2004){Cox}, {Primack}, {Jonsson}, \&
  {Somerville}}]{2004ApJ...607L..87C}
{Cox}, T.~J., {Primack}, J., {Jonsson}, P., \& {Somerville}, R.~S. 2004, \apjl,
  607, L87

\bibitem[{{Cox} {et~al.}(2008{\natexlab{b}}){Cox}, {Younger}, {Hernquist}, \&
  {Hopkins}}]{2008IAUS..245...63C}
{Cox}, T.~J., {Younger}, J., {Hernquist}, L., \& {Hopkins}, P.~F.
  2008{\natexlab{b}}, in IAU Symposium, Vol. 245, IAU Symposium, ed.
  M.~{Bureau}, E.~{Athanassoula}, \& B.~{Barbuy}, 63--66

\bibitem[{{Cuesta-Bolao} \& {Serna}(2003)}]{2003A&A...405..917C}
{Cuesta-Bolao}, M.~J. \& {Serna}, A. 2003, \aap, 405, 917

\bibitem[{{Dahlen} {et~al.}(2007){Dahlen}, {Mobasher}, {Dickinson}, {Ferguson},
  {Giavalisco}, {Kretchmer}, \& {Ravindranath}}]{2007ApJ...654..172D}
{Dahlen}, T., {Mobasher}, B., {Dickinson}, M., {et~al.} 2007, \apj, 654, 172

\bibitem[{{Darg} {et~al.}(2010){Darg}, {Kaviraj}, {Lintott}, {Schawinski},
  {Sarzi}, {Bamford}, {Silk}, {Andreescu}, {Murray}, {Nichol}, {Raddick},
  {Slosar}, {Szalay}, {Thomas}, \& {Vandenberg}}]{2010MNRAS.401.1552D}
{Darg}, D.~W., {Kaviraj}, S., {Lintott}, C.~J., {et~al.} 2010, \mnras, 401,
  1552

\bibitem[{{Davies} {et~al.}(2009){Davies}, {Gilbank}, {Glazebrook}, {Bower},
  {Baldry}, {Balogh}, {Hau}, {Li}, {McCarthy}, \&
  {Savaglio}}]{2009MNRAS.395L..76D}
{Davies}, G.~T., {Gilbank}, D.~G., {Glazebrook}, K., {et~al.} 2009, \mnras,
  395, L76

\bibitem[{{de Lapparent}(2003)}]{2003A&A...408..845D}
{de Lapparent}, V. 2003, \aap, 408, 845

\bibitem[{{De Lucia} \& {Blaizot}(2007)}]{2007MNRAS.375....2D}
{De Lucia}, G. \& {Blaizot}, J. 2007, \mnras, 375, 2

\bibitem[{{De Lucia} {et~al.}(2006){De Lucia}, {Springel}, {White}, {Croton},
  \& {Kauffmann}}]{2006MNRAS.366..499D}
{De Lucia}, G., {Springel}, V., {White}, S.~D.~M., {Croton}, D., \&
  {Kauffmann}, G. 2006, \mnras, 366, 499

\bibitem[{{De Propris} {et~al.}(2007){De Propris}, {Conselice}, {Liske},
  {Driver}, {Patton}, {Graham}, \& {Allen}}]{2007ApJ...666..212D}
{De Propris}, R., {Conselice}, C.~J., {Liske}, J., {et~al.} 2007, \apj, 666,
  212

\bibitem[{{De Propris} {et~al.}(2005){De Propris}, {Liske}, {Driver}, {Allen},
  \& {Cross}}]{2005AJ....130.1516D}
{De Propris}, R., {Liske}, J., {Driver}, S.~P., {Allen}, P.~D., \& {Cross},
  N.~J.~G. 2005, \aj, 130, 1516

\bibitem[{{de Ravel} {et~al.}(2009){de Ravel}, {Le F{\`e}vre}, {Tresse},
  {Bottini}, {Garilli}, {Le Brun}, {Maccagni}, {Scaramella}, {Scodeggio},
  {Vettolani}, {Zanichelli}, {Adami}, {Arnouts}, {Bardelli}, {Bolzonella},
  {Cappi}, {Charlot}, {Ciliegi}, {Contini}, {Foucaud}, {Franzetti},
  {Gavignaud}, {Guzzo}, {Ilbert}, {Iovino}, {Lamareille}, {McCracken},
  {Marano}, {Marinoni}, {Mazure}, {Meneux}, {Merighi}, {Paltani}, {Pell{\`o}},
  {Pollo}, {Pozzetti}, {Radovich}, {Vergani}, {Zamorani}, {Zucca}, {Bondi},
  {Bongiorno}, {Brinchmann}, {Cucciati}, {de La Torre}, {Gregorini}, {Memeo},
  {Perez-Montero}, {Mellier}, {Merluzzi}, \& {Temporin}}]{2009A&A...498..379D}
{de Ravel}, L., {Le F{\`e}vre}, O., {Tresse}, L., {et~al.} 2009, \aap, 498, 379

\bibitem[{{Dekel} {et~al.}(2009{\natexlab{a}}){Dekel}, {Birnboim}, {Engel},
  {Freundlich}, {Goerdt}, {Mumcuoglu}, {Neistein}, {Pichon}, {Teyssier}, \&
  {Zinger}}]{2009Natur.457..451D}
{Dekel}, A., {Birnboim}, Y., {Engel}, G., {et~al.} 2009{\natexlab{a}}, \nat,
  457, 451

\bibitem[{{Dekel} {et~al.}(2009{\natexlab{b}}){Dekel}, {Sari}, \&
  {Ceverino}}]{2009ApJ...703..785D}
{Dekel}, A., {Sari}, R., \& {Ceverino}, D. 2009{\natexlab{b}}, \apj, 703, 785

\bibitem[{{di Matteo} {et~al.}(2008){di Matteo}, {Bournaud}, {Martig},
  {Combes}, {Melchior}, \& {Semelin}}]{2008A&A...492...31D}
{di Matteo}, P., {Bournaud}, F., {Martig}, M., {et~al.} 2008, \aap, 492, 31

\bibitem[{{Domingue} {et~al.}(2009){Domingue}, {Xu}, {Jarrett}, \&
  {Cheng}}]{2009arXiv0901.4545D}
{Domingue}, D.~L., {Xu}, C.~K., {Jarrett}, T.~H., \& {Cheng}, Y. 2009, ArXiv
  e-prints

\bibitem[{{Donovan} {et~al.}(2007){Donovan}, {Hibbard}, \& {van
  Gorkom}}]{2007AJ....134.1118D}
{Donovan}, J.~L., {Hibbard}, J.~E., \& {van Gorkom}, J.~H. 2007, \aj, 134, 1118

\bibitem[{{Drory} \& {Alvarez}(2008)}]{2008ApJ...680...41D}
{Drory}, N. \& {Alvarez}, M. 2008, \apj, 680, 41

\bibitem[{{Eggen} {et~al.}(1962){Eggen}, {Lynden-Bell}, \&
  {Sandage}}]{1962ApJ...136..748E}
{Eggen}, O.~J., {Lynden-Bell}, D., \& {Sandage}, A.~R. 1962, \apj, 136, 748

\bibitem[{{Eliche-Moral} {et~al.}(2006{\natexlab{a}}){Eliche-Moral},
  {Balcells}, {Aguerri}, \& {Gonz{\'a}lez-Garc{\'{\i}}a}}]{2006A&A...457...91E}
{Eliche-Moral}, M.~C., {Balcells}, M., {Aguerri}, J.~A.~L., \&
  {Gonz{\'a}lez-Garc{\'{\i}}a}, A.~C. 2006{\natexlab{a}}, \aap, 457, 91

\bibitem[{{Eliche-Moral} {et~al.}(2006{\natexlab{b}}){Eliche-Moral},
  {Balcells}, {Prieto}, {Garc{\'{\i}}a-Dab{\'o}}, {Erwin}, \&
  {Crist{\'o}bal-Hornillos}}]{2006ApJ...639..644E}
{Eliche-Moral}, M.~C., {Balcells}, M., {Prieto}, M., {et~al.}
  2006{\natexlab{b}}, \apj, 639, 644

\bibitem[{{Elmegreen}(2009)}]{2009arXiv0903.1937E}
{Elmegreen}, B.~G. 2009, ArXiv e-prints

\bibitem[{{Elston} {et~al.}(1988){Elston}, {Rieke}, \&
  {Rieke}}]{1988ApJ...331L..77E}
{Elston}, R., {Rieke}, G.~H., \& {Rieke}, M.~J. 1988, \apjl, 331, L77

\bibitem[{{Emonts} {et~al.}(2006){Emonts}, {Morganti}, {Tadhunter}, {Holt},
  {Oosterloo}, {van der Hulst}, \& {Wills}}]{2006A&A...454..125E}
{Emonts}, B.~H.~C., {Morganti}, R., {Tadhunter}, C.~N., {et~al.} 2006, \aap,
  454, 125

\bibitem[{{Emsellem} {et~al.}(2007){Emsellem}, {Cappellari}, {Krajnovi{\'c}},
  {van de Ven}, {Bacon}, {Bureau}, {Davies}, {de Zeeuw}, {Falc{\'o}n-Barroso},
  {Kuntschner}, {McDermid}, {Peletier}, \& {Sarzi}}]{2007MNRAS.379..401E}
{Emsellem}, E., {Cappellari}, M., {Krajnovi{\'c}}, D., {et~al.} 2007, \mnras,
  379, 401

\bibitem[{{Faber} {et~al.}(2007){Faber}, {Willmer}, {Wolf}, {Koo}, {Weiner},
  {Newman}, {Im}, {Coil}, {Conroy}, {Cooper}, {Davis}, {Finkbeiner}, {Gerke},
  {Gebhardt}, {Groth}, {Guhathakurta}, {Harker}, {Kaiser}, {Kassin},
  {Kleinheinrich}, {Konidaris}, {Kron}, {Lin}, {Luppino}, {Madgwick},
  {Meisenheimer}, {Noeske}, {Phillips}, {Sarajedini}, {Schiavon}, {Simard},
  {Szalay}, {Vogt}, \& {Yan}}]{2007ApJ...665..265F}
{Faber}, S.~M., {Willmer}, C.~N.~A., {Wolf}, C., {et~al.} 2007, \apj, 665, 265

\bibitem[{{Falkenberg} {et~al.}(2009){Falkenberg}, {Kotulla}, \&
  {Fritze}}]{2009MNRAS.397.1940F}
{Falkenberg}, M.~A., {Kotulla}, R., \& {Fritze}, U. 2009, \mnras, 397, 1940

\bibitem[{{Feldmann} {et~al.}(2008){Feldmann}, {Mayer}, \&
  {Carollo}}]{2008ApJ...684.1062F}
{Feldmann}, R., {Mayer}, L., \& {Carollo}, C.~M. 2008, \apj, 684, 1062

\bibitem[{{Ferreiro} \& {Pastoriza}(2004)}]{2004A&A...428..837F}
{Ferreiro}, D.~L. \& {Pastoriza}, M.~G. 2004, \aap, 428, 837

\bibitem[{{Ferreras} {et~al.}(2009{\natexlab{a}}){Ferreras}, {Lisker},
  {Pasquali}, \& {Kaviraj}}]{2009MNRAS.395..554F}
{Ferreras}, I., {Lisker}, T., {Pasquali}, A., \& {Kaviraj}, S.
  2009{\natexlab{a}}, \mnras, 395, 554

\bibitem[{{Ferreras} {et~al.}(2009{\natexlab{b}}){Ferreras}, {Lisker},
  {Pasquali}, {Khochfar}, \& {Kaviraj}}]{2009MNRAS.396.1573F}
{Ferreras}, I., {Lisker}, T., {Pasquali}, A., {Khochfar}, S., \& {Kaviraj}, S.
  2009{\natexlab{b}}, \mnras, 396, 1573

\bibitem[{{Fontana} {et~al.}(2009){Fontana}, {Santini}, {Grazian},
  {Pentericci}, {Fiore}, {Castellano}, {Giallongo}, {Menci}, {Salimbeni},
  {Cristiani}, {Nonino}, \& {Vanzella}}]{2009A&A...501...15F}
{Fontana}, A., {Santini}, P., {Grazian}, A., {et~al.} 2009, \aap, 501, 15

\bibitem[{{Fontanot} {et~al.}(2009){Fontanot}, {De Lucia}, {Monaco},
  {Somerville}, \& {Santini}}]{2009MNRAS.397.1776F}
{Fontanot}, F., {De Lucia}, G., {Monaco}, P., {Somerville}, R.~S., \&
  {Santini}, P. 2009, \mnras, 397, 1776

\bibitem[{{F{\"o}rster Schreiber} {et~al.}(2009){F{\"o}rster Schreiber},
  {Genzel}, {Bouch{\'e}}, {Cresci}, {Davies}, {Buschkamp}, {Shapiro},
  {Tacconi}, {Hicks}, {Genel}, {Shapley}, {Erb}, {Steidel}, {Lutz},
  {Eisenhauer}, {Gillessen}, {Sternberg}, {Renzini}, {Cimatti}, {Daddi},
  {Kurk}, {Lilly}, {Kong}, {Lehnert}, {Nesvadba}, {Verma}, {McCracken},
  {Arimoto}, {Mignoli}, \& {Onodera}}]{2009ApJ...706.1364F}
{F{\"o}rster Schreiber}, N.~M., {Genzel}, R., {Bouch{\'e}}, N., {et~al.} 2009,
  \apj, 706, 1364

\bibitem[{{Franceschini} {et~al.}(2006){Franceschini}, {Rodighiero}, {Cassata},
  {Berta}, {Vaccari}, {Nonino}, {Vanzella}, {Hatziminaoglou}, {Antichi}, \&
  {Cristiani}}]{2006A&A...453..397F}
{Franceschini}, A., {Rodighiero}, G., {Cassata}, P., {et~al.} 2006, \aap, 453,
  397

\bibitem[{{Francis} {et~al.}(1996){Francis}, {Woodgate}, {Warren}, {Moller},
  {Mazzolini}, {Bunker}, {Lowenthal}, {Williams}, {Minezaki}, {Kobayashi}, \&
  {Yoshii}}]{1996ApJ...457..490F}
{Francis}, P.~J., {Woodgate}, B.~E., {Warren}, S.~J., {et~al.} 1996, \apj, 457,
  490

\bibitem[{{Franx} {et~al.}(2008){Franx}, {van Dokkum}, {Schreiber}, {Wuyts},
  {Labb{\'e}}, \& {Toft}}]{2008ApJ...688..770F}
{Franx}, M., {van Dokkum}, P.~G., {Schreiber}, N.~M.~F., {et~al.} 2008, \apj,
  688, 770

\bibitem[{{Franzetti} {et~al.}(2007){Franzetti}, {Scodeggio}, {Garilli},
  {Vergani}, {Maccagni}, {Guzzo}, {Tresse}, {Ilbert}, {Lamareille}, {Contini},
  {Le F{\`e}vre}, {Zamorani}, {Brinchmann}, {Charlot}, {Bottini}, {Le Brun},
  {Picat}, {Scaramella}, {Vettolani}, {Zanichelli}, {Adami}, {Arnouts},
  {Bardelli}, {Bolzonella}, {Cappi}, {Ciliegi}, {Foucaud}, {Gavignaud},
  {Iovino}, {McCracken}, {Marano}, {Marinoni}, {Mazure}, {Meneux}, {Merighi},
  {Paltani}, {Pell{\`o}}, {Pollo}, {Pozzetti}, {Radovich}, {Zucca}, {Cucciati},
  \& {Walcher}}]{2007A&A...465..711F}
{Franzetti}, P., {Scodeggio}, M., {Garilli}, B., {et~al.} 2007, \aap, 465, 711

\bibitem[{{Gallazzi} {et~al.}(2005){Gallazzi}, {Charlot}, {Brinchmann},
  {White}, \& {Tremonti}}]{2005MNRAS.362...41G}
{Gallazzi}, A., {Charlot}, S., {Brinchmann}, J., {White}, S.~D.~M., \&
  {Tremonti}, C.~A. 2005, \mnras, 362, 41

\bibitem[{{Gardner}(1998)}]{1998PASP..110..291G}
{Gardner}, J.~P. 1998, \pasp, 110, 291

\bibitem[{{Geach} {et~al.}(2009){Geach}, {Smail}, {Coppin}, {Moran}, {Edge}, \&
  {Ellis}}]{2009MNRAS.395L..62G}
{Geach}, J.~E., {Smail}, I., {Coppin}, K., {et~al.} 2009, \mnras, 395, L62

\bibitem[{{Genel} {et~al.}(2009){Genel}, {Genzel}, {Bouch{\'e}}, {Naab}, \&
  {Sternberg}}]{2009ApJ...701.2002G}
{Genel}, S., {Genzel}, R., {Bouch{\'e}}, N., {Naab}, T., \& {Sternberg}, A.
  2009, \apj, 701, 2002

\bibitem[{{Genzel} {et~al.}(2008){Genzel}, {Burkert}, {Bouch{\'e}}, {Cresci},
  {F{\"o}rster Schreiber}, {Shapley}, {Shapiro}, {Tacconi}, {Buschkamp},
  {Cimatti}, {Daddi}, {Davies}, {Eisenhauer}, {Erb}, {Genel}, {Gerhard},
  {Hicks}, {Lutz}, {Naab}, {Ott}, {Rabien}, {Renzini}, {Steidel}, {Sternberg},
  \& {Lilly}}]{2008ApJ...687...59G}
{Genzel}, R., {Burkert}, A., {Bouch{\'e}}, N., {et~al.} 2008, \apj, 687, 59

\bibitem[{{Gilbank} {et~al.}(2003){Gilbank}, {Smail}, {Ivison}, \&
  {Packham}}]{2003MNRAS.346.1125G}
{Gilbank}, D.~G., {Smail}, I., {Ivison}, R.~J., \& {Packham}, C. 2003, \mnras,
  346, 1125

\bibitem[{{Gimeno} {et~al.}(2007){Gimeno}, {Dottori}, {D{\'{\i}}az},
  {Rodrigues}, \& {Carranza}}]{2007AJ....133.2327G}
{Gimeno}, G.~N., {Dottori}, H.~A., {D{\'{\i}}az}, R.~J., {Rodrigues}, I., \&
  {Carranza}, G.~J. 2007, \aj, 133, 2327

\bibitem[{{Glazebrook} {et~al.}(2004){Glazebrook}, {Abraham}, {McCarthy},
  {Savaglio}, {Chen}, {Crampton}, {Murowinski}, {J{\o}rgensen}, {Roth}, {Hook},
  {Marzke}, \& {Carlberg}}]{2004Natur.430..181G}
{Glazebrook}, K., {Abraham}, R.~G., {McCarthy}, P.~J., {et~al.} 2004, \nat,
  430, 181

\bibitem[{{Gonz{\'a}lez-Garc{\'{\i}}a} \&
  {Balcells}(2005)}]{2005MNRAS.357..753G}
{Gonz{\'a}lez-Garc{\'{\i}}a}, A.~C. \& {Balcells}, M. 2005, \mnras, 357, 753

\bibitem[{{Gonz{\'a}lez-Garc{\'{\i}}a}
  {et~al.}(2006){Gonz{\'a}lez-Garc{\'{\i}}a}, {Balcells}, \&
  {Olshevsky}}]{2006MNRAS.372L..78G}
{Gonz{\'a}lez-Garc{\'{\i}}a}, A.~C., {Balcells}, M., \& {Olshevsky}, V.~S.
  2006, \mnras, 372, L78

\bibitem[{{Governato} {et~al.}(2009){Governato}, {Brook}, {Brooks}, {Mayer},
  {Willman}, {Jonsson}, {Stilp}, {Pope}, {Christensen}, {Wadsley}, \&
  {Quinn}}]{2009MNRAS.398..312G}
{Governato}, F., {Brook}, C.~B., {Brooks}, A.~M., {et~al.} 2009, \mnras, 398,
  312

\bibitem[{{Hammer} {et~al.}(2005){Hammer}, {Flores}, {Elbaz}, {Zheng}, {Liang},
  \& {Cesarsky}}]{2005A&A...430..115H}
{Hammer}, F., {Flores}, H., {Elbaz}, D., {et~al.} 2005, \aap, 430, 115

\bibitem[{{Hammer} {et~al.}(2009{\natexlab{a}}){Hammer}, {Flores}, {Puech},
  {Athanassoula}, {Rodrigues}, {Yang}, \&
  {Delgado-Serrano}}]{2009arXiv0903.3962H}
{Hammer}, F., {Flores}, H., {Puech}, M., {et~al.} 2009{\natexlab{a}}, ArXiv
  e-prints

\bibitem[{{Hammer} {et~al.}(2009{\natexlab{b}}){Hammer}, {Flores}, {Yang},
  {Athanassoula}, {Puech}, {Rodrigues}, \& {Peirani}}]{2009A&A...496..381H}
{Hammer}, F., {Flores}, H., {Yang}, Y.~B., {et~al.} 2009{\natexlab{b}}, \aap,
  496, 381

\bibitem[{{Hearn} \& {Lamb}(2001)}]{2001ApJ...551..651H}
{Hearn}, N.~C. \& {Lamb}, S.~A. 2001, \apj, 551, 651

\bibitem[{{Heavens} {et~al.}(2004){Heavens}, {Panter}, {Jimenez}, \&
  {Dunlop}}]{2004Natur.428..625H}
{Heavens}, A., {Panter}, B., {Jimenez}, R., \& {Dunlop}, J. 2004, \nat, 428,
  625

\bibitem[{{Heckman} {et~al.}(1999){Heckman}, {Armus}, {Weaver}, \&
  {Wang}}]{1999ApJ...517..130H}
{Heckman}, T.~M., {Armus}, L., {Weaver}, K.~A., \& {Wang}, J. 1999, \apj, 517,
  130

\bibitem[{{Hern{\'a}ndez-Toledo} {et~al.}(2006){Hern{\'a}ndez-Toledo},
  {Avila-Reese}, {Salazar-Contreras}, \& {Conselice}}]{2006AJ....132...71H}
{Hern{\'a}ndez-Toledo}, H.~M., {Avila-Reese}, V., {Salazar-Contreras}, J.~R.,
  \& {Conselice}, C.~J. 2006, \aj, 132, 71

\bibitem[{{Hetznecker} \& {Burkert}(2006)}]{2006MNRAS.370.1905H}
{Hetznecker}, H. \& {Burkert}, A. 2006, \mnras, 370, 1905

\bibitem[{{Hopkins} \& {Beacom}(2006)}]{2006ApJ...651..142H}
{Hopkins}, A.~M. \& {Beacom}, J.~F. 2006, \apj, 651, 142

\bibitem[{{Hopkins} {et~al.}(2008){Hopkins}, {Cox}, {Kere{\v s}}, \&
  {Hernquist}}]{2008ApJS..175..390H}
{Hopkins}, P.~F., {Cox}, T.~J., {Kere{\v s}}, D., \& {Hernquist}, L. 2008,
  \apjs, 175, 390

\bibitem[{{Hopkins} {et~al.}(2009){Hopkins}, {Cox}, {Younger}, \&
  {Hernquist}}]{2009ApJ...691.1168H}
{Hopkins}, P.~F., {Cox}, T.~J., {Younger}, J.~D., \& {Hernquist}, L. 2009,
  \apj, 691, 1168

\bibitem[{{Huertas-Company} {et~al.}(2010){Huertas-Company}, {Aguerri},
  {Tresse}, {Bolzonella}, {Koekemoer}, \& {Maier}}]{2010arXiv1002.3076H}
{Huertas-Company}, M., {Aguerri}, J.~A.~L., {Tresse}, L., {et~al.} 2010, ArXiv
  e-prints

\bibitem[{{Huertas-Company} {et~al.}(2009{\natexlab{a}}){Huertas-Company},
  {Foex}, {Soucail}, \& {Pell{\'o}}}]{2009A&A...505...83H}
{Huertas-Company}, M., {Foex}, G., {Soucail}, G., \& {Pell{\'o}}, R.
  2009{\natexlab{a}}, \aap, 505, 83

\bibitem[{{Huertas-Company} {et~al.}(2009{\natexlab{b}}){Huertas-Company},
  {Tasca}, {Rouan}, {Pelat}, {Kneib}, {Le F{\`e}vre}, {Capak}, {Kartaltepe},
  {Koekemoer}, {McCracken}, {Salvato}, {Sanders}, \&
  {Willott}}]{2009A&A...497..743H}
{Huertas-Company}, M., {Tasca}, L., {Rouan}, D., {et~al.} 2009{\natexlab{b}},
  \aap, 497, 743

\bibitem[{{Hughes} {et~al.}(1998){Hughes}, {Serjeant}, {Dunlop},
  {Rowan-Robinson}, {Blain}, {Mann}, {Ivison}, {Peacock}, {Efstathiou}, {Gear},
  {Oliver}, {Lawrence}, {Longair}, {Goldschmidt}, \&
  {Jenness}}]{1998Natur.394..241H}
{Hughes}, D.~H., {Serjeant}, S., {Dunlop}, J., {et~al.} 1998, \nat, 394, 241

\bibitem[{{Ilbert} {et~al.}(2006){Ilbert}, {Lauger}, {Tresse}, {Buat},
  {Arnouts}, {Le F{\`e}vre}, {Burgarella}, {Zucca}, {Bardelli}, {Zamorani},
  {Bottini}, {Garilli}, {Le Brun}, {Maccagni}, {Picat}, {Scaramella},
  {Scodeggio}, {Vettolani}, {Zanichelli}, {Adami}, {Arnaboldi}, {Bolzonella},
  {Cappi}, {Charlot}, {Contini}, {Foucaud}, {Franzetti}, {Gavignaud}, {Guzzo},
  {Iovino}, {McCracken}, {Marano}, {Marinoni}, {Mathez}, {Mazure}, {Meneux},
  {Merighi}, {Paltani}, {Pello}, {Pollo}, {Pozzetti}, {Radovich}, {Bondi},
  {Bongiorno}, {Busarello}, {Ciliegi}, {Mellier}, {Merluzzi}, {Ripepi}, \&
  {Rizzo}}]{2006A&A...453..809I}
{Ilbert}, O., {Lauger}, S., {Tresse}, L., {et~al.} 2006, \aap, 453, 809

\bibitem[{{Ilbert} {et~al.}(2010){Ilbert}, {Salvato}, {Le Floc'h}, {Aussel},
  {Capak}, {McCracken}, {Mobasher}, {Kartaltepe}, {Scoville}, {Sanders},
  {Arnouts}, {Bundy}, {Cassata}, {Kneib}, {Koekemoer}, {Le F{\`e}vre}, {Lilly},
  {Surace}, {Taniguchi}, {Tasca}, {Thompson}, {Tresse}, {Zamojski}, {Zamorani},
  \& {Zucca}}]{2010ApJ...709..644I}
{Ilbert}, O., {Salvato}, M., {Le Floc'h}, E., {et~al.} 2010, \apj, 709, 644

\bibitem[{{Im} {et~al.}(1996){Im}, {Griffiths}, {Ratnatunga}, \&
  {Sarajedini}}]{1996ApJ...461L..79I}
{Im}, M., {Griffiths}, R.~E., {Ratnatunga}, K.~U., \& {Sarajedini}, V.~L. 1996,
  \apjl, 461, L79

\bibitem[{{Imanishi}(2009)}]{2009ApJ...694..751I}
{Imanishi}, M. 2009, \apj, 694, 751

\bibitem[{{Iono} {et~al.}(2009){Iono}, {Wilson}, {Yun}, {Baker}, {Petitpas},
  {Peck}, {Krips}, {Cox}, {Matsushita}, {Mihos}, \&
  {Pihlstrom}}]{2009ApJ...695.1537I}
{Iono}, D., {Wilson}, C.~D., {Yun}, M.~S., {et~al.} 2009, \apj, 695, 1537

\bibitem[{{Jimenez} {et~al.}(2007){Jimenez}, {Bernardi}, {Haiman}, {Panter}, \&
  {Heavens}}]{2007ApJ...669..947J}
{Jimenez}, R., {Bernardi}, M., {Haiman}, Z., {Panter}, B., \& {Heavens}, A.~F.
  2007, \apj, 669, 947

\bibitem[{{Jogee} {et~al.}(2009){Jogee}, {Miller}, {Penner}, {Skelton},
  {Conselice}, {Somerville}, {Bell}, {Zheng}, {Rix}, {Robaina}, {Barazza},
  {Barden}, {Borch}, {Beckwith}, {Caldwell}, {Peng}, {Heymans}, {McIntosh},
  {H{\"a}u{\ss}ler}, {Jahnke}, {Meisenheimer}, {Sanchez}, {Wisotzki}, {Wolf},
  \& {Papovich}}]{2009ApJ...697.1971J}
{Jogee}, S., {Miller}, S.~H., {Penner}, K., {et~al.} 2009, \apj, 697, 1971

\bibitem[{{Johansson} {et~al.}(2009){Johansson}, {Naab}, \&
  {Burkert}}]{2009ApJ...690..802J}
{Johansson}, P.~H., {Naab}, T., \& {Burkert}, A. 2009, \apj, 690, 802

\bibitem[{{Joseph} \& {Wright}(1985)}]{1985MNRAS.214...87J}
{Joseph}, R.~D. \& {Wright}, G.~S. 1985, \mnras, 214, 87

\bibitem[{{Kang} \& {Im}(2009)}]{2009ApJ...691L..33K}
{Kang}, E. \& {Im}, M. 2009, \apjl, 691, L33

\bibitem[{{Kannappan} {et~al.}(2004){Kannappan}, {Jansen}, \&
  {Barton}}]{2004AJ....127.1371K}
{Kannappan}, S.~J., {Jansen}, R.~A., \& {Barton}, E.~J. 2004, \aj, 127, 1371

\bibitem[{{Kauffmann} {et~al.}(2003){Kauffmann}, {Heckman}, {White}, {Charlot},
  {Tremonti}, {Brinchmann}, {Bruzual}, {Peng}, {Seibert}, {Bernardi},
  {Blanton}, {Brinkmann}, {Castander}, {Cs{\'a}bai}, {Fukugita}, {Ivezic},
  {Munn}, {Nichol}, {Padmanabhan}, {Thakar}, {Weinberg}, \&
  {York}}]{2003MNRAS.341...33K}
{Kauffmann}, G., {Heckman}, T.~M., {White}, S.~D.~M., {et~al.} 2003, \mnras,
  341, 33

\bibitem[{{Kauffmann} \& {White}(1993)}]{1993MNRAS.261..921K}
{Kauffmann}, G. \& {White}, S.~D.~M. 1993, \mnras, 261, 921

\bibitem[{{Kauffmann} {et~al.}(2004){Kauffmann}, {White}, {Heckman},
  {M{\'e}nard}, {Brinchmann}, {Charlot}, {Tremonti}, \&
  {Brinkmann}}]{2004MNRAS.353..713K}
{Kauffmann}, G., {White}, S.~D.~M., {Heckman}, T.~M., {et~al.} 2004, \mnras,
  353, 713

\bibitem[{{Kaviraj} {et~al.}(2009{\natexlab{a}}){Kaviraj}, {Devriendt},
  {Ferreras}, {Yi}, \& {Silk}}]{2009A&A...503..445K}
{Kaviraj}, S., {Devriendt}, J.~E.~G., {Ferreras}, I., {Yi}, S.~K., \& {Silk},
  J. 2009{\natexlab{a}}, \aap, 503, 445

\bibitem[{{Kaviraj} {et~al.}(2009{\natexlab{b}}){Kaviraj}, {Peirani},
  {Khochfar}, {Silk}, \& {Kay}}]{2009MNRAS.394.1713K}
{Kaviraj}, S., {Peirani}, S., {Khochfar}, S., {Silk}, J., \& {Kay}, S.
  2009{\natexlab{b}}, \mnras, 394, 1713

\bibitem[{{Keel}(2008)}]{2008ASPC..396..243K}
{Keel}, W.~C. 2008, in Astronomical Society of the Pacific Conference Series,
  Vol. 396, Astronomical Society of the Pacific Conference Series, ed. J.~G.
  {Funes} \& E.~M. {Corsini}, 243

\bibitem[{{Keel} \& {Wu}(1995)}]{1995AJ....110..129K}
{Keel}, W.~C. \& {Wu}, W. 1995, \aj, 110, 129

\bibitem[{{Kennicutt}(1989)}]{1989ApJ...344..685K}
{Kennicutt}, Jr., R.~C. 1989, \apj, 344, 685

\bibitem[{{Khochfar} \& {Burkert}(2001)}]{2001ApJ...561..517K}
{Khochfar}, S. \& {Burkert}, A. 2001, \apj, 561, 517

\bibitem[{{Khochfar} \& {Burkert}(2003)}]{2003ApJ...597L.117K}
{Khochfar}, S. \& {Burkert}, A. 2003, \apjl, 597, L117

\bibitem[{{Khochfar} \& {Silk}(2006)}]{2006MNRAS.370..902K}
{Khochfar}, S. \& {Silk}, J. 2006, \mnras, 370, 902

\bibitem[{{Kim} {et~al.}(2009){Kim}, {Wise}, \& {Abel}}]{2009ApJ...694L.123K}
{Kim}, J.-h., {Wise}, J.~H., \& {Abel}, T. 2009, \apjl, 694, L123

\bibitem[{{Kitzbichler} \& {White}(2006)}]{2006MNRAS.366..858K}
{Kitzbichler}, M.~G. \& {White}, S.~D.~M. 2006, \mnras, 366, 858

\bibitem[{{Kodama} {et~al.}(2007){Kodama}, {Tanaka}, {Kajisawa}, {Kurk},
  {Venemans}, {De Breuck}, {Vernet}, \& {Lidman}}]{2007MNRAS.377.1717K}
{Kodama}, T., {Tanaka}, I., {Kajisawa}, M., {et~al.} 2007, \mnras, 377, 1717

\bibitem[{{Kormendy} \& {Fisher}(2008)}]{2008ASPC..396..297K}
{Kormendy}, J. \& {Fisher}, D.~B. 2008, in Astronomical Society of the Pacific
  Conference Series, Vol. 396, Astronomical Society of the Pacific Conference
  Series, ed. J.~G. {Funes} \& E.~M. {Corsini}, 297

\bibitem[{{Kuchinski} {et~al.}(2000){Kuchinski}, {Freedman}, {Madore},
  {Trewhella}, {Bohlin}, {Cornett}, {Fanelli}, {Marcum}, {Neff}, {O'Connell},
  {Roberts}, {Smith}, {Stecher}, \& {Waller}}]{2000ApJS..131..441K}
{Kuchinski}, L.~E., {Freedman}, W.~L., {Madore}, B.~F., {et~al.} 2000, \apjs,
  131, 441

\bibitem[{{Lanzoni} {et~al.}(2005){Lanzoni}, {Guiderdoni}, {Mamon},
  {Devriendt}, \& {Hatton}}]{2005MNRAS.361..369L}
{Lanzoni}, B., {Guiderdoni}, B., {Mamon}, G.~A., {Devriendt}, J., \& {Hatton},
  S. 2005, \mnras, 361, 369

\bibitem[{{Law} {et~al.}(2009){Law}, {Steidel}, {Erb}, {Larkin}, {Pettini},
  {Shapley}, \& {Wright}}]{2009ApJ...697.2057L}
{Law}, D.~R., {Steidel}, C.~C., {Erb}, D.~K., {et~al.} 2009, \apj, 697, 2057

\bibitem[{{Le F{\`e}vre} {et~al.}(2000){Le F{\`e}vre}, {Abraham}, {Lilly},
  {Ellis}, {Brinchmann}, {Schade}, {Tresse}, {Colless}, {Crampton},
  {Glazebrook}, {Hammer}, \& {Broadhurst}}]{2000MNRAS.311..565L}
{Le F{\`e}vre}, O., {Abraham}, R., {Lilly}, S.~J., {et~al.} 2000, \mnras, 311,
  565

\bibitem[{{Leitherer}(2000)}]{2000ESASP.445...37L}
{Leitherer}, C. 2000, in ESA Special Publication, Vol. 445, Star Formation from
  the Small to the Large Scale, ed. F.~{Favata}, A.~{Kaas}, \& A.~{Wilson}, 37

\bibitem[{{Leitherer}(2001)}]{2001ASPC..245..390L}
{Leitherer}, C. 2001, in Astronomical Society of the Pacific Conference Series,
  Vol. 245, Astrophysical Ages and Times Scales, ed. T.~{von Hippel},
  C.~{Simpson}, \& N.~{Manset}, 390

\bibitem[{{Leitherer} {et~al.}(1996){Leitherer}, {Alloin},
  {Fritze-v.~Alvensleben}, {Gallagher}, {Huchra}, {Matteucci}, {O'Connell},
  {Beckman}, {Bertelli}, {Bica}, {Boisson}, {Bonatto}, {Bothun}, {Bressan},
  {Brodie}, {Bruzual}, {Burstein}, {Buser}, {Caldwell}, {Casuso},
  {Cervi{\~n}o}, {Charlot}, {Chavez}, {Chiosi}, {Christian}, {Cuisinier},
  {Dallier}, {de Koter}, {Delisle}, {Diaz}, {Dopita}, {Dorman}, {Fagotto},
  {Fanelli}, {Fioc}, {Garcia-Vargas}, {Girardi}, {Goldader}, {Hardy},
  {Heckman}, {Iglesias}, {Jablonka}, {Joly}, {Jones}, {Kurth}, {Lancon},
  {Lejeune}, {Loxen}, {Maeder}, {Malagnini}, {Marigo}, {Mas-Hesse}, {Meynet},
  {Moller}, {Molla}, {Morossi}, {Nasi}, {Nichols}, {Odegaard}, {Parker},
  {Pastoriza}, {Peletier}, {Robert}, {Rocca-Volmerange}, {Schaerer}, {Schmidt},
  {Schmitt}, {Schommer}, {Schmutz}, {Roos}, {Silva}, {Stasi{\'n}ska},
  {Sutherland}, {Tantalo}, {Traat}, {Vallenari}, {Vazdekis}, {Walborn},
  {Worthey}, \& {Wu}}]{1996PASP..108..996L}
{Leitherer}, C., {Alloin}, D., {Fritze-v.~Alvensleben}, U., {et~al.} 1996,
  \pasp, 108, 996

\bibitem[{{Lin} {et~al.}(2008){Lin}, {Patton}, {Koo}, {Casteels}, {Conselice},
  {Faber}, {Lotz}, {Willmer}, {Hsieh}, {Chiueh}, {Newman}, {Novak}, {Weiner},
  \& {Cooper}}]{2008ApJ...681..232L}
{Lin}, L., {Patton}, D.~R., {Koo}, D.~C., {et~al.} 2008, \apj, 681, 232

\bibitem[{{Liu} {et~al.}(2000){Liu}, {Dey}, {Graham}, {Bundy}, {Steidel},
  {Adelberger}, \& {Dickinson}}]{2000AJ....119.2556L}
{Liu}, M.~C., {Dey}, A., {Graham}, J.~R., {et~al.} 2000, \aj, 119, 2556

\bibitem[{{L{\'o}pez-Sanjuan} {et~al.}(2009{\natexlab{a}}){L{\'o}pez-Sanjuan},
  {Balcells}, {Garc{\'{\i}}a-Dab{\'o}}, {Prieto}, {Crist{\'o}bal-Hornillos},
  {Eliche-Moral}, {Abreu}, {Erwin}, \& {Guzm{\'a}n}}]{2009ApJ...694..643L}
{L{\'o}pez-Sanjuan}, C., {Balcells}, M., {Garc{\'{\i}}a-Dab{\'o}}, C.~E.,
  {et~al.} 2009{\natexlab{a}}, \apj, 694, 643

\bibitem[{{L{\'o}pez-Sanjuan} {et~al.}(2009{\natexlab{b}}){L{\'o}pez-Sanjuan},
  {Balcells}, {P{\'e}rez-Gonz{\'a}lez}, {Barro}, {Garc{\'{\i}}a-Dab{\'o}},
  {Gallego}, \& {Zamorano}}]{2009A&A...501..505L}
{L{\'o}pez-Sanjuan}, C., {Balcells}, M., {P{\'e}rez-Gonz{\'a}lez}, P.~G.,
  {et~al.} 2009{\natexlab{b}}, \aap, 501, 505

\bibitem[{{L{\'o}pez-Sanjuan} {et~al.}(2010){L{\'o}pez-Sanjuan}, {Balcells},
  {P{\'e}rez-Gonz{\'a}lez}, {Barro}, {Garc{\'{\i}}a-Dab{\'o}}, {Gallego}, \&
  {Zamorano}}]{2010ApJ...710.1170L}
{L{\'o}pez-Sanjuan}, C., {Balcells}, M., {P{\'e}rez-Gonz{\'a}lez}, P.~G.,
  {et~al.} 2010, \apj, 710, 1170

\bibitem[{{L{\'o}pez-Sanjuan} {et~al.}(2008){L{\'o}pez-Sanjuan},
  {Garc{\'{\i}}a-Dab{\'o}}, \& {Balcells}}]{2008PASP..120..571L}
{L{\'o}pez-Sanjuan}, C., {Garc{\'{\i}}a-Dab{\'o}}, C.~E., \& {Balcells}, M.
  2008, \pasp, 120, 571

\bibitem[{{Lotz} {et~al.}(2008{\natexlab{a}}){Lotz}, {Davis}, {Faber},
  {Guhathakurta}, {Gwyn}, {Huang}, {Koo}, {Le Floc'h}, {Lin}, {Newman},
  {Noeske}, {Papovich}, {Willmer}, {Coil}, {Conselice}, {Cooper}, {Hopkins},
  {Metevier}, {Primack}, {Rieke}, \& {Weiner}}]{2008ApJ...672..177L}
{Lotz}, J.~M., {Davis}, M., {Faber}, S.~M., {et~al.} 2008{\natexlab{a}}, \apj,
  672, 177

\bibitem[{{Lotz} {et~al.}(2008{\natexlab{b}}){Lotz}, {Jonsson}, {Cox}, \&
  {Primack}}]{2008MNRAS.391.1137L}
{Lotz}, J.~M., {Jonsson}, P., {Cox}, T.~J., \& {Primack}, J.~R.
  2008{\natexlab{b}}, \mnras, 391, 1137

\bibitem[{{Lotz} {et~al.}(2004){Lotz}, {Primack}, \&
  {Madau}}]{2004AJ....128..163L}
{Lotz}, J.~M., {Primack}, J., \& {Madau}, P. 2004, \aj, 128, 163

\bibitem[{{Maier} {et~al.}(2009){Maier}, {Lilly}, {Zamorani}, {Scodeggio},
  {Lamareille}, {Contini}, {Sargent}, {Scarlata}, {Oesch}, {Carollo},
  {LeF{\`e}vre}, {Renzini}, {Kneib}, {Mainieri}, {Bardelli}, {Bolzonella},
  {Bongiorno}, {Caputi}, {Coppa}, {Cucciati}, {de la Torre}, {de Ravel},
  {Franzetti}, {Garilli}, {Iovino}, {Kampczyk}, {Knobel}, {Kova{\v c}},
  {LeBorgne}, {LeBrun}, {Mignoli}, {Pello}, {Peng}, {Montero}, {Ricciardelli},
  {Silverman}, {Tanaka}, {Tasca}, {Tresse}, {Vergani}, {Zucca}, {Abbas},
  {Bottini}, {Cappi}, {Cassata}, {Cimatti}, {Fumana}, {Guzzo}, {Halliday},
  {Koekemoer}, {Leauthaud}, {Maccagni}, {Marinoni}, {McCracken}, {Memeo},
  {Meneux}, {Porciani}, {Pozzetti}, \& {Scaramella}}]{2009ApJ...694.1099M}
{Maier}, C., {Lilly}, S.~J., {Zamorani}, G., {et~al.} 2009, \apj, 694, 1099

\bibitem[{{Mannucci} {et~al.}(2001){Mannucci}, {Basile}, {Poggianti},
  {Cimatti}, {Daddi}, {Pozzetti}, \& {Vanzi}}]{2001MNRAS.326..745M}
{Mannucci}, F., {Basile}, F., {Poggianti}, B.~M., {et~al.} 2001, \mnras, 326,
  745

\bibitem[{{Mart{\'{\i}}nez-Delgado} {et~al.}(2009){Mart{\'{\i}}nez-Delgado},
  {Pohlen}, {Gabany}, {Majewski}, {Pe{\~n}arrubia}, \&
  {Palma}}]{2009ApJ...692..955M}
{Mart{\'{\i}}nez-Delgado}, D., {Pohlen}, M., {Gabany}, R.~J., {et~al.} 2009,
  \apj, 692, 955

\bibitem[{{Maslanka}(1984)}]{1984Ap&SS.100..407M}
{Maslanka}, K.~D. 1984, \apss, 100, 407

\bibitem[{{Matsuoka} \& {Kawara}(2010)}]{2010MNRAS.tmp..534M}
{Matsuoka}, Y. \& {Kawara}, K. 2010, \mnras, 534

\bibitem[{{Mihos} \& {Hernquist}(1996)}]{1996ApJ...464..641M}
{Mihos}, J.~C. \& {Hernquist}, L. 1996, \apj, 464, 641

\bibitem[{{Moustakas} {et~al.}(2004){Moustakas}, {Casertano}, {Conselice},
  {Dickinson}, {Eisenhardt}, {Ferguson}, {Giavalisco}, {Grogin}, {Koekemoer},
  {Lucas}, {Mobasher}, {Papovich}, {Renzini}, {Somerville}, \&
  {Stern}}]{2004ApJ...600L.131M}
{Moustakas}, L.~A., {Casertano}, S., {Conselice}, C.~J., {et~al.} 2004, \apjl,
  600, L131

\bibitem[{{Mu{\~n}oz-Mateos} {et~al.}(2009){Mu{\~n}oz-Mateos}, {Gil de Paz},
  {Boissier}, {Zamorano}, {Dale}, {P{\'e}rez-Gonz{\'a}lez}, {Gallego},
  {Madore}, {Bendo}, {Thornley}, {Draine}, {Boselli}, {Buat}, {Calzetti},
  {Moustakas}, \& {Kennicutt}}]{2009ApJ...701.1965M}
{Mu{\~n}oz-Mateos}, J.~C., {Gil de Paz}, A., {Boissier}, S., {et~al.} 2009,
  \apj, 701, 1965

\bibitem[{{Naab} \& {Burkert}(2003)}]{2003ApJ...597..893N}
{Naab}, T. \& {Burkert}, A. 2003, \apj, 597, 893

\bibitem[{{Naab} {et~al.}(2009){Naab}, {Johansson}, \&
  {Ostriker}}]{2009ApJ...699L.178N}
{Naab}, T., {Johansson}, P.~H., \& {Ostriker}, J.~P. 2009, \apjl, 699, L178

\bibitem[{{Naab} {et~al.}(2006){Naab}, {Khochfar}, \&
  {Burkert}}]{2006ApJ...636L..81N}
{Naab}, T., {Khochfar}, S., \& {Burkert}, A. 2006, \apjl, 636, L81

\bibitem[{{Nakamura} {et~al.}(2003){Nakamura}, {Fukugita}, {Yasuda}, {Loveday},
  {Brinkmann}, {Schneider}, {Shimasaku}, \& {SubbaRao}}]{2003AJ....125.1682N}
{Nakamura}, O., {Fukugita}, M., {Yasuda}, N., {et~al.} 2003, \aj, 125, 1682

\bibitem[{{Navarro} \& {Steinmetz}(2000)}]{2000ApJ...528..607N}
{Navarro}, J.~F. \& {Steinmetz}, M. 2000, \apj, 528, 607

\bibitem[{{Neistein} {et~al.}(2006){Neistein}, {van den Bosch}, \&
  {Dekel}}]{2006MNRAS.372..933N}
{Neistein}, E., {van den Bosch}, F.~C., \& {Dekel}, A. 2006, \mnras, 372, 933

\bibitem[{{Nolan}(2004)}]{2004cgpc.sympE..38N}
{Nolan}, L.~A. 2004, in Clusters of Galaxies: Probes of Cosmological Structure
  and Galaxy Evolution, ed. J.~S. {Mulchaey}, A.~{Dressler}, \& A.~{Oemler}

\bibitem[{{Odewahn} {et~al.}(1996){Odewahn}, {Windhorst}, {Driver}, \&
  {Keel}}]{1996ApJ...472L..13O}
{Odewahn}, S.~C., {Windhorst}, R.~A., {Driver}, S.~P., \& {Keel}, W.~C. 1996,
  \apjl, 472, L13

\bibitem[{{Oesch} {et~al.}(2009){Oesch}, {Carollo}, {Feldmann}, {Hahn},
  {Lilly}, {Sargent}, {Scarlata}, {Aller}, {Aussel}, {Bolzonella}, {Bschorr},
  {Bundy}, {Capak}, {Ilbert}, {Kneib}, {Koekemoer}, {Kovac}, {Leauthaud}, {Le
  Floc'h}, {Massey}, {McCracken}, {Pozzetti}, {Renzini}, {Rhodes}, {Salvato},
  {Sanders}, {Scoville}, {Sheth}, {Taniguchi}, \&
  {Thompson}}]{2009arXiv0911.1126O}
{Oesch}, P.~A., {Carollo}, C.~M., {Feldmann}, R., {et~al.} 2009, ArXiv e-prints

\bibitem[{{Papovich} {et~al.}(2005){Papovich}, {Dickinson}, {Giavalisco},
  {Conselice}, \& {Ferguson}}]{2005ApJ...631..101P}
{Papovich}, C., {Dickinson}, M., {Giavalisco}, M., {Conselice}, C.~J., \&
  {Ferguson}, H.~C. 2005, \apj, 631, 101

\bibitem[{{Parkinson} {et~al.}(2008){Parkinson}, {Cole}, \&
  {Helly}}]{2008MNRAS.383..557P}
{Parkinson}, H., {Cole}, S., \& {Helly}, J. 2008, \mnras, 383, 557

\bibitem[{{Patel} {et~al.}(2009){Patel}, {Holden}, {Kelson}, {Illingworth}, \&
  {Franx}}]{2009ApJ...705L..67P}
{Patel}, S.~G., {Holden}, B.~P., {Kelson}, D.~D., {Illingworth}, G.~D., \&
  {Franx}, M. 2009, \apjl, 705, L67

\bibitem[{{Patton} {et~al.}(2002){Patton}, {Pritchet}, {Carlberg}, {Marzke},
  {Yee}, {Hall}, {Lin}, {Morris}, {Sawicki}, {Shepherd}, \&
  {Wirth}}]{2002ApJ...565..208P}
{Patton}, D.~R., {Pritchet}, C.~J., {Carlberg}, R.~G., {et~al.} 2002, \apj,
  565, 208

\bibitem[{{P{\'e}rez-Gonz{\'a}lez} {et~al.}(2005){P{\'e}rez-Gonz{\'a}lez},
  {Rieke}, {Egami}, {Alonso-Herrero}, {Dole}, {Papovich}, {Blaylock}, {Jones},
  {Rieke}, {Rigby}, {Barmby}, {Fazio}, {Huang}, \&
  {Martin}}]{2005ApJ...630...82P}
{P{\'e}rez-Gonz{\'a}lez}, P.~G., {Rieke}, G.~H., {Egami}, E., {et~al.} 2005,
  \apj, 630, 82

\bibitem[{{P{\'e}rez-Gonz{\'a}lez} {et~al.}(2008){P{\'e}rez-Gonz{\'a}lez},
  {Rieke}, {Villar}, {Barro}, {Blaylock}, {Egami}, {Gallego}, {Gil de Paz},
  {Pascual}, {Zamorano}, \& {Donley}}]{2008ApJ...675..234P}
{P{\'e}rez-Gonz{\'a}lez}, P.~G., {Rieke}, G.~H., {Villar}, V., {et~al.} 2008,
  \apj, 675, 234

\bibitem[{{Pierre} {et~al.}(2001){Pierre}, {Lidman}, {Hunstead}, {Alloin},
  {Casali}, {Cesarsky}, {Chanial}, {Duc}, {Fadda}, {Flores}, {Madden}, \&
  {Vigroux}}]{2001A&A...372L..45P}
{Pierre}, M., {Lidman}, C., {Hunstead}, R., {et~al.} 2001, \aap, 372, L45

\bibitem[{{Poggianti} {et~al.}(2009{\natexlab{a}}){Poggianti},
  {Arag{\'o}n-Salamanca}, {Zaritsky}, {DeLucia}, {Milvang-Jensen}, {Desai},
  {Jablonka}, {Halliday}, {Rudnick}, {Varela}, {Bamford}, {Best}, {Clowe},
  {Noll}, {Saglia}, {Pell{\'o}}, {Simard}, {von der Linden}, \&
  {White}}]{2009ApJ...693..112P}
{Poggianti}, B.~M., {Arag{\'o}n-Salamanca}, A., {Zaritsky}, D., {et~al.}
  2009{\natexlab{a}}, \apj, 693, 112

\bibitem[{{Poggianti} {et~al.}(2009{\natexlab{b}}){Poggianti}, {Fasano},
  {Bettoni}, {Cava}, {Dressler}, {Vanzella}, {Varela}, {Couch}, {D'Onofrio},
  {Fritz}, {Kjaergaard}, {Moles}, \& {Valentinuzzi}}]{2009ApJ...697L.137P}
{Poggianti}, B.~M., {Fasano}, G., {Bettoni}, D., {et~al.} 2009{\natexlab{b}},
  \apjl, 697, L137

\bibitem[{{Pollack} {et~al.}(2007){Pollack}, {Max}, \&
  {Schneider}}]{2007ApJ...660..288P}
{Pollack}, L.~K., {Max}, C.~E., \& {Schneider}, G. 2007, \apj, 660, 288

\bibitem[{{Pozzetti} {et~al.}(2003){Pozzetti}, {Cimatti}, {Zamorani}, {Daddi},
  {Menci}, {Fontana}, {Renzini}, {Mignoli}, {Poli}, {Saracco}, {Broadhurst},
  {Cristiani}, {D'Odorico}, {Giallongo}, \& {Gilmozzi}}]{2003A&A...402..837P}
{Pozzetti}, L., {Cimatti}, A., {Zamorani}, G., {et~al.} 2003, \aap, 402, 837

\bibitem[{{Puech} {et~al.}(2009{\natexlab{a}}){Puech}, {Hammer}, {Flores},
  {Delgado-Serrano}, {Rodrigues}, \& {Yang}}]{2009arXiv0903.3961P}
{Puech}, M., {Hammer}, F., {Flores}, H., {et~al.} 2009{\natexlab{a}}, ArXiv
  e-prints

\bibitem[{{Puech} {et~al.}(2009{\natexlab{b}}){Puech}, {Hammer}, {Flores},
  {Neichel}, \& {Yang}}]{2009A&A...493..899P}
{Puech}, M., {Hammer}, F., {Flores}, H., {Neichel}, B., \& {Yang}, Y.
  2009{\natexlab{b}}, \aap, 493, 899

\bibitem[{{Ravindranath} {et~al.}(2006){Ravindranath}, {Giavalisco},
  {Ferguson}, {Conselice}, {Katz}, {Weinberg}, {Lotz}, {Dickinson}, {Fall},
  {Mobasher}, \& {Papovich}}]{2006ApJ...652..963R}
{Ravindranath}, S., {Giavalisco}, M., {Ferguson}, H.~C., {et~al.} 2006, \apj,
  652, 963

\bibitem[{{Rawat} {et~al.}(2009){Rawat}, {Wadadekar}, \&
  {DeMello}}]{2009ApJ...695.1315R}
{Rawat}, A., {Wadadekar}, Y., \& {DeMello}, D. 2009, \apj, 695, 1315

\bibitem[{{Rines} {et~al.}(2007){Rines}, {Finn}, \&
  {Vikhlinin}}]{2007ApJ...665L...9R}
{Rines}, K., {Finn}, R., \& {Vikhlinin}, A. 2007, \apjl, 665, L9

\bibitem[{{Robaina} {et~al.}(2009){Robaina}, {Bell}, {Skelton}, {Mc Intosh},
  {Somerville}, {Zheng}, {Rix}, {Bacon}, {Balogh}, {Barazza}, {Barden},
  {B{\"o}hm}, {Caldwell}, {Gallazzi}, {Gray}, {H{\"a}ussler}, {Heymans},
  {Jahnke}, {Jogee}, {van Kampen}, {Lane}, {Meisenheimer}, {Papovich}, {Peng},
  {S{\'a}nchez}, {Skibba}, {Taylor}, {Wisotzki}, \&
  {Wolf}}]{2009ApJ...704..324R}
{Robaina}, A.~R., {Bell}, E.~F., {Skelton}, R.~E., {et~al.} 2009, \apj, 704,
  324

\bibitem[{{Roberts}(1975)}]{1975dgs..conf..113R}
{Roberts}, Jr., W.~W. 1975, in La Dynamique des galaxies spirales, ed.
  L.~{Weliachew}, 113

\bibitem[{{Robertson} {et~al.}(2006){Robertson}, {Hernquist}, {Cox}, {Di
  Matteo}, {Hopkins}, {Martini}, \& {Springel}}]{2006ApJ...641...90R}
{Robertson}, B., {Hernquist}, L., {Cox}, T.~J., {et~al.} 2006, \apj, 641, 90

\bibitem[{{Roche} {et~al.}(2006){Roche}, {Dunlop}, {Caputi}, {McLure},
  {Willott}, \& {Crampton}}]{2006MNRAS.370...74R}
{Roche}, N.~D., {Dunlop}, J., {Caputi}, K.~I., {et~al.} 2006, \mnras, 370, 74

\bibitem[{{Rossa} {et~al.}(2007){Rossa}, {Laine}, {van der Marel}, {Mihos},
  {Hibbard}, {B{\"o}ker}, \& {Zabludoff}}]{2007AJ....134.2124R}
{Rossa}, J., {Laine}, S., {van der Marel}, R.~P., {et~al.} 2007, \aj, 134, 2124

\bibitem[{{Rudnick} {et~al.}(2009){Rudnick}, {von der Linden}, {Pell{\'o}},
  {Arag{\'o}n-Salamanca}, {Marchesini}, {Clowe}, {De Lucia}, {Halliday},
  {Jablonka}, {Milvang-Jensen}, {Poggianti}, {Saglia}, {Simard}, {White}, \&
  {Zaritsky}}]{2009ApJ...700.1559R}
{Rudnick}, G., {von der Linden}, A., {Pell{\'o}}, R., {et~al.} 2009, \apj, 700,
  1559

\bibitem[{{Ryan} {et~al.}(2007){Ryan}, {Hathi}, {Cohen}, {Malhotra}, {Rhoads},
  {Windhorst}, {Budav{\'a}ri}, {Pirzkal}, {Xu}, {Panagia}, {Moustakas}, {di
  Serego Alighieri}, \& {Yan}}]{2007ApJ...668..839R}
{Ryan}, Jr., R.~E., {Hathi}, N.~P., {Cohen}, S.~H., {et~al.} 2007, \apj, 668,
  839

\bibitem[{{Saracco} {et~al.}(2009){Saracco}, {Longhetti}, \&
  {Andreon}}]{2009MNRAS.392..718S}
{Saracco}, P., {Longhetti}, M., \& {Andreon}, S. 2009, \mnras, 392, 718

\bibitem[{{Scarlata} {et~al.}(2007{\natexlab{a}}){Scarlata}, {Carollo},
  {Lilly}, {Sargent}, {Feldmann}, {Kampczyk}, {Porciani}, {Koekemoer},
  {Scoville}, {Kneib}, {Leauthaud}, {Massey}, {Rhodes}, {Tasca}, {Capak},
  {Maier}, {McCracken}, {Mobasher}, {Renzini}, {Taniguchi}, {Thompson},
  {Sheth}, {Ajiki}, {Aussel}, {Murayama}, {Sanders}, {Sasaki}, {Shioya}, \&
  {Takahashi}}]{2007ApJS..172..406S}
{Scarlata}, C., {Carollo}, C.~M., {Lilly}, S., {et~al.} 2007{\natexlab{a}},
  \apjs, 172, 406

\bibitem[{{Scarlata} {et~al.}(2007{\natexlab{b}}){Scarlata}, {Carollo},
  {Lilly}, {Feldmann}, {Kampczyk}, {Renzini}, {Cimatti}, {Halliday}, {Daddi},
  {Sargent}, {Koekemoer}, {Scoville}, {Kneib}, {Leauthaud}, {Massey}, {Rhodes},
  {Tasca}, {Capak}, {McCracken}, {Mobasher}, {Taniguchi}, {Thompson}, {Ajiki},
  {Aussel}, {Murayama}, {Sanders}, {Sasaki}, {Shioya}, \&
  {Takahashi}}]{2007ApJS..172..494S}
{Scarlata}, C., {Carollo}, C.~M., {Lilly}, S.~J., {et~al.} 2007{\natexlab{b}},
  \apjs, 172, 494

\bibitem[{{Schawinski} {et~al.}(2010){Schawinski}, {Dowlin}, {Thomas}, {Urry},
  \& {Edmondson}}]{2010ApJ...714L.108S}
{Schawinski}, K., {Dowlin}, N., {Thomas}, D., {Urry}, C.~M., \& {Edmondson}, E.
  2010, \apjl, 714, L108

\bibitem[{{Schawinski} {et~al.}(2007){Schawinski}, {Thomas}, {Sarzi},
  {Maraston}, {Kaviraj}, {Joo}, {Yi}, \& {Silk}}]{2007MNRAS.382.1415S}
{Schawinski}, K., {Thomas}, D., {Sarzi}, M., {et~al.} 2007, \mnras, 382, 1415

\bibitem[{{Schurer} {et~al.}(2009){Schurer}, {Calura}, {Silva}, {Pipino},
  {Granato}, {Matteucci}, \& {Maiolino}}]{2009MNRAS.394.2001S}
{Schurer}, A., {Calura}, F., {Silva}, L., {et~al.} 2009, \mnras, 394, 2001

\bibitem[{{Schweizer}(2005)}]{2005ASSL..329..143S}
{Schweizer}, F. 2005, in Astrophysics and Space Science Library, Vol. 329,
  Starbursts: From 30 Doradus to Lyman Break Galaxies, ed. R.~{de Grijs} \&
  R.~M. {Gonz{\'a}lez Delgado}, 143

\bibitem[{{Schweizer} \& {Seitzer}(2007)}]{2007AJ....133.2132S}
{Schweizer}, F. \& {Seitzer}, P. 2007, \aj, 133, 2132

\bibitem[{{Scoville} \& {Young}(1983)}]{1983ApJ...265..148S}
{Scoville}, N. \& {Young}, J.~S. 1983, \apj, 265, 148

\bibitem[{{Sobral} {et~al.}(2009){Sobral}, {Best}, {Geach}, {Smail}, {Kurk},
  {Cirasuolo}, {Casali}, {Ivison}, {Coppin}, \& {Dalton}}]{2009MNRAS.398...75S}
{Sobral}, D., {Best}, P.~N., {Geach}, J.~E., {et~al.} 2009, \mnras, 398, 75

\bibitem[{{Springel} \& {Hernquist}(2005)}]{2005ApJ...622L...9S}
{Springel}, V. \& {Hernquist}, L. 2005, \apjl, 622, L9

\bibitem[{{Springel} {et~al.}(2005){Springel}, {White}, {Jenkins}, {Frenk},
  {Yoshida}, {Gao}, {Navarro}, {Thacker}, {Croton}, {Helly}, {Peacock}, {Cole},
  {Thomas}, {Couchman}, {Evrard}, {Colberg}, \& {Pearce}}]{2005Natur.435..629S}
{Springel}, V., {White}, S.~D.~M., {Jenkins}, A., {et~al.} 2005, \nat, 435, 629

\bibitem[{{Statler} \& {McNamara}(2002)}]{2002ApJ...581.1032S}
{Statler}, T.~S. \& {McNamara}, B.~R. 2002, \apj, 581, 1032

\bibitem[{{Stewart} {et~al.}(2009{\natexlab{a}}){Stewart}, {Bullock}, {Barton},
  \& {Wechsler}}]{2009ApJ...702.1005S}
{Stewart}, K.~R., {Bullock}, J.~S., {Barton}, E.~J., \& {Wechsler}, R.~H.
  2009{\natexlab{a}}, \apj, 702, 1005

\bibitem[{{Stewart} {et~al.}(2009{\natexlab{b}}){Stewart}, {Bullock},
  {Wechsler}, \& {Maller}}]{2009ApJ...702..307S}
{Stewart}, K.~R., {Bullock}, J.~S., {Wechsler}, R.~H., \& {Maller}, A.~H.
  2009{\natexlab{b}}, \apj, 702, 307

\bibitem[{{Stewart} {et~al.}(2008){Stewart}, {Bullock}, {Wechsler}, {Maller},
  \& {Zentner}}]{2008ApJ...683..597S}
{Stewart}, K.~R., {Bullock}, J.~S., {Wechsler}, R.~H., {Maller}, A.~H., \&
  {Zentner}, A.~R. 2008, \apj, 683, 597

\bibitem[{{Strateva} {et~al.}(2001){Strateva}, {Ivezi{\'c}}, {Knapp},
  {Narayanan}, {Strauss}, {Gunn}, {Lupton}, {Schlegel}, {Bahcall}, {Brinkmann},
  {Brunner}, {Budav{\'a}ri}, {Csabai}, {Castander}, {Doi}, {Fukugita}, {Gy{\H
  o}ry}, {Hamabe}, {Hennessy}, {Ichikawa}, {Kunszt}, {Lamb}, {McKay},
  {Okamura}, {Racusin}, {Sekiguchi}, {Schneider}, {Shimasaku}, \&
  {York}}]{2001AJ....122.1861S}
{Strateva}, I., {Ivezi{\'c}}, {\v Z}., {Knapp}, G.~R., {et~al.} 2001, \aj, 122,
  1861

\bibitem[{{Surace} \& {Sanders}(2000)}]{2000AJ....120..604S}
{Surace}, J.~A. \& {Sanders}, D.~B. 2000, \aj, 120, 604

\bibitem[{{Surace} {et~al.}(2000){Surace}, {Sanders}, \&
  {Evans}}]{2000ApJ...529..170S}
{Surace}, J.~A., {Sanders}, D.~B., \& {Evans}, A.~S. 2000, \apj, 529, 170

\bibitem[{{Tacconi} {et~al.}(2008){Tacconi}, {Genzel}, {Smail}, {Neri},
  {Chapman}, {Ivison}, {Blain}, {Cox}, {Omont}, {Bertoldi}, {Greve},
  {F{\"o}rster Schreiber}, {Genel}, {Lutz}, {Swinbank}, {Shapley}, {Erb},
  {Cimatti}, {Daddi}, \& {Baker}}]{2008ApJ...680..246T}
{Tacconi}, L.~J., {Genzel}, R., {Smail}, I., {et~al.} 2008, \apj, 680, 246

\bibitem[{{Tasca} {et~al.}(2009){Tasca}, {Kneib}, {Iovino}, {Le F{\`e}vre},
  {Kova{\v c}}, {Bolzonella}, {Lilly}, {Abraham}, {Cassata}, {Cucciati},
  {Guzzo}, {Tresse}, {Zamorani}, {Capak}, {Garilli}, {Scodeggio}, {Sheth},
  {Zucca}, {Carollo}, {Contini}, {Mainieri}, {Renzini}, {Bardelli},
  {Bongiorno}, {Caputi}, {Coppa}, {de La Torre}, {de Ravel}, {Franzetti},
  {Kampczyk}, {Knobel}, {Koekemoer}, {Lamareille}, {Le Borgne}, {Le Brun},
  {Maier}, {Mignoli}, {Pello}, {Peng}, {Perez Montero}, {Ricciardelli},
  {Silverman}, {Vergani}, {Tanaka}, {Abbas}, {Bottini}, {Cappi}, {Cimatti},
  {Ilbert}, {Leauthaud}, {Maccagni}, {Marinoni}, {McCracken}, {Memeo},
  {Meneux}, {Oesch}, {Porciani}, {Pozzetti}, {Scaramella}, \&
  {Scarlata}}]{2009A&A...503..379T}
{Tasca}, L.~A.~M., {Kneib}, J., {Iovino}, A., {et~al.} 2009, \aap, 503, 379

\bibitem[{{Temi} {et~al.}(2009){Temi}, {Brighenti}, \&
  {Mathews}}]{2009ApJ...695....1T}
{Temi}, P., {Brighenti}, F., \& {Mathews}, W.~G. 2009, \apj, 695, 1

\bibitem[{{Thomas} {et~al.}(2005){Thomas}, {Maraston}, {Bender}, \& {Mendes de
  Oliveira}}]{2005ApJ...621..673T}
{Thomas}, D., {Maraston}, C., {Bender}, R., \& {Mendes de Oliveira}, C. 2005,
  \apj, 621, 673

\bibitem[{{Tinsley}(1980)}]{1980ApJ...241...41T}
{Tinsley}, B.~M. 1980, \apj, 241, 41

\bibitem[{{Tojeiro} \& {Percival}(2010)}]{2010arXiv1001.2015T}
{Tojeiro}, R. \& {Percival}, W.~J. 2010, ArXiv e-prints

\bibitem[{{Toomre}(1977)}]{1977egsp.conf..401T}
{Toomre}, A. 1977, in Evolution of Galaxies and Stellar Populations, ed. B.~M.
  {Tinsley} \& R.~B. {Larson}, 401

\bibitem[{{Tosi}(2009)}]{2009arXiv0901.1090T}
{Tosi}, M. 2009, ArXiv e-prints

\bibitem[{{Tran} {et~al.}(2005){Tran}, {van Dokkum}, {Franx}, {Illingworth},
  {Kelson}, \& {Schreiber}}]{2005ApJ...627L..25T}
{Tran}, K.-V.~H., {van Dokkum}, P., {Franx}, M., {et~al.} 2005, \apjl, 627, L25

\bibitem[{{Trewhella}(1998)}]{1998MNRAS.297..807T}
{Trewhella}, M. 1998, \mnras, 297, 807

\bibitem[{{V{\"a}is{\"a}nen} {et~al.}(2008){V{\"a}is{\"a}nen}, {Mattila},
  {Kniazev}, {Adamo}, {Efstathiou}, {Farrah}, {Johansson}, {{\"O}stlin},
  {Buckley}, {Burgh}, {Crause}, {Hashimoto}, {Lira}, {Loaring}, {Nordsieck},
  {Romero-Colmenero}, {Ryder}, {Still}, \& {Zijlstra}}]{2008MNRAS.384..886V}
{V{\"a}is{\"a}nen}, P., {Mattila}, S., {Kniazev}, A., {et~al.} 2008, \mnras,
  384, 886

\bibitem[{{van der Wel} {et~al.}(2007){van der Wel}, {Holden}, {Franx},
  {Illingworth}, {Postman}, {Kelson}, {Labb{\'e}}, {Wuyts}, {Blakeslee}, \&
  {Ford}}]{2007ApJ...670..206V}
{van der Wel}, A., {Holden}, B.~P., {Franx}, M., {et~al.} 2007, \apj, 670, 206

\bibitem[{{van Dokkum}(2005)}]{2005AJ....130.2647V}
{van Dokkum}, P.~G. 2005, \aj, 130, 2647

\bibitem[{{van Dokkum} {et~al.}(2000){van Dokkum}, {Franx}, {Fabricant},
  {Illingworth}, \& {Kelson}}]{2000ApJ...541...95V}
{van Dokkum}, P.~G., {Franx}, M., {Fabricant}, D., {Illingworth}, G.~D., \&
  {Kelson}, D.~D. 2000, \apj, 541, 95

\bibitem[{{van Dokkum} {et~al.}(2006){van Dokkum}, {Quadri}, {Marchesini},
  {Rudnick}, {Franx}, {Gawiser}, {Herrera}, {Wuyts}, {Lira}, {Labb{\'e}},
  {Maza}, {Illingworth}, {F{\"o}rster Schreiber}, {Kriek}, {Rix}, {Taylor},
  {Toft}, {Webb}, \& {Yi}}]{2006ApJ...638L..59V}
{van Dokkum}, P.~G., {Quadri}, R., {Marchesini}, D., {et~al.} 2006, \apjl, 638,
  L59

\bibitem[{{Villar} {et~al.}(2008){Villar}, {Gallego}, {P{\'e}rez-Gonz{\'a}lez},
  {Pascual}, {Noeske}, {Koo}, {Barro}, \& {Zamorano}}]{2008ApJ...677..169V}
{Villar}, V., {Gallego}, J., {P{\'e}rez-Gonz{\'a}lez}, P.~G., {et~al.} 2008,
  \apj, 677, 169

\bibitem[{{Waddington} {et~al.}(2002){Waddington}, {Windhorst}, {Cohen},
  {Dunlop}, {Peacock}, {Jimenez}, {McLure}, {Bunker}, {Spinrad}, {Dey}, \&
  {Stern}}]{2002MNRAS.336.1342W}
{Waddington}, I., {Windhorst}, R.~A., {Cohen}, S.~H., {et~al.} 2002, \mnras,
  336, 1342

\bibitem[{{Walcher} {et~al.}(2008){Walcher}, {Lamareille}, {Vergani},
  {Arnouts}, {Buat}, {Charlot}, {Tresse}, {Le F{\`e}vre}, {Bolzonella},
  {Brinchmann}, {Pozzetti}, {Zamorani}, {Bottini}, {Garilli}, {Le Brun},
  {Maccagni}, {Milliard}, {Scaramella}, {Scodeggio}, {Vettolani}, {Zanichelli},
  {Adami}, {Bardelli}, {Cappi}, {Ciliegi}, {Contini}, {Franzetti}, {Foucaud},
  {Gavignaud}, {Guzzo}, {Ilbert}, {Iovino}, {McCracken}, {Marano}, {Marinoni},
  {Mazure}, {Meneux}, {Merighi}, {Paltani}, {Pell{\`o}}, {Pollo}, {Radovich},
  {Zucca}, {Lonsdale}, \& {Martin}}]{2008A&A...491..713W}
{Walcher}, C.~J., {Lamareille}, F., {Vergani}, D., {et~al.} 2008, \aap, 491,
  713

\bibitem[{{Wang}(1991)}]{1991ApJ...383L..37W}
{Wang}, B. 1991, \apjl, 383, L37

\bibitem[{{Weiner} {et~al.}(2005){Weiner}, {Phillips}, {Faber}, {Willmer},
  {Vogt}, {Simard}, {Gebhardt}, {Im}, {Koo}, {Sarajedini}, {Wu}, {Forbes},
  {Gronwall}, {Groth}, {Illingworth}, {Kron}, {Rhodes}, {Szalay}, \&
  {Takamiya}}]{2005ApJ...620..595W}
{Weiner}, B.~J., {Phillips}, A.~C., {Faber}, S.~M., {et~al.} 2005, \apj, 620,
  595

\bibitem[{{Weinzirl} {et~al.}(2009){Weinzirl}, {Jogee}, {Khochfar}, {Burkert},
  \& {Kormendy}}]{2009ApJ...696..411W}
{Weinzirl}, T., {Jogee}, S., {Khochfar}, S., {Burkert}, A., \& {Kormendy}, J.
  2009, \apj, 696, 411

\bibitem[{{Westra} {et~al.}(2010){Westra}, {Geller}, {Kurtz}, {Fabricant}, \&
  {Dell'Antonio}}]{2010ApJ...708..534W}
{Westra}, E., {Geller}, M.~J., {Kurtz}, M.~J., {Fabricant}, D.~G., \&
  {Dell'Antonio}, I. 2010, \apj, 708, 534

\bibitem[{{White} \& {Rees}(1978)}]{1978MNRAS.183..341W}
{White}, S.~D.~M. \& {Rees}, M.~J. 1978, \mnras, 183, 341

\bibitem[{{Wild} {et~al.}(2009{\natexlab{a}}){Wild}, {Walcher}, \&
  {Johansson}}]{2009arXiv0910.1598W}
{Wild}, V., {Walcher}, C.~J., \& {Johansson}, P.~H. 2009{\natexlab{a}}, ArXiv
  e-prints

\bibitem[{{Wild} {et~al.}(2009{\natexlab{b}}){Wild}, {Walcher}, {Johansson},
  {Tresse}, {Charlot}, {Pollo}, {Le F{\`e}vre}, \& {de
  Ravel}}]{2009MNRAS.395..144W}
{Wild}, V., {Walcher}, C.~J., {Johansson}, P.~H., {et~al.} 2009{\natexlab{b}},
  \mnras, 395, 144

\bibitem[{{Wolf} {et~al.}(2009){Wolf}, {Arag{\'o}n-Salamanca}, {Balogh},
  {Barden}, {Bell}, {Gray}, {Peng}, {Bacon}, {Barazza}, {B{\"o}hm}, {Caldwell},
  {Gallazzi}, {H{\"a}u{\ss}ler}, {Heymans}, {Jahnke}, {Jogee}, {van Kampen},
  {Lane}, {McIntosh}, {Meisenheimer}, {Papovich}, {S{\'a}nchez}, {Taylor},
  {Wisotzki}, \& {Zheng}}]{2009MNRAS.393.1302W}
{Wolf}, C., {Arag{\'o}n-Salamanca}, A., {Balogh}, M., {et~al.} 2009, \mnras,
  393, 1302

\bibitem[{{Wu} {et~al.}(2001){Wu}, {Faber}, \& {Lauer}}]{2001defi.conf..170W}
{Wu}, K.~L., {Faber}, S.~M., \& {Lauer}, T.~R. 2001, in Deep Fields, ed.
  S.~{Cristiani}, A.~{Renzini}, \& R.~E. {Williams}, 170

\bibitem[{{Xu} {et~al.}(2007){Xu}, {Zhang}, {Chang}, \&
  {Liu}}]{2007astro.ph..1490X}
{Xu}, L., {Zhang}, C., {Chang}, B., \& {Liu}, H. 2007, ArXiv Astrophysics
  e-prints

\bibitem[{{Yan} \& {Thompson}(2003)}]{2003ApJ...586..765Y}
{Yan}, L. \& {Thompson}, D. 2003, \apj, 586, 765

\bibitem[{{Yang} {et~al.}(2009){Yang}, {Hammer}, {Flores}, {Puech}, \&
  {Rodrigues}}]{2009A&A...501..437Y}
{Yang}, Y., {Hammer}, F., {Flores}, H., {Puech}, M., \& {Rodrigues}, M. 2009,
  \aap, 501, 437

\bibitem[{{Yoshii} \& {Takahara}(1988)}]{1988ApJ...326....1Y}
{Yoshii}, Y. \& {Takahara}, F. 1988, \apj, 326, 1

\bibitem[{{Young} \& {Scoville}(1982)}]{1982ApJ...260L..41Y}
{Young}, J.~S. \& {Scoville}, N. 1982, \apjl, 260, L41

\bibitem[{{Zezas} {et~al.}(2003){Zezas}, {Ward}, \&
  {Murray}}]{2003ApJ...594L..31Z}
{Zezas}, A., {Ward}, M.~J., \& {Murray}, S.~S. 2003, \apjl, 594, L31

\bibitem[{{Zhao} \& {Newberg}(2006)}]{2006astro.ph.12034Z}
{Zhao}, C. \& {Newberg}, H.~J. 2006, ArXiv Astrophysics e-prints

\bibitem[{{Zucca} {et~al.}(2006){Zucca}, {Ilbert}, {Bardelli}, {Tresse},
  {Zamorani}, {Arnouts}, {Pozzetti}, {Bolzonella}, {McCracken}, {Bottini},
  {Garilli}, {Le Brun}, {Le F{\`e}vre}, {Maccagni}, {Picat}, {Scaramella},
  {Scodeggio}, {Vettolani}, {Zanichelli}, {Adami}, {Arnaboldi}, {Cappi},
  {Charlot}, {Ciliegi}, {Contini}, {Foucaud}, {Franzetti}, {Gavignaud},
  {Guzzo}, {Iovino}, {Marano}, {Marinoni}, {Mazure}, {Meneux}, {Merighi},
  {Paltani}, {Pell{\`o}}, {Pollo}, {Radovich}, {Bondi}, {Bongiorno},
  {Busarello}, {Cucciati}, {Gregorini}, {Lamareille}, {Mathez}, {Mellier},
  {Merluzzi}, {Ripepi}, \& {Rizzo}}]{2006A&A...455..879Z}
{Zucca}, E., {Ilbert}, O., {Bardelli}, S., {et~al.} 2006, \aap, 455, 879

\end{thebibliography}

\end{document}